\documentclass[aasmacros]{aa}
\usepackage{amsmath}
\usepackage{wasysym}
\usepackage{marvosym}
\usepackage{txfonts}
\usepackage{placeins}
\usepackage{siunitx}
\usepackage{enumitem}
\DeclareSIUnit[]\astronomicalunit{\text{au}}

\usepackage{natbib}
\bibpunct{(}{)}{;}{a}{}{,} % to follow the A&A style

\usepackage{graphicx}
\usepackage[dvipsnames,table]{xcolor}
\usepackage{subcaption}

\usepackage{scalerel}
\usepackage{tikz}
\usetikzlibrary{svg.path}

\definecolor{orcidlogocol}{HTML}{A6CE39}
\tikzset{
  orcidlogo/.pic={
    \fill[orcidlogocol] svg{M256,128c0,70.7-57.3,128-128,128C57.3,256,0,198.7,0,128C0,57.3,57.3,0,128,0C198.7,0,256,57.3,256,128z};
    \fill[white] svg{M86.3,186.2H70.9V79.1h15.4v48.4V186.2z}
                 svg{M108.9,79.1h41.6c39.6,0,57,28.3,57,53.6c0,27.5-21.5,53.6-56.8,53.6h-41.8V79.1z M124.3,172.4h24.5c34.9,0,42.9-26.5,42.9-39.7c0-21.5-13.7-39.7-43.7-39.7h-23.7V172.4z}
                 svg{M88.7,56.8c0,5.5-4.5,10.1-10.1,10.1c-5.6,0-10.1-4.6-10.1-10.1c0-5.6,4.5-10.1,10.1-10.1C84.2,46.7,88.7,51.3,88.7,56.8z};
  }
}

\newcommand\orcidicon[1]{\href{https://orcid.org/#1}{\mbox{\scalerel*{
\begin{tikzpicture}[yscale=-1,transform shape]
\pic{orcidlogo};
\end{tikzpicture}
}{|}}}}

\usepackage{hyperref}
\hypersetup{draft=False, colorlinks=true, citecolor=blue, urlcolor=blue, linkcolor=red, linktoc=all}

\usepackage{pifont}

\DeclareMathOperator\erf{erf}
\DeclareMathOperator{\sign}{sign}

\def\mearth{M_\oplus}

\def\msun{M_\odot}

\def\f1{f_{\rm I}}
\def\mstar{M_*}

\def\beq{\begin{equation}}
\def\eeq{\end{equation}}

\def\t2{\tau_{\rm II}}

\def\sigmas0{\Sigma_{\rm s,0}}
\def\mj{M_{\textrm{\tiny \jupiter }}}

\def\s0{S_0}

\def\apjl{ApJ}                % Astrophysical Journal, Letters
\def\apjs{ApJS}               % Astrophysical Journal, Supplement
\def\ao{Appl.Optics}          % Applied Optics
\def\aap{A\&A}                % Astronomy and Astrophysics
\def\({\left(}
\def\){\right)}
\def\<{\left<}
\def\>{\right>}

\defcitealias{2018MNRAS.475.5618B}{B18}
\defcitealias{Kimura2016}{K16}
\defcitealias{2017MNRAS.469.3813H}{HK17}
\defcitealias{2008ApJ...685..560D}{DL08}
\defcitealias{2010ApJ...724..730D}{DL10}
\defcitealias{2019MNRAS.486.4398F}{FNS19}
\defcitealias{1986ApJ...309..846L}{LP86}
\defcitealias{2002MNRAS.334..248A}{A02}
\defcitealias{2015A&A...582A.112B}{BLJ15}

\begin{document}

\title{Calibrated gas accretion and orbital migration of protoplanets in 1D disc models$^\star$}

   \author{O. Schib\inst{\ref{bern},\ref{uzh}}
          \and
          C.~Mordasini\inst{\ref{bern}}
          \and
          R.~Helled\inst{\ref{uzh}}
          }

   \institute{
Physikalisches Institut,
Universit\"at Bern, Gesellschaftsstrasse 6, 3012 Bern, Switzerland\\
\email{oliver.schib@space.unibe.ch}
\label{bern}
\and
Center for Theoretical Astrophysics and Cosmology,
Institute for Computational Science,\\
Universit\"at Z\"urich, Winterthurerstrasse~190, 8057~Z\"urich, Switzerland
\label{uzh}
             }

\date{Received 29 July 2021 / Accepted May 3 2022}
%\textcolor{ForestGreen}{}

\abstract
%Context
{Orbital migration and gas accretion are two interdependent key processes that govern the evolution of planets in protoplanetary discs. The final planetary properties such as masses and orbital periods strongly depend  on the treatment of those two processes.}
%Aims
{Our aim is to develop a simple prescription for migration and accretion in 1D disc models, calibrated with results of 3D hydrodynamic simulations. Our focus lies on non-self-gravitating discs, but we also discuss to what degree our prescription could be applied when the discs are self-gravitating.}
%Methods
{We studied migration using torque densities. Our model for the torque density is based on existing fitting formulas, which we subsequently modify to prevent premature gap-opening. At higher planetary masses, we also apply torque densities from hydrodynamic simulations directly to our 1D model. These torque densities allow us to model the orbital evolution of an initially low-mass planet that undergoes runaway-accretion to become a massive planet. The two-way exchange of angular momentum between disc and planet is included. This leads to a self-consistent treatment of gap formation that only relies on directly accessible disc parameters.

We present a formula for Bondi  and Hill  gas accretion in the disc-limited regime. This formula is self-consistent in the sense that mass is removed from the disc in the location from where it is accreted. The prescription is  appropriate when the planet is smaller than, comparable to, or larger than the disc scale height.}
%Results
{We find that the resulting evolution in mass and semi-major axis in the 1D framework is in good agreement with those from 3D hydrodynamical simulations for a range of parameters.}
%Conclusions
{Our prescription is valuable for simultaneously modelling migration and accretion in 1D models, which allows  a planet's evolution to be followed over the entire lifetime of a disc. It is applicable also in situations where the surface density is significantly disturbed by multiple gap-opening planets or processes like infall.
We conclude that it is appropriate and beneficial to apply torque densities from hydrodynamic simulations in 1D models, at least in the parameter space we study here.
More work is needed  in order to determine whether our approach is also applicable in an even wider parameter space and in situations with more complex disc thermodynamics, or when the disc is self-gravitating.}

\keywords{protoplanetary disks -- accretion, accretion disks -- planets and satellites: formation} 

\titlerunning{Migration and Accretion in 1D models}

\maketitle

\section{Introduction}\label{sect:Introduction}

\begingroup
\renewcommand*{\thefootnote}{$\star$}
\footnotetext[1]{Torque data are only available at the CDS via anonymous ftp to \href{http://cdsarc.u-strasbg.fr}{cdsarc.u-strasbg.fr} (\href{ftp://130.79.128.5}{130.79.128.5}) or via \href{http://cdsarc.u-strasbg.fr/viz-bin/cat/J/A+A/664/A138}{http://cdsarc.u-strasbg.fr/viz-bin/cat/J/A+A/664/A138}}
\endgroup
Many exoplanets are observed in locations where they likely did not form \citep{2012ARA&A..50..211K}. The dynamical evolution of planets embedded in protoplanetary discs is a key process in understanding planet formation \citep{1997Icar..126..261W}. For example, the existence of giant planets in mean-motion resonance implies that migration takes place \citep{2013LNP...861..201B}.
In addition, the detection of hot Jupiters like  51 Pegasus b \citep{1995Natur.378..355M} has raised the question of  how they got to their present day locations as in situ formation is very challenging.
However the formation of hot Jupiters is debated. Gas disc migration may explain the formation of parts of the observed population.
Orbital migration is also expected to be important for the formation of warm Jupiters, giant planets close to their host star, yet further away than hot Jupiters  \cite[e.g.][]{2018arXiv181209264A,2018ARA&A..56..175D}. 
Fast migration, however, is not sufficient to explain the observed exoplanet locations. Processes that can slow down or halt migration are equally important, otherwise planets would tend to migrate all the way inwards and merge with their host star.  One important process that can slow down orbital migration is gap formation. By interacting tidally with the disc a planet can reduce the surface density in its orbit substantially, which in turn decreases  its migration rate.
Broadly speaking, migration can be divided into two major regimes. As long as the planet has a mass lower than $\approx$ \SI{10}{\mearth} to \SI{100}{\mearth} (the exact value  depends on the disc's properties and the stellar mass), it does not perturb the disc significantly. These low masses correspond to  type~I migration, which can be very fast (timescales of \num{e3} to \SI{e4}{yr}). More massive planets perturb the disc, pushing disc material away from their orbits, and therefore  open a gap in the disc. Gap opening results in slower migration, a regime known as type~II migration. In this regime, migration proceeds more slowly, typically on a \SI{e5}{yr} timescale.

This is, however, a strongly simplified picture. A deeper understanding of planetary migration can only be gained by studying further processes. In the type~I regime the total torque can be written as a sum of Lindblad and corotation torques. The Lindblad torque results from the exchange of angular momentum at Lindblad resonances in the disc and usually leads to inward migration \citep{goldreichward1973,2002ApJ...565.1257T}. Corotation torques are a result of the gravitational interaction of the planet with disc material in the corotation region. They can be directed both inwards and outwards, and can even be strong enough to cause a net outward migration of the planet. Here the thermodynamics of the disc are important. If the disc does not cool efficiently, the net torque can be outward \citep{2006A&A...459L..17P}. This is relevant only for lower mass planets, however, since the corotation torque saturates at sufficiently high masses and the outward migration cannot be sustained \citep{Paardekooper2011b}. These effects can be studied by calculating torques in isothermal \citep{2002ApJ...565.1257T} or non-isothermal \citep{Paardekooper2010} discs, and the resulting torques can also be applied  in 1D models. It is necessary, however,  to confirm these results in 2D and 3D hydrodynamic simulations. In the following we give some examples.
\citet{massetcasoli2010} calculate formulae of the saturated type~I torque and perform 2D simulations to verify their results.
\citet{Paardekooper2011b} provide a formula for the  non-isothermal type~I torque that includes the effects of diffusion and show that the torque agrees well with their 2D hydrodynamic simulations.
\citet{2008A&A...487L...9K} perform 2D simulations of embedded planets and include heating, cooling, and radiative diffusion to study the  magnitude and direction of migration.
\citet{2011MNRAS.416.1971B} perform 2D hydrodynamic simulations of migrating planets in self-gravitating discs. They show that planets formed by fragmentation in the outer disc are likely to migrate inwards very rapidly in type~I.
\citet{2006A&A...459L..17P} perform 3D hydrodynamic simulations to study the effect of a proper energy balance on the interaction of a low-mass planet with its disc. They demonstrate that migration can be directed outwards if the disc's opacity  is high enough.
Another detailed 3D numerical study of the effect of outward migration can be found in \citet{kleybitsch2009}. The authors demonstrate that this effect can be even stronger than in 2D when the same opacity law is used.

There is a wealth of  literature on these topics. Reviews can be found in \citet{lubowida2010}, \citet{2012ARA&A..50..211K}, \citet{2013LNP...861..201B}, \citet{Baruteau2014}, and \citet{2016SSRv..205...77B}.

Migration is an inherently three-dimensional process and it has been investigated extensively including 2D and 3D hydrodynamic simulations,\ as described. The various analytic formulae for the migration rate that have resulted from these studies can be used in 1D models. In the type~I regime, these typically depend on the slopes of surface density and temperature in the disc. In a disc with multiple planets carving deep or partial gaps, these quantities are very hard to determine. Furthermore, the torque exerted on the disc material by the planet is not included in this picture. While this is a good approximation for low-mass planets, it does not allow for a gap to be formed. In our work we model the gap formation explicitly by modelling the evolution of the disc's surface density in reaction to the presence of a growing massive planet.

In the classical picture of type~II migration, planets migrate on a viscous timescale. This is not always the case, however. Mass can also cross the gap and migration may be faster or slower than the viscous inflow velocity \citep{2015A&A...574A..52D}.  It is still very challenging to  model  the transition between type~I and type~II migration. While conditions for gap opening have been derived (e.g. \citealt{2006Icar..181..587C}), if the planet's migration timescale is shorter than the gap-opening timescale, a gap may never open. In massive discs, an additional very rapid runaway type of migration may arise: type~III migration (\citealt{2003ApJ...588..494M}, see also \citealt{2015ApJ...802...56M}).

A key feature of the observed exoplanet population is its diversity. Any successful planet formation theory must explain this diversity, which is only possible statistically. For computational reasons, the necessary parameter studies and population synthesis calculations can only be performed using low-dimensional models. One class of such models that is often often applied are the 1D vertically integrated models (e.g. \citealt{1994ApJ...421..640N,2005A&A...442..703H}). The advantage of these models is their low computational cost; the evolution of protoplanetary discs can be studied from formation to dispersal in a large parameter space.
Modelling migration in 1D has its limitations.  Dynamical processes, like the co-orbital dynamics, cannot be modelled directly in 1D since they are inherently three-dimensional. Therefore, it may be necessary to introduce additional approximations or parametrisations that are valid only in some regimes.
At the same time, the location in the disc where gaps form is paramount for the predicted population of planets.
A few simple prescriptions, for example for the depth and shape of the gap \citep{2017PASJ...69...97K,2014ApJ...782...88F} and the migration process \citep{2020MNRAS.494.5666I} have been suggested. The difficulty with these approaches is that they often rely on the knowledge of the unperturbed surface density near the planet. In practice, this property is only known when the disc evolution starts from well-defined initial conditions. After a period of evolution, or when the disc formation is included through infall from the molecular cloud core, or when several gap-opening planets perturb the disc, the unperturbed surface density is meaningless or difficult to determine.

The goal of this paper is to derive a prescription for migration and accretion in 1D disc models, calibrated by the results of 3D hydrodynamic simulations. In order to overcome the difficulties described above, this prescription uses torque densities and relies only on directly accessible disc parameters, and self-consistently models the exchange of mass and angular momentum. We do not attempt to reproduce the exact gap shapes found in hydrodynamic simulations. To what degree this is possible when considering torque densities in the context of accreting migrating planets is currently unclear. We hope to address this important topic in the future.

The paper is organised as follows. In Sect.~\ref{sect:Model} we describe the numerical  set-up. Section~\ref{sec:param} discusses the investigated parameter space and our initial conditions. In Sect.~\ref{sect:Results} we present our results in comparison with those from hydrodynamic simulations.
In Sect. \ref{Sect:CompLit} we perform a comparison with a different prescription found in the literature. Section~\ref{sect:Discussion} contains a discussion of the results and the limitations of the model. In Sect.~\ref{sect:Conclusions} we summarise and  conclude.

\section{Model description}\label{sect:Model}
We apply the disc formation and evolution model from \citet{2021A&A...645A..43S}. It describes the temporal evolution of a rotationally symmetric 1D disc of gas, described by the vertically integrated surface density $\Sigma \equiv \Sigma(r,t)$.
Here we introduce a planet on a circular orbit.
We use cylindrical polar coordinates, with $r$ denoting the radial direction. The disc's mid-plane is located at $z=\SI{0}{au}$. In this section we briefly review the model in the form in which it is applied in this work.
 We note that we use the term `planet' throughout this work for simplicity, even though `protoplanet' or `proto-brown dwarf' may be more appropriate in some cases.

\subsection{Disc evolution}

The evolution of a  protoplanetary disc is described by the viscous evolution equation (\citealp{1952ZNatA...7...87L} and \citealp{1974MNRAS.168..603L}), adding the effects of angular momentum injection by a planet and mass sinks
\begin{equation}
\label{eq_evo}
\frac{\partial \Sigma}{\partial t} = \frac{3}{r} \frac{\partial}{\partial r} \left[ r^{1/2} \frac{\partial}{\partial r} \left( \nu \Sigma r^{1/2} \right) - \frac{2 \Lambda \Sigma}{\Omega}\right] - S_\mathrm{acc},
\end{equation}
where $S_\mathrm{acc}$ is a sink term for the mass accreted by planets (other sink terms such as photoevaporation are not considered here) and  $\Omega$ denotes the Keplerian angular frequency of the disc. The expression for $S_\mathrm{acc}$ is given in Sect.~\ref{subs_acc}. The term $2 \Lambda \Sigma / \Omega$  describes the gravitational interaction between the planet and the disc, leading to angular momentum exchange. The torque density distribution $\Lambda \equiv - d \mathcal{T} / d m$ used in this work is discussed in Sect.~\ref{subs_mig}. In our convention, $\mathcal{T} = 2 \pi \int_0^\infty\!\frac{d\mathcal{T}}{dm}(r)\Sigma(r) r\,\mathrm{d}r$ is the total torque on the planet.
Equation~\ref{eq_evo} is solved on a grid of \num{2800} logarithmically spaced cells extending from \num{0.03} to \SI{30000}{au}. We performed a resolution test where we increase or decrease the resolution by a factor of two and found that it has a negligible effect on our results. We use the solver from \citet{2010A&A...513A..79B}.

We assume hydrostatic equilibrium in the vertical direction and define the pressure scale height $H$ through
\begin{equation}
\label{eq_vert}
    \rho(r,z)=\rho_0(r)\, \mathrm{exp}\,\left(-\frac{z^2}{2H(r)^2}\right).
\end{equation}
Here, $\rho$ is the density in the disc, and its midplane value $\rho_0(r)$ is related to $\Sigma(r)$ through
\begin{equation}
    \Sigma(r) = \rho_0(r) H(r) \sqrt{2\pi}.
\end{equation}
In this paper we assume a constant aspect ratio $H/r$. The disc's scale height is related to the isothermal sound speed
\begin{equation}
    c_\mathrm{s}=\sqrt{\frac{k_\mathrm{B}T}{\mu \mathrm{u}}}
\label{eq_cs}
\end{equation}
through
\begin{equation}
    H = \frac{c_\mathrm{s}}{\Omega}.
\label{eq_H}
\end{equation}
In Eq.~\ref{eq_cs} $k_\mathrm{B}$ is the Boltzmann constant, $\mu \equiv 2.3$ the mean molecular weight, and $\mathrm{u}$ the atomic mass unit.

\subsection{Gas accretion}\label{subs_acc}

We study gas accretion in the disc-limited regime (accretion is limited by the disc's mass reservoir, but not the planet's ability to accrete).
For low-mass planets embedded in a disc, their Kelvin-Helmholtz contraction \citep{2001ApJ...553..999I} limits gas accretion (for example Sect.~3.8.1 in \citealt{mordasinialibert2012b}). In this phase our model represents an upper limit.

Our accretion model is based on the Bondi and Hill accretion described in \citealt{2008ApJ...685..560D} (DL08). The authors study migrating and accreting planets using 3D nested-grid hydrodynamical simulations of locally isothermal discs. We note that the locally isothermal assumption is likely to influence the gas accretion. However, it is still unclear whether
a departure from isothermality increases or  decreases  the accretion rate. For example, \citet{ayliffebate2009} find that accretion rates are highest in a locally isothermal disc, while \citet{machidakokubo2010} find the opposite, although both of these studies perform radiation hydrodynamics simulations of accretion.
Based on their numerical simulations, \citealt{2008ApJ...685..560D} parametrise the accretion rate onto a planet with mass $M_p$ in their Eq.~9:
\begin{equation}\label{eq_macc}
    \dot{M}_p \sim \frac{\Sigma}{H}\Omega R_f^3.
\end{equation}
Here $R_f$ is the feeding zone radius, taken to be either the Bondi radius $R_\mathrm{B} = G M_p/c_\mathrm{s}^2$ or the Hill radius $R_\mathrm{H} = r_p \left[ M_p/\left(3 \mstar\right)\right]^{1/3}$, where the sound speed is evaluated at the planet's semi-major axis  $r_p$. We assume circular orbits. The inverse growth timescales $\dot{M}_p/M_p$ (equal to $\tau_\mathrm{B}^{-1}$ in the Bondi regime, and $\tau_\mathrm{H}^{-1}$ in the  Hill regime) then become
\begin{equation}\label{eq_Bondi}
    \frac{1}{\tau_\mathrm{B}} = C_\mathrm{B} \Omega \frac{\Sigma r^2}{\mstar}\left(\frac{r_p}{H}\right)^7\left(\frac{M_p}{\mstar}\right)^2,
\end{equation}
\begin{equation}\label{eq_Hill}
    \frac{1}{\tau_\mathrm{H}} = \frac{1}{3}C_\mathrm{H}\Omega \frac{\Sigma r^2}{\mstar}\left(\frac{r_p}{H}\right),
\end{equation}
where $C_\mathrm{B}$ and $C_\mathrm{H}$ are dimensionless coefficients of order unity.
\citetalias{2008ApJ...685..560D} found $C_\mathbf{B} = \num{2.6}$ and $C_\mathrm{H} = \num{0.89}$ to agree best with their hydrodynamic simulations. Since our prescription of accretion differs from theirs, we  have to use different values for $C_\mathrm{B}$ and $C_\mathrm{H}$, as discussed later.
The overall inverse growth timescale is defined as
\begin{equation}
    1/\tau_G = 
    \begin{cases}
        1/\tau_\mathrm{B} & M_p < M_t \\
        1/\tau_\mathrm{H} & M_p \ge M_t,
    \end{cases}
\end{equation}
where
\begin{equation}\label{eq_mt}
    M_t = \frac{\mstar}{\sqrt{3}}\sqrt{\frac{C_\mathrm{H}}{C_\mathrm{B}}}\left(\frac{H}{r_p}\right)^3
\end{equation}
is the transition mass when $\tau_\mathrm{H}=\tau_\mathrm{B}$.

\citet{2012ApJ...746..110Z} perform 2D  hydrodynamic simulations of self-gravitating protostellar discs with infall.
They find that the accretion rate $\dot{M}_C$ onto their clumps agrees reasonably well with their Eq.~14:
\begin{equation}
\label{eq_macc2d}
    \dot{M}_c = 4 \Sigma \Omega R_\mathrm{H}^2.
\end{equation}
This expression agrees with Eq.~\ref{eq_Hill} up to a factor $4/C_\mathrm{H}$ when $R_\mathrm{H}=H$.
We can therefore use this accretion scheme for clumps and  for compact planets, and irrespective of whether the disc is self-gravitating and/or subject to infall. We note, however, that the size of the feeding zone is likely to be smaller in the self-gravitating regime \citep{2013ApJ...767...63S}.

We next apply Eq.~\ref{eq_macc} to our 1D model, refining the approach. Instead of using global values of $\Sigma$ and $\Omega$, we calculate the contributions from each grid cell inside the feeding zone separately. The accreted mass is then removed self-consistently from the disc at the correct location. We obtain the following accretion rate:
\begin{equation}\label{eq_mdot}
\begin{aligned}
    \dot{M}_\mathrm{gas} & = 2\,C_\mathrm{B,H} \int_{r_p-R_f}^{r_p+R_f}\!\int_{0}^{\sqrt{R_f^2-(r-r_p)^2}}\!\rho(r,z)\,\mathrm{d}z\,v_\mathrm{rel}\,\mathrm{d}r\\
    &= \sqrt{2 \pi}\,C_\mathrm{B,H} \int_{r_p-R_f}^{r_p+R_f}\!\rho_0(r) H(r) \erf{\left(\frac{\sqrt{R_f^2-(r-r_p)^2}}{\sqrt{2} H(r)}\right)}v_\mathrm{rel}\,\mathrm{d}r.
\end{aligned}
\end{equation}
In the last step we substituted $\rho(r,z)$ from Eq.~\ref{eq_vert}.
The factor $2$ at the beginning takes into account that we only integrate over half of the circular feeding area. The coefficient $C_\mathrm{B,H}$ is equal to $C_\mathrm{B}$ in the Bondi regime and $C_\mathrm{H}$ in the Hill regime.
The expression $\sqrt{R_f^2-(r-r_p)^2}$ denotes the distance from the mid-plane to the top or bottom of the feeding zone in $z$-direction. 

\begin{figure}
  \includegraphics[width=\linewidth]{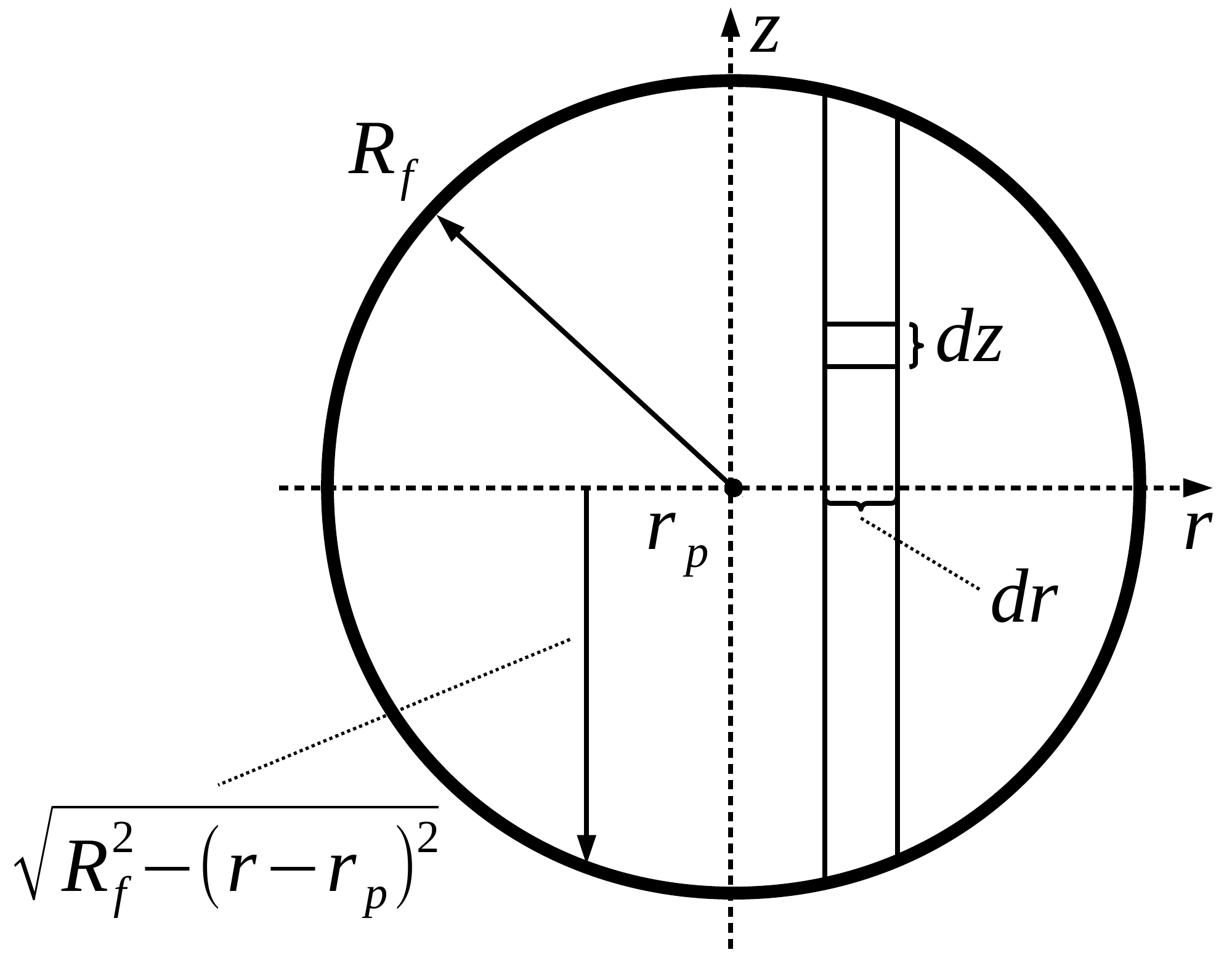}
  \caption{Schematic of the accretion geometry. The feeding zone is centred at the location of the planet $r_p$, at $z=0$. The direction tangential to the planet's motion is perpendicular to the page.}
    \label{fig_schem}
\end{figure}

A schematic of the geometry is given in Fig.~\ref{fig_schem}. The quantity 
$v_\mathrm{rel} = \left|r \Omega - r_p \Omega_p\right|$ is the gas' velocity  relative to the planet. The presence of the planet introduces a perturbation to the Keplerian flow of the gas. Therefore, using $\Omega$ in the expression for $v_\mathrm{rel}$ is potentially problematic near the planet (e.g. \citealp{lubowseibert1999}). However, the form of Eqs.~\ref{eq_macc} and \ref{eq_macc2d}, and the agreement of these expressions with hydrodynamic simulations, suggests that the total accretion rate should not be very different. When the planet mass is low there is no strong perturbation of the disc. For massive planets the surface density near the planet is depleted and does not contribute significantly to the accretion. Therefore,  this assumption is not expected to affect  our results. We note that if we tried to reproduce gap shapes from hydrodynamic simulations accurately, these aspects would become important. The remaining integral in Eq.~\ref{eq_mdot} is evaluated numerically. The corresponding mass is removed from the disc and added to the planet in every grid cell separately at each time step.
Comparing Eq.~\ref{eq_mdot} with the requirement $\dot{M}_\mathrm{gas} = 2 \pi \int_{r_p-R_f}^{r_p+R_f} \! r S_\mathrm{acc} \, \mathrm{d}r$ gives the sink term in Eq.~\ref{eq_evo}:
\begin{equation}
        S_\mathrm{acc}(r,t) = \frac{1}{\sqrt{2 \pi}} \frac{H}{r} \rho_0(r) C_\mathrm{B,H} \erf{\left(\frac{\sqrt{R_f^2-(r-r_p)^2}}{\sqrt{2} H}\right)} v_\mathrm{rel}.
\end{equation}

Equations \ref{eq_macc} \citepalias{2008ApJ...685..560D} and \ref{eq_macc2d} \citep{2012ApJ...746..110Z} make the implicit assumption that the planet's radius is much smaller or larger, respectively, than the disc's scale height. In the first case the cross section is $\pi R_f^2$; in the second it is $2 R_f H$. These expressions do not apply when the planet's size is comparable to the scale height. We solve this problem by using the analytic expression for the vertical structure (Eq.~\ref{eq_vert}) in the calculation of $\dot{M}_\mathrm{gas}$. Equation~\ref{eq_mdot} is thus valid when the planet is smaller, larger, or comparable to the disc scale height.

In the Hill regime we studied two different configurations for the feeding zone radius: $R_f = R_\mathrm{H}$ and $R_f = 3 R_\mathrm{H}$. We found that $R_f = R_\mathrm{H}$ depletes the feeding zone too quickly, and that $R_f = 3 R_\mathrm{H}$ reduces this effect (see Sect.~\ref{sec_resbase}). Using a different feeding zone radius means the $C_\mathrm{H}$ also needs to be different in order to obtain the same inverse growth timescale. The values we adopted for the different configurations studied are given in Table~\ref{table_config}.
In order to prevent unphysical effects due to a discontinuous jump in $R_f$, we apply a smooth transition between the Bondi and Hill regimes.

\begin{table}
\centering
\begin{tabular}{ccccc}
\hline\hline
\begin{tabular}[c]{@{}c@{}}Configuration\end{tabular} &
\begin{tabular}[c]{@{}c@{}}$R_{f,\mathrm{Hill}}$\end{tabular} &
\begin{tabular}[c]{@{}c@{}}$C_\mathrm{B}$\end{tabular} &
\begin{tabular}[c]{@{}c@{}}$C_\mathrm{H}$\end{tabular} &
\begin{tabular}[c]{@{}c@{}}Torque model\end{tabular} \\
\hline
\citetalias{2010ApJ...724..730D}-torque & $R_\mathrm{H}$ & 10.0 & 4.3 & $\mathcal{F}(x,0.5,1)$  \\
$R_f = 3 R_\mathrm{H}$ & $3 R_\mathrm{H}$ & 10.0 & 0.19 & $\mathcal{F}(x,0.5,1)$ \\
Torque mod             & $3 R_\mathrm{H}$ & 10.0 & 0.19 & $\mathcal{F},$ modified  \\
\rowcolor{CornflowerBlue!40}
High mass torque       & $3 R_\mathrm{H}$ & 10.0 & 0.19 & \citetalias{2010ApJ...724..730D} (interpol.)  \\
\citetalias{1986ApJ...309..846L}-formula & $3 R_\mathrm{H}$ & 10.0 & 0.19 & Lin\&Papa. 1986  \\
\citetalias{2002MNRAS.334..248A}-formula & $3 R_\mathrm{H}$ & 10.0 & 0.205 & Armitage+ 2002  \\
Type~I & $R_\mathrm{H}$ & 10.0 & 4.1 & Tanaka+ 2002  \\
\hline
\end{tabular}
\caption{Different configurations for accretion and migration used in this work. $R_{f,\mathrm{Hill}}$ is the feeding zone radius in the Hill regime, $C_\mathrm{B}$ and $C_\mathrm{H}$ are dimensionless coefficients we determine by comparing our inverse growth timescales with those obtained in \citetalias{2008ApJ...685..560D}. For practical use of our prescription, we recommend using our High mass torque model flagged in light blue.}
\label{table_config}
\end{table}

\subsection{Migration}
\label{subs_mig}

Our description of migration is motivated by the impulse approximation \citep{1979MNRAS.188..191L,1979MNRAS.186..799L,1986ApJ...309..846L}. This ensures a two-way angular momentum exchange between disc and planet. Instead of using the classical (analytical) impulse approximation, we apply a more modern formalism. We investigated how the classical impulse approximation, as well as a modified version, behave in Appendix~\ref{app_ia}. \citealt{2010ApJ...724..730D} (DL10) perform 3D hydrodynamic simulations of locally isothermal discs to study the type~I migration torque. We follow them in assuming that the torque distribution per unit disc mass has the form
\begin{equation}
        \frac{d \mathcal{T}}{dm} = \mathcal{F} \left(x, \beta, \zeta\right) \Omega(r_p)^2 r_p^2 \left(\frac{M_p}{\mstar}\right)^2 \left(\frac{r_p}{H(r_p)}\right)^4,
\label{eq_torque}
\end{equation}
where $\mathcal{F}$ is a dimensionless function describing the torque's shape, and the radial scaling factor is
\begin{equation}
x = \frac{r-r_p}{\mathrm{max}(H,R_\mathrm{H})}.
\end{equation}
The parameters $\beta$ and $\zeta$ are the gradients of the surface density and temperature, respectively, such that $\Sigma \propto r^{-\beta}$ and $T \propto r^{-\zeta}$.
\citetalias{2010ApJ...724..730D} find an analytical fit for $\mathcal{F}$:
\begin{equation}
\label{eq_f}
        \mathcal{F}(x,\beta,\zeta) = \left[ p_1 e^{-(x+p_2)^2/p_3^2} + p_4 e^{-(x-p_5)^2/p_6^2} \right] \tanh{(p_7-p_8 x)}.
\end{equation}
In Eq.~\ref{eq_f}  $p_i\equiv p_i(\beta,\zeta)$, $i \in \lbrace 1...8 \rbrace$ are parameters that \citetalias{2010ApJ...724..730D} find by fitting to their hydrodynamic simulations for a large grid of disc parameters. The values we use in this study are listed in Table~\ref{table_p}. The first column for $(\beta,\zeta) = (0.5,1)$ contains the values appropriate for our comparison with \citetalias{2008ApJ...685..560D} (see Sect.~\ref{sec:param}). The second column for $(\beta,\zeta) = (1,0.5)$ is obtained using a linear interpolation from the values for a combination of $(\beta, \zeta)$ of $(1,0)$, $(0.5,1),$ and $(1.5,1)$ since no parameters are available for $(1,0.5)$ in \citetalias{2010ApJ...724..730D}. We used this second set of parameters for our comparison with \citet{2019MNRAS.486.4398F} (FNS19) in Sect.~\ref{sec_reslase}.

Using these fitting parameters seems to defeat our goal to find a prescription of migration independent of the slopes in $\Sigma$ and $T$. However, in practice it is sufficient to choose the fitting parameters approximately adequate to the simulation performed since the torque density distributions do not depend very sensitively on the precise values of the slopes (see Section~6 in \citetalias{2010ApJ...724..730D}). We also tested whether this is the case by varying either $\beta$ or $\zeta$ by $\pm 50 \%$ or $25 \%$, respectively and found that the difference in semi-major axis migrated is at most $12 \%$. This sensitivity calculation can be found in Appendix~\ref{app_slope}. Therefore, our prescription can also be used  for discs with with $(\beta, \zeta) \ne (1,0)$, $(0.5,1)$, as long as the difference does not exceed a few tens of percent. This also depends   on the required precision. As an example, Appendix~\ref{fig_app_slopetest} discusses the impact of changing $\beta$ or $\zeta$ by \num{0.25} in either direction.
We note that in non-isothermal discs the dependency of migration on  the temperature slope, for example,  is no longer small. Unfortunately, grids of torque densities like those given in \citetalias{2010ApJ...724..730D} for locally isothermal discs are not available for non-isothermal discs. We plan to improve our prescription if such torque densities become available in the future.
\begin{table}
\centering
\begin{tabular}{cll}
\hline\hline
\begin{tabular}[c]{@{}c@{}}($\beta$,$\zeta$)\end{tabular} &
\begin{tabular}[c]{@{}c@{}}(0.5,1)\end{tabular} &
\begin{tabular}[c]{@{}c@{}}(1,0.5)\end{tabular} \\
\hline
$p_1$ & 0.0297597 & ~0.02936775  \\
$p_2$ & 1.09770   & ~1.1394975   \\
$p_3$ & 0.938567  & ~0.92180175  \\
$p_4$ & 0.0421186 & ~0.0426663   \\
$p_5$ & 0.902328  & ~0.8643815   \\
$p_6$ & 1.03579   & ~1.1014675   \\
$p_7$ & 0.0981183 & -0.123028675  \\
$p_8$ & 4.68108  & ~3.72573     \\
\hline
\end{tabular}
\caption{Values used for the $p_i$ parameters in Eq.~\ref{eq_f} for Sect.~\ref{sec_resbase}\protect\nobreakdash-\ref{sec_resuhsg} and \ref{sec_reslase}, respectively.}
\label{table_p}
\end{table}

Using torque densities in 1D models may result in premature gap-opening, as described in \citet{2017MNRAS.469.3813H}.
The authors study gap formation in 1D and 2D models. They propose modifying the torque density, setting it to zero for $\lvert r-r_0 \rvert < \num{0.95} R_\mathrm{H}$, with a sharp transition for $\num{0.95} < \lvert r-r_0 \rvert < \num{1.05} R_\mathrm{H}$ in order to prevent the formation of gaps that are  too deep. We followed this approach, but made the region where the torque density is truncated larger, choosing $\num{1.8} R_\mathrm{H}$ instead of $R_\mathrm{H}$. This is necessary because leaving it at $1 R_\mathrm{H}$ still caused a gap to open too soon compared to the hydrodynamic simulations (Sect.~\ref{sec_resbase}). We call this modified torque density `torque mod'. The key here is that we only used the modified torque density in the torque acting on the disc. The migration rate is still calculated based on the unmodified torque. This is important since using the modified torque also for the calculation of the migration rate would again result in too slow migration.\footnote{\citet{2017MNRAS.469.3813H} study planets on a fixed orbit, so the torque from the disc on the planet and the planet's migration are not investigated.} The advantage of this approach is that it can be applied to the entire range of torque densities given in \citetalias{2010ApJ...724..730D}. There is a drawback: using different planet--disc and disc--planet torques violates the conservation of angular momentum. For this reason, we also investigated a different approach for the torque density.
This torque model, which we call `high mass torque' is based on Fig.~15 in \citetalias{2010ApJ...724..730D}. In this figure, the authors show torque density distributions for a planet of increasing mass from their 3D hydrodynamic simulations. The distributions are scaled by $\Omega^2 r_p^2 (M_p/M_*)^2 (r_p/H)^4$ and demonstrate the influence of the planet's mass. The figure reveals that the scaled torque density changes and decreases in magnitude with increasing planet mass, in particular close to the planet. Starting at approximately $\SI{1}{\mj}$ (\num{1} Jupiter mass), an inversion appears: the torque density takes on negative values inside the planet location and then reaches positive values outside the planet location. These features are also seen in Fig.~\ref{fig_app_torque} in  Appendix~\ref{app_prac}, where we plot the scaled  high mass torque for various masses.
As pointed out by \citetalias{2010ApJ...724..730D}, these subtleties are not important once a deep gap has been opened in the disc. However, the inversion is key during the gap formation process, as we  discuss in Sect.~\ref{sec_resbase}. The torque densities given in Fig.~15 of \citetalias{2010ApJ...724..730D} are for $(\beta, \zeta) =(0.5,1)$. We use $\mathcal{F} (x,0.5,1)$ for a planet of zero mass and interpolate linearly in mass using the (digitised) torque densities from the figure. The numerical data we used for the interpolation is available in tabulated form at the CDS.
Figure~\ref{fig_torque} shows the scaled torque densities for the three cases we discussed. The most massive planet for which a torque density is given in \citetalias{2010ApJ...724..730D}'s Figure~15 has a mass of \SI{2}{\mj}. We use this torque density also for more massive planets (no extrapolation). The inversion, as it develops with increasing planet mass, is additionally shown in Fig.~\ref{fig_app_torque} of  Appendix~\ref{app_prac}. Figure~\ref{fig_app_torque} also shows the \citetalias{1986ApJ...309..846L} formula and the \citetalias{2002MNRAS.334..248A} formula (scaled for visibility) that are discussed in Appendix~\ref{app_ia}. 

\begin{figure}
  \includegraphics[width=\linewidth]{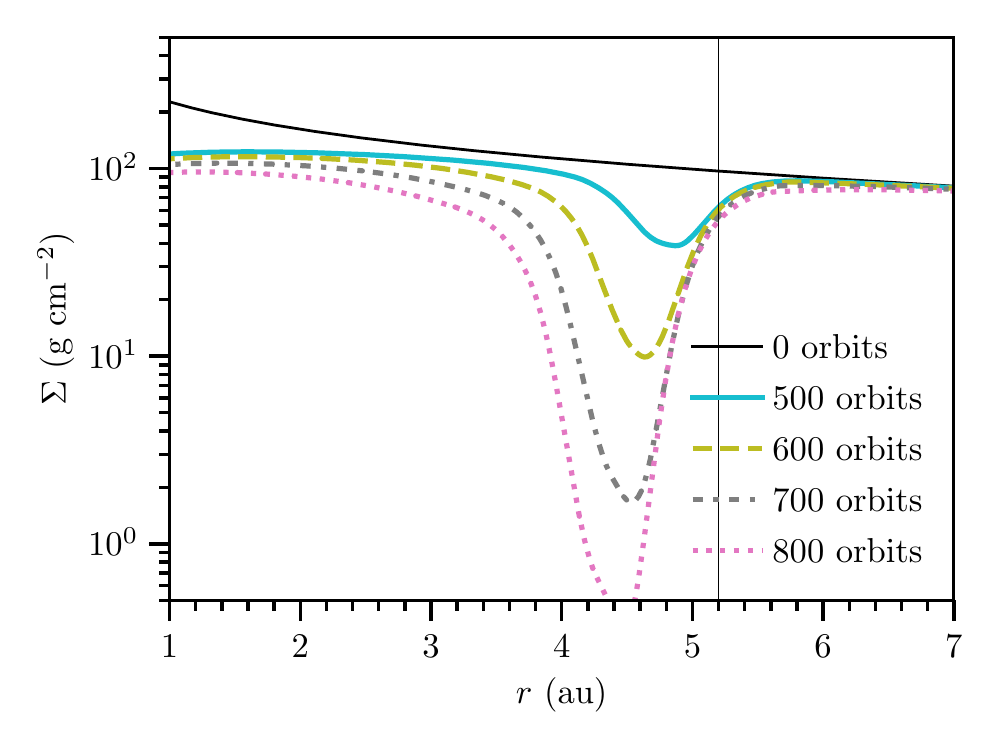}
  \caption{Surface density around an accreting, migrating planet at different times. At the beginning, the planet has a low mass ($\SI{5}{\mearth}$) and does not perturb the surface density appreciably. As the planet grows, it changes the surface density in its vicinity both by accreting gas and by pushing disc material away through tidal interaction. After \num{1000} orbits, the planet has migrated to $\approx \SI{4.2}{au}$, reached a mass of $\SI{1}{\mj}$ and opened a gap in the disc. The planet's initial location ($\SI{5.2}{au}$) is indicated with a thin black vertical line. This simulation is described in more detail in Sect.~\ref{sec_resbase}.}
    \label{fig_surfdens}
\end{figure}

\begin{figure}
  \includegraphics[width=\linewidth]{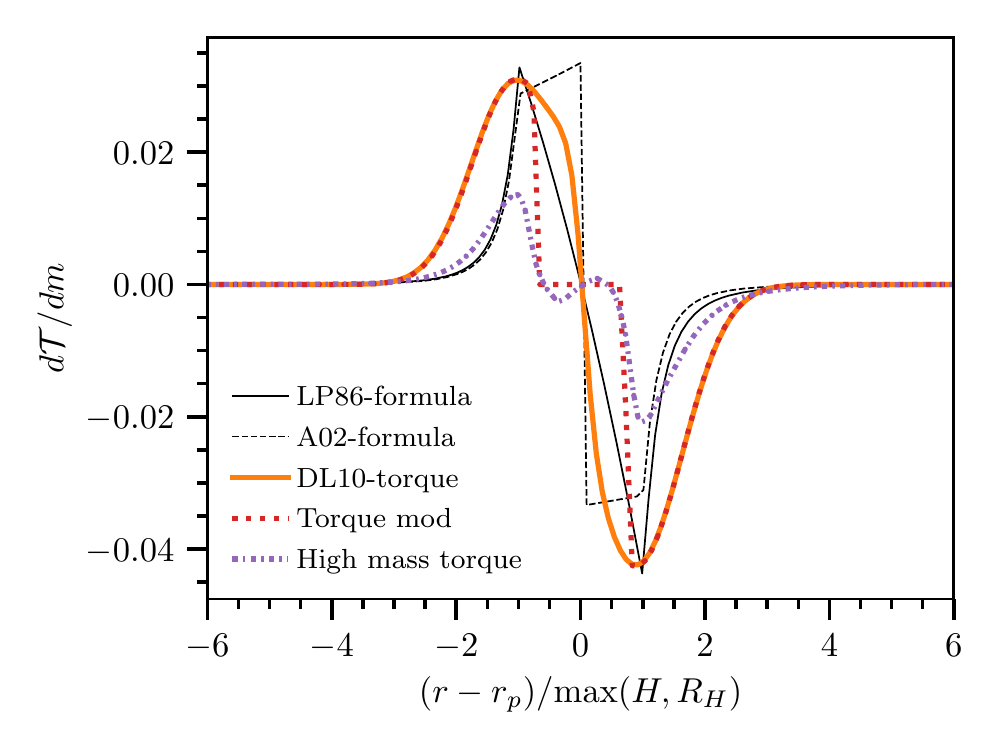}
  \caption{Torque densities used in this work: $\mathcal{F} (0.5,1)$ \citepalias{2010ApJ...724..730D}, a modified version (torque mod) where $\mathcal{F}$ is truncated near the planet when applied to the disc, and a variant model (high mass torque) based on an interpolation of torque densities obtained from a hydrodynamical simulation. The last case is for a $\SI{1.5}{\mj}$ planet.
  The torque densities are scaled by $\Omega^2 r_p^2 (M_p/M_*)^2 (r_p/H)^4$. Additionally, the \citetalias{1986ApJ...309..846L} formula (thin black solid line) and the \citetalias{2002MNRAS.334..248A} formula (thin black dashed line) are shown (see Appendix~\ref{app_ia}). These are divided by 3 and 15, respectively for better visibility. }
    \label{fig_torque}
\end{figure}

As an example, we show the time evolution of the surface density of an accreting, migrating planet in Fig.~\ref{fig_surfdens}. It illustrates, how the planet interacts with the disc over time. The ``High mass torque'' is used in this example, and the initial conditions are the ones from our baseline case that we will discuss in Sect.~\ref{sec_resbase}. The evolution of the surface density in the other cases we studied is presented in Appendix~\ref{app_sd}. In this study we do not discuss the precise shape of the gap since this data is not available from  \citetalias{2008ApJ...685..560D}.

\subsection{Autogravitation}
In the present study the disc mass is low compared to the stellar mass. In this scenario the disc's self-gravity (autogravitation) can be neglected. If the disc is comparable in mass to its host star, the disc's scale height and angular frequency will change \citep{2005A&A...442..703H}. We already studied this effect in detail in \citet{2021A&A...645A..43S}. Our prescriptions presented here can in principle still be applied. However, some of the equations given in Section~\ref{sect:Model} need to be modified. In particular, $H$ needs to be replaced by $c_\mathrm{s} / \Omega$ in Eq.~\ref{eq_torque}.
The expressions for accretion also change slightly due to the different vertical structure. The details are given in Appendix~\ref{app_ag}.

\section{Investigated parameter space and initial conditions}\label{sec:param}
In this section we describe the set-up of our simulations for the different comparisons we performed.

For the comparison with \citetalias{2008ApJ...685..560D} (Sects.~\ref{sec_resbase} - \ref{sec_reshvsc}) we chose initial conditions  similar to theirs. Our simulations start with a surface density profile of the form
\begin{equation}
    \Sigma(r) = \Sigma_0 \left(\frac{r}{\SI{5.2}{au}}\right)^{\beta_\Sigma} \exp{\left[ \left(\frac{\SI{5.2}{au}}{r_\mathrm{max}}\right)^{2+\beta_\Sigma} \right]} \exp{\left[ - \left(\frac{r}{r_\mathrm{max}}\right)^{2+\beta_\Sigma}\right]},
\label{eq_sigma}
\end{equation}
where $\beta_\Sigma = -0.5$ is the surface density radial slope (the same as in \citetalias{2008ApJ...685..560D}), and  $r_\mathrm{max}$ is the radius of the outer exponential cutoff, for which we chose $\SI{30}{au}$. The inner disc is truncated at $\SI{0.05}{au}$. We note that for numerical reasons, \citetalias{2008ApJ...685..560D} study only a part of the disc's radial extent (\SIrange[]{2.08}{13}{au} for the cases relevant here). Due to the different nature of their 3D simulation, their boundary conditions are also different (see their Sect.~2.1.3). Since the viscosity is low in the majority of the simulations we performed, the effect of the different boundary conditions is not important. This changes in the case with higher viscosity, where the boundary conditions play a significant role, as we discuss in Sect.~\ref{sec_reshvsc}. We checked that increasing or decreasing our $r_\mathrm{max}$ by one-third has a very small influence on our results.
The initial value of the surface density at $\SI{5.2}{au}$, $\Sigma_0$ is chosen between $\SI{100}{g.cm^{-2}}$ and $\SI{500}{g.cm^{-2}}$ according to the different cases in \citetalias{2008ApJ...685..560D}.
Using Eq.~\ref{eq_cs} we get a temperature profile proportional to $r^{-1}$:
\begin{equation}
    T(r) = \left(\frac{H}{r}\right)^2\frac{\mu \mathrm{u} G \mstar}{k_\mathrm{B}}r^{-1}.
\label{eq_t}
\end{equation}
We set $(\beta,\zeta) = (0.5,1)$ in Eq.~\ref{eq_f}, as suggested by the slopes of $\Sigma$ and $T$ in Eqs.~\ref{eq_sigma} and \ref{eq_t}.
A constant kinematic viscosity of the same magnitude as \citetalias{2008ApJ...685..560D} was also used (see Sect.~\ref{sect:Results}).

For our comparison with \citetalias{2019MNRAS.486.4398F} in Sect. \ref{sec_reslase}, we used the initial surface density given in their Fig.~1 (digitised), and the temperature profile given in their Eq.~1:
\begin{equation}
    T(r) = \SI{200}{K} \left(\frac{r}{\SI{1}{au}}\right)^{-1}.
\end{equation}
This temperature profile agrees well with the initial profile in \citetalias{2019MNRAS.486.4398F}. However, while our temperatures stay constant in time, their temperatures start to deviate inside of $\sim \SI{30}{au}$ due to artificial viscosity heating. We expect that this will slow down migration somewhat inside of $\sim \SI{40}{au}$.
Instead of using a constant viscosity as above, we use the alpha-viscosity
\begin{equation}
    \nu = \alpha \frac{c_\mathrm{s}^2}{\Omega}
\end{equation}
\citep{1973A&A....24..337S} with $\alpha = \num{3e-2}$. This value approximately corresponds to the artificial viscosity of the codes used in \citetalias{2019MNRAS.486.4398F} (their Section~3.3.1). Hydrodynamic codes such as the ones used in their study typically produce artificial viscosities comparable to this value. 

\section{Results}\label{sect:Results}
In this section we calibrate our model and compare our results to  those inferred using hydrodynamic simulations, first to \citetalias{2008ApJ...685..560D} and then   to \citetalias{2019MNRAS.486.4398F}.
We use the baseline case (see Sect.~\ref{sec_resbase}) from \citetalias{2008ApJ...685..560D} to calibrate the parameters $C_\mathrm{B}$, $C_\mathrm{H}$, the width of the feeding zone $R_f$, and the truncation width for the torque mod (see Sect.~\ref{sect:Model}).
Once these parameters are calibrated  they remain unchanged. For practical use we recommend our high mass torque model with the parameters listed in Table~\ref{table_config}.

In the following sections we compare the results for a number of different sets of initial conditions.
Table~\ref{table_comp} gives an overview of the cases studied in the following sections.
\begin{table}
\centering
\begin{tabular}{lccc}
\hline\hline
\begin{tabular}[c]{@{}c@{}}Section\end{tabular} &
\begin{tabular}[c]{@{}c@{}}$\Sigma_{5.2} (\SI{}{g.cm^{-2}})$\end{tabular} &
\begin{tabular}[c]{@{}c@{}}$H/r$\end{tabular} &
\begin{tabular}[c]{@{}c@{}}$\nu (\SI{}{cm^2.s^{-1}})$\end{tabular} \\
\hline
\ref{sec_resbase} Baseline & 100 & 0.05 & $\num{e15}$\\
\ref{sec_reshsig} Higher $\Sigma$& 300 & 0.05 & $\num{e15}$ \\
\ref{sec_resuhsg} Very high $\Sigma$& 500 & 0.05 & $\num{e15}$ \\
\ref{sec_resltmp} Low $T$& 100 & 0.04 & $\num{e15}$ \\
\ref{sec_reshvsc} High $\nu$& 100 & 0.05 & $\num{e16}$ \\
\hline
\end{tabular}
\caption{Parameters used in the comparisons in Sects.~\ref{sec_resbase} to \ref{sec_reshvsc}: Initial surface density at $\SI{5.2}{au}$, aspect ratio, and kinematic vicosity. The initial planetary mass is \SI{5}{\mearth} in all cases.}
\label{table_comp}
\end{table}

\subsection{Baseline case: Calibration of the 1D model}
\label{sec_resbase}
In this simulation a $\SI{5}{\mearth}$ planet is inserted at $\SI{5.2}{au}$ into a disc with an initial surface density of $\SI{100}{g.cm^{-2}}$ at $\SI{5.2}{au}$ with a constant kinematic viscosity of $\SI{e15}{cm^2.s^{-1}}$ (corresponding to $\alpha = 0.004$ at \SI{5.2}{au}) and a constant aspect ratio of $H/r = 0.05$ (corresponding to an initial temperature of $\approx \SI{118}{K}$ at \SI{5.2}{au}). These parameters are identical to those used by \citetalias{2008ApJ...685..560D}.

The planet is then allowed to migrate and accrete gas.
We note that   the study by \citetalias{2008ApJ...685..560D} and our work both focus on the `disc-limited' mode of gas accretion (Sect.~\ref{subs_acc}): it is assumed that the planet can accrete all the gas available from the disc (also known as the rapid gas accretion phase). To what degree these accretion rates can be absorbed depends very much on the planetary  properties. For example, in the core accretion scenario, gas only starts to be accreted rapidly once the critical core mass has been reached \citep{Pollack1996,2021A&A...656A..69E}. The critical core mass is of the order of $\SI{10}{\mearth}$. The disc-limited regime is typically reached at total planet masses of \SIrange[]{50}{100}{\mearth} \citep{movshovitzbodenheimer2010a,2014A&A...566A.141M}.

\begin{figure*}
  \begin{subfigure}[pt]{0.49\textwidth}
  \includegraphics[width=\linewidth]{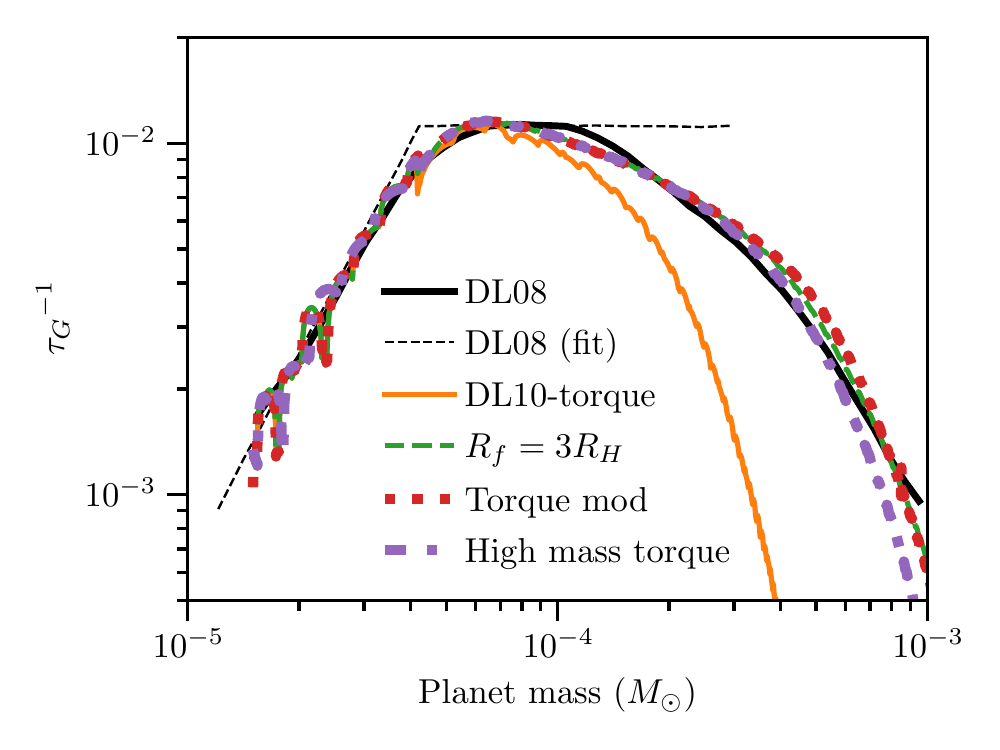}
  \end{subfigure}
  \begin{subfigure}[pt]{0.49\textwidth}
  \includegraphics[width=\linewidth]{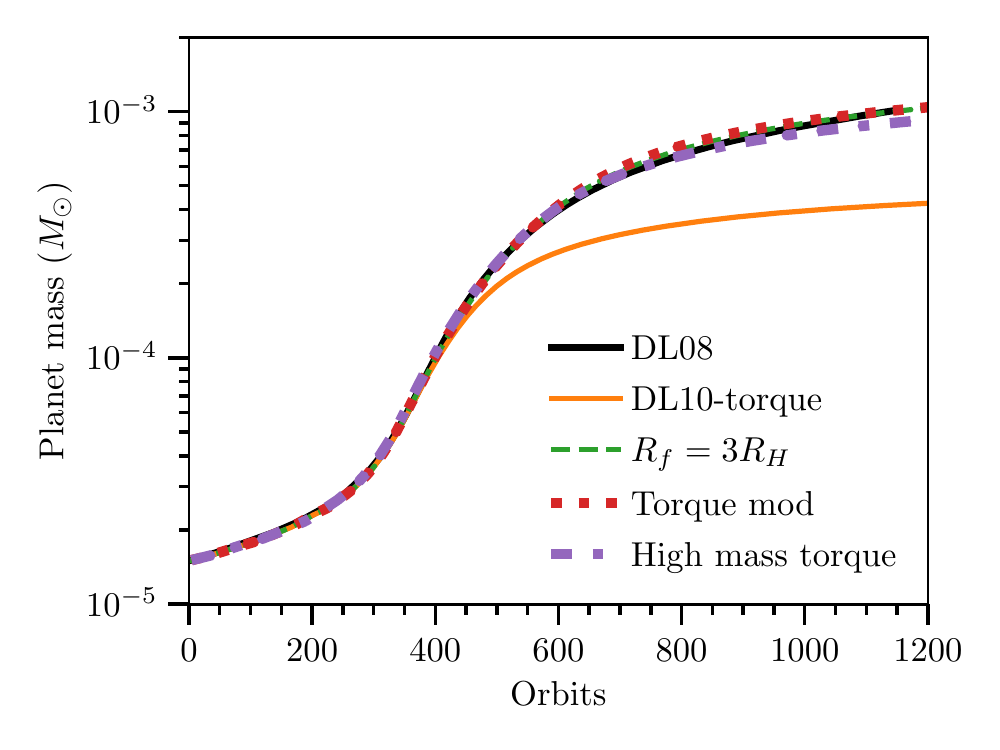}
  \end{subfigure}
  \begin{subfigure}[pt]{0.49\textwidth}
  \includegraphics[width=\linewidth]{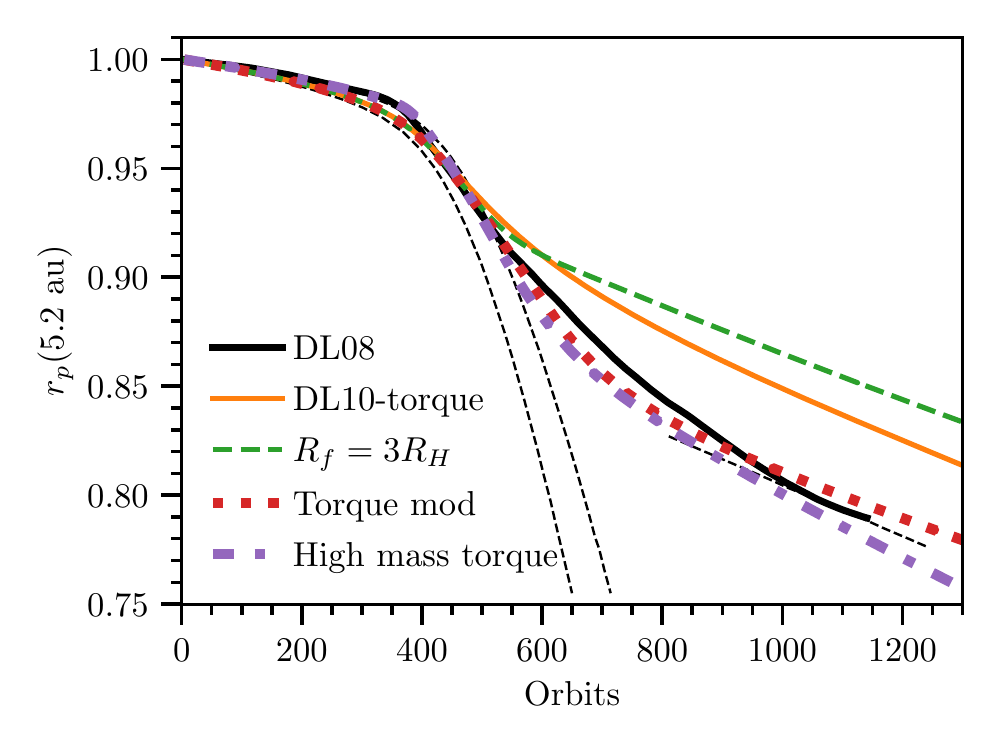}
  \end{subfigure}
  \begin{subfigure}[pt]{0.49\textwidth}
  \includegraphics[width=\linewidth]{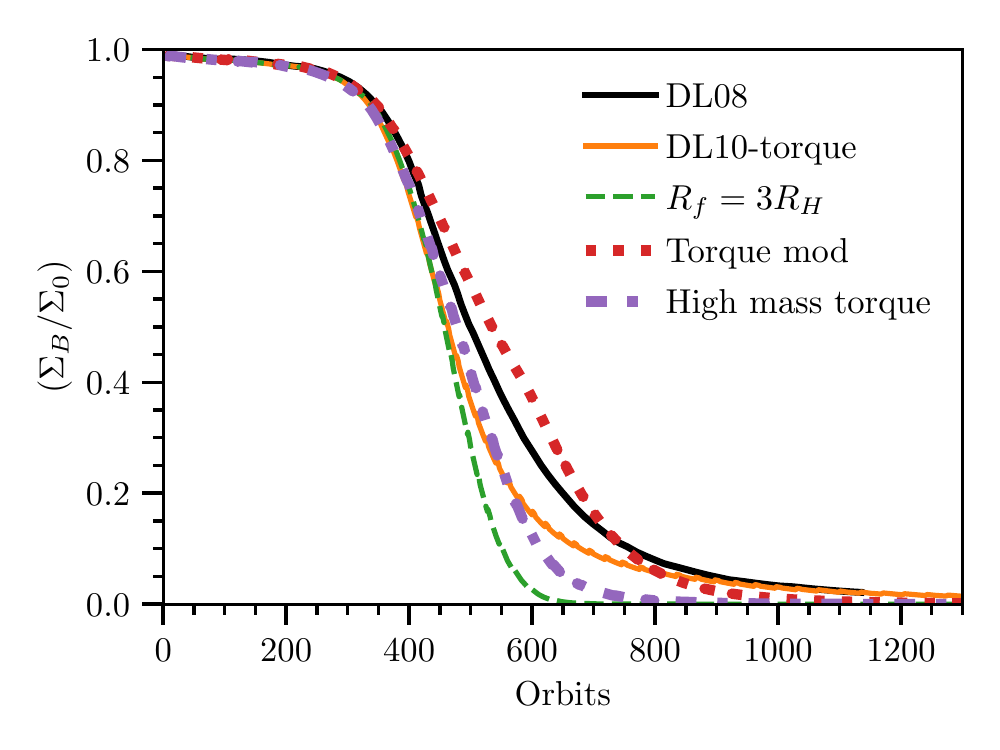}
  \end{subfigure}
  \caption{Time evolution of the baseline  case with an initial surface density of \SI{100}{g.cm^{-2}} at \SI{5.2}{au}. The solid black line represents the 3D result from \citetalias{2008ApJ...685..560D}. The solid orange line corresponds to our 1D result when using the \citetalias{2010ApJ...724..730D} torque. The green dashed line gives again the result with the \citetalias{2010ApJ...724..730D} torque, but with the feeding zone radius increased threefold. The red dotted line shows a calculation where, additionally, the torque density has been modified. The purple dash-dotted line shows the variant model, where interpolated torque densities for planets of different masses from \citetalias{2010ApJ...724..730D} were applied. The torque densities are described in Sect.~\ref{subs_mig}.
  Top left: Inverse growth timescale in units of the orbital timescale at the initial location ($\approx$ \SI{12}{yr}). The thin black dashed line represents the fit given by Eq.~\ref{eq_Bondi} and Eq.~\ref{eq_Hill}.
  Top right: Mass evolution.
  Bottom left: Orbital migration. The thin dashed black lines represent (from left to right) predictions for saturated and unsaturated type~I, and type~II theory.
  Bottom right: Evolution of the mean surface density around the planet. Time is given in orbits at the planet's initial separation ($\SI{5.2}{au}$).}
  \label{fig_base}
\end{figure*}

The top left panel in Fig.~\ref{fig_base} shows the inverse growth timescale $\tau_G^{-1}$ as a function of planet mass when applying the \citetalias{2010ApJ...724..730D} torque with a feeding zone radius of $1 R_\mathrm{H}$ (orange solid line). The accretion agrees well with that from  \citetalias{2008ApJ...685..560D} in the Bondi regime (we chose $C_\mathrm{B}$ accordingly). For the Hill regime we chose $C_\mathrm{H}$ such that the inverse growth timescale agrees with the analytical fit at early times (corresponding to a planet mass of $\approx \SI{20}{\mearth}$). However, the accretion rate drops too soon. This happens because a gap opens very rapidly. This effect is seen in the bottom right panel of Fig.~\ref{fig_base}. The figure shows $\Sigma_B$, the surface density near the planet, as a fraction of the initial $\Sigma_0$. The surface density is averaged over a radial band of width $\SI{2}{H}$, centred on the planet. The value of  $\Sigma_B$ drops somewhat faster in our calculation (orange solid line) compared to the hydrodynamic calculation.
In order to prevent the premature depletion of the feeding zone, we increased its radius by a factor of three (and reduced $C_\mathrm{H}$ accordingly) since this gives the best agreement with the hydrodynamical simulation. The result (green dashed line in Fig.~\ref{fig_base}) is now very similar to that from \citetalias{2008ApJ...685..560D}. Consequently, the mass evolution is now also very similar (top right panel in the figure). The planet reaches $\SI{300}{\mearth}$ at the end of the hydrodynamic simulation after $\approx \num{1200}$ orbits.

We next investigate the behaviour of migration in our model. The bottom left panel of Fig.~\ref{fig_base} shows the planet's semi-major axis versus  time.
It reveals that the migration rate in the \citetalias{2010ApJ...724..730D} torque case agrees well for the first $\sim \num{500}$ orbits, but slows down  too much when approaching the type~II regime (orange solid line). This slowdown is even stronger with the increased feeding zone radius (green dashed line).
While the increased feeding zone gives a better agreement for the accretion, it has an opposite effect on the migration: a gap is opened even more quickly due to the faster increase in mass, and the  migration is too slow. This is again seen in the bottom right panel of the figure, where $\Sigma_B$ now drops even more quickly and reaches almost zero after $\num{600}$ orbits. The premature gap opening in 1D models is a well-known result. It is studied in \citet{2017MNRAS.469.3813H}.

Now we consider the torque mod described in Sect.~\ref{subs_mig}. The effect is again seen in Fig.~\ref{fig_base} (red dotted line). The agreement with \citetalias{2008ApJ...685..560D} is very good both in terms of mass evolution (top right panel) and migration (bottom left panel). The decrease in surface density is slightly slower than in the hydrodynamic calculation.
Finally, applying the high mass torque described in Sect.~\ref{subs_mig} gives a very similar result. The reduction of the torque density in this case, in combination with the inversion near the planet, produces fast migration (as in \citetalias{2008ApJ...685..560D}) and angular momentum is conserved. The surface density near the planet is still depleted more quickly compared to \citetalias{2008ApJ...685..560D}, but it does not inhibit mass growth or slow down migration significantly. The planet mass reached at the end of the hydrodynamic simulation is $\approx 1 \%$ higher (torque mod) and $\approx 10 \%$ lower (high mass torque) respectively, compared to \citetalias{2008ApJ...685..560D}. The bottom left panel also shows predictions from saturated and unsaturated type~I theory,  and type~II theory (thin black dashed lines). As expected, the planet initially follows the type~I prediction well, but slows down as it grows in mass and starts perturbing the disc, approaching the type~II expectation. The theoretical type~I tracks were calculated by \citetalias{2008ApJ...685..560D} based on the initial value of the surface density.

We find that when we calibrate $C_\mathrm{B}$, $C_\mathrm{H}$, $R_f$ and (for the torque mod prescription) the width of the truncation region, the agreement with the hydrodynamic simulation is good. In the following sections we investigate whether this agreement persists in the other cases studied by \citetalias{2008ApJ...685..560D}.

\subsection{Higher surface density}
\label{sec_reshsig}
\begin{figure*}
  \begin{subfigure}[pt]{0.49\textwidth}
  \includegraphics[width=\linewidth]{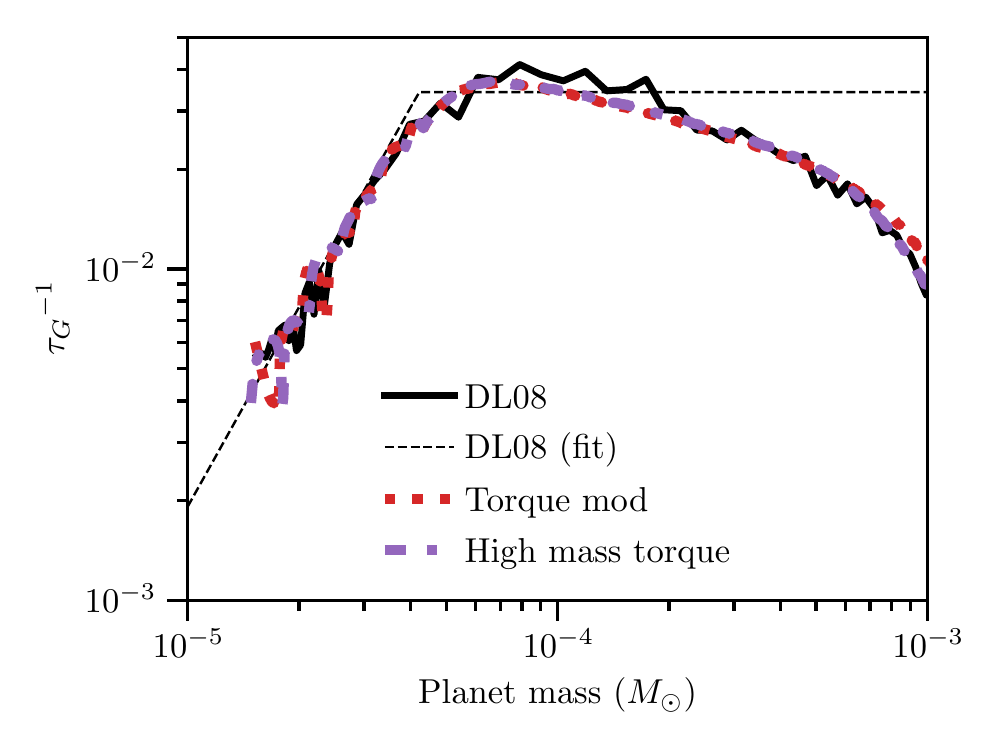}
  \end{subfigure}
  \begin{subfigure}[pt]{0.49\textwidth}
  \includegraphics[width=\linewidth]{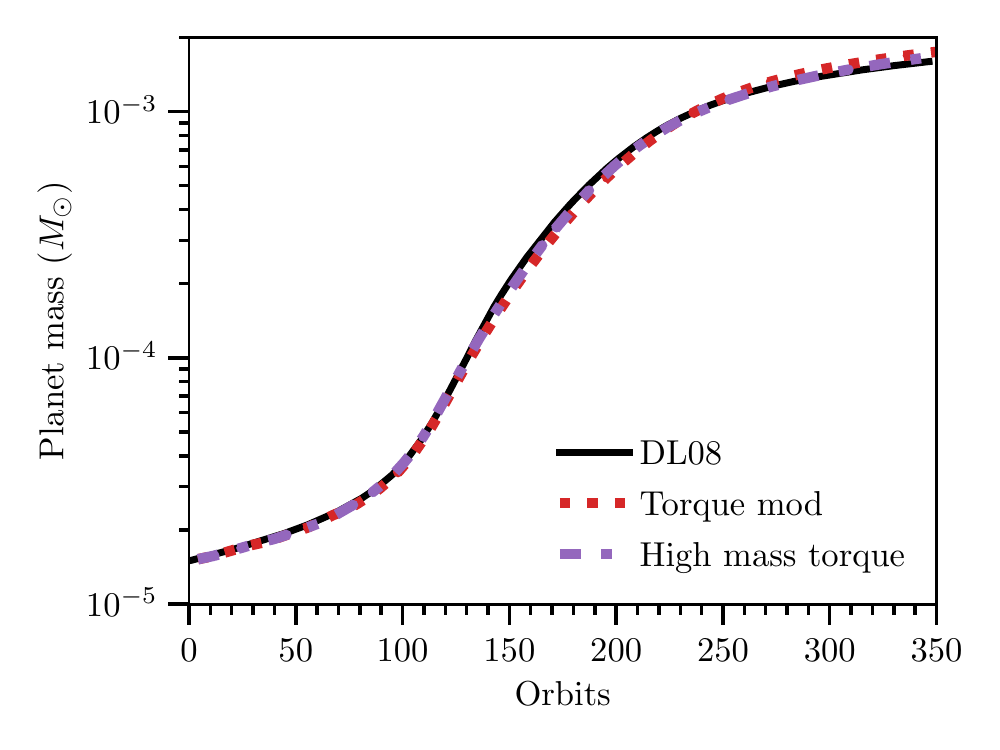}
  \end{subfigure}
  \begin{subfigure}[pt]{0.49\textwidth}
  \includegraphics[width=\linewidth]{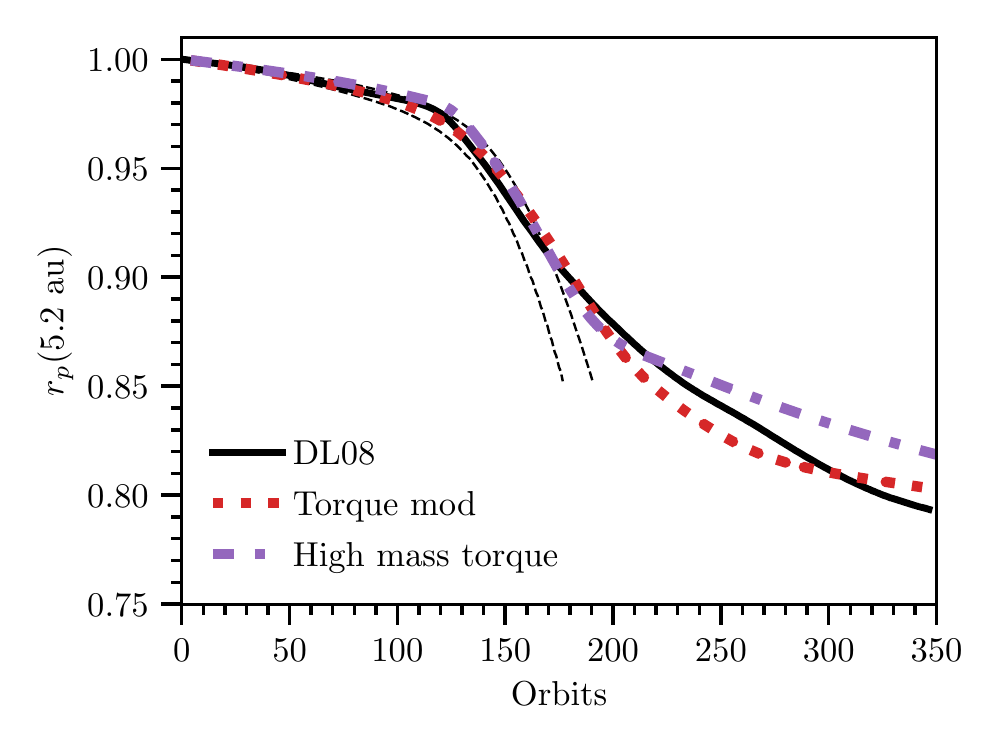}
  \end{subfigure}
  \begin{subfigure}[pt]{0.49\textwidth}
  \includegraphics[width=\linewidth]{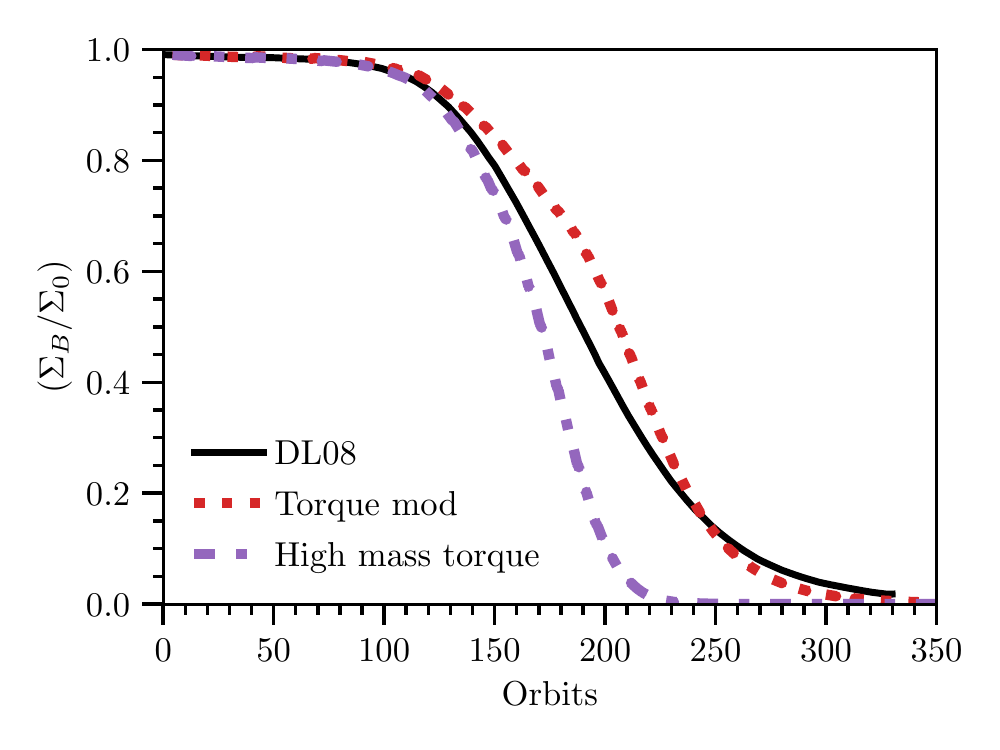}
  \end{subfigure}
  \caption{Same as  Fig.~\ref{fig_base}, but for  increased surface density (\SI{300}{g.cm^{-2}} at \SI{5.2}{au}).}
  \label{fig_hsig}
\end{figure*}

Here we consider, as in \citetalias{2008ApJ...685..560D}, a case where the initial surface density is increased by a factor of three relative to the baseline case ($\SI{300}{g.cm^{-2}}$ at $\SI{5.2}{au}$), while all the other conditions were left the same. The results are shown in  Fig.~\ref{fig_hsig}. For clarity, we omit the \citetalias{2010ApJ...724..730D} torque and $R_f = 3 R_\mathrm{H}$ cases from now on.

We find a similar behaviour to  the baseline case. The increased feeding zone radius, combined with the modified torque density, results in a migration rate (bottom left panel) and an inverse growth timescale (top left panel) very similar to that found by \citetalias{2008ApJ...685..560D}. 
This is true for both variant torque models we considered and is true despite the fact that, in this calculation, the planet is growing much faster, reaching $\SI{300}{\mearth}$ already after $\num{220}$ orbits. The masses we find are $\approx 5 \%$ (high mass torque) and $\approx 10 \%$ (torque mod) higher than what \citetalias{2008ApJ...685..560D} found after 350 orbits.

The decrease in surface density is again somewhat slower in the torque mod case and somewhat faster in the high mass torque case, compared to the hydrodynamic simulation (bottom right panel). We conclude that the calibrations done with the baseline case lead to good agreement of 1D and 3D simulations also when at a higher initial surface density.

\subsection{Very high surface density}
\label{sec_resuhsg}

\begin{figure*}
  \begin{subfigure}[pt]{0.49\textwidth}
  \includegraphics[width=\linewidth]{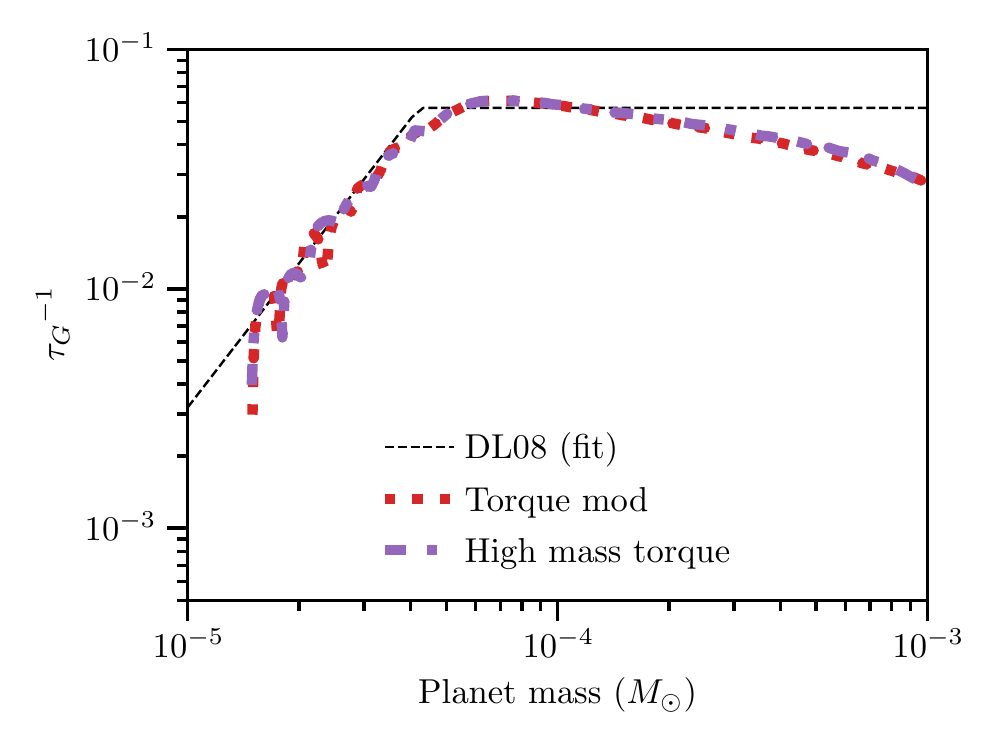}
  \end{subfigure}
  \begin{subfigure}[pt]{0.49\textwidth}
  \includegraphics[width=\linewidth]{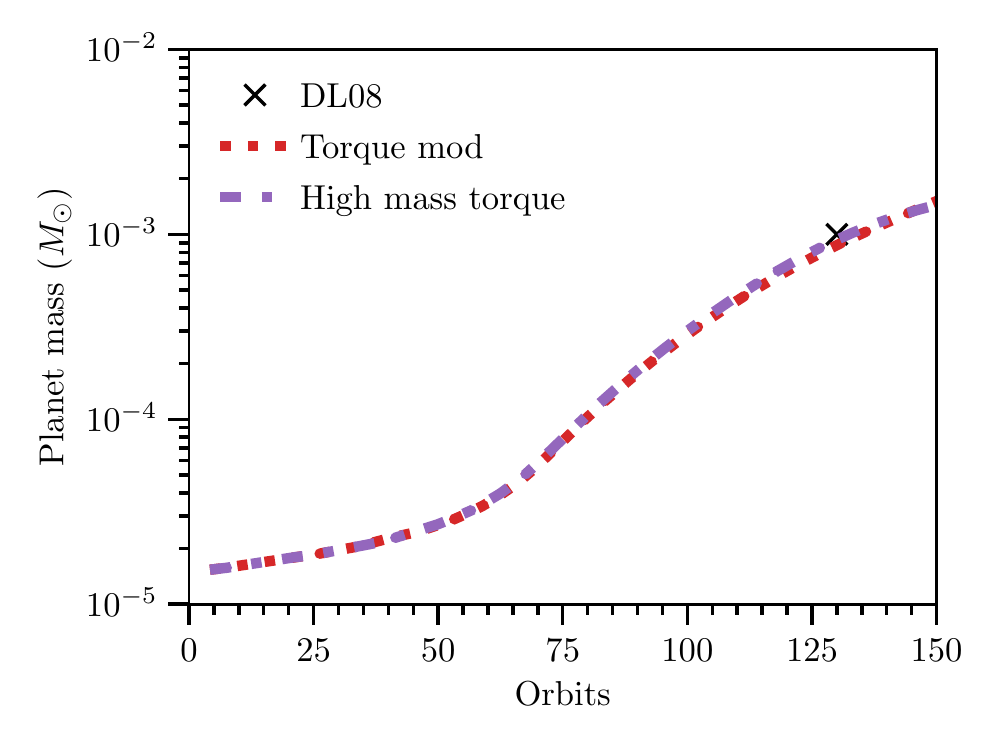}
  \end{subfigure}
  \begin{subfigure}[pt]{0.49\textwidth}
  \includegraphics[width=\linewidth]{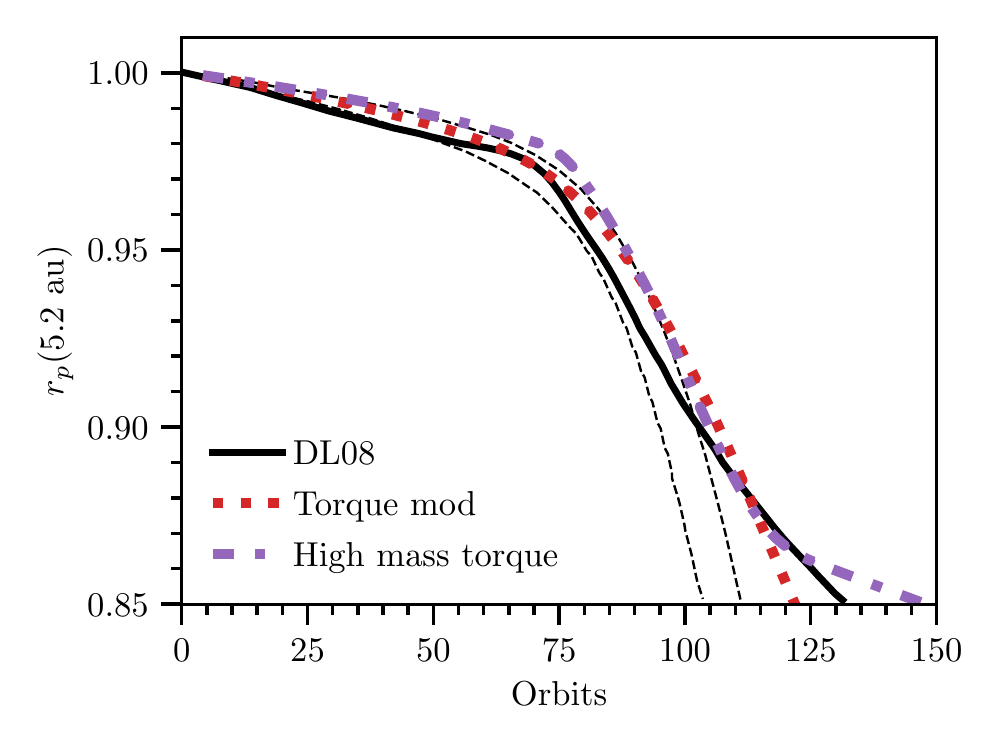}
  \end{subfigure}
  \begin{subfigure}[pt]{0.49\textwidth}
  \includegraphics[width=\linewidth]{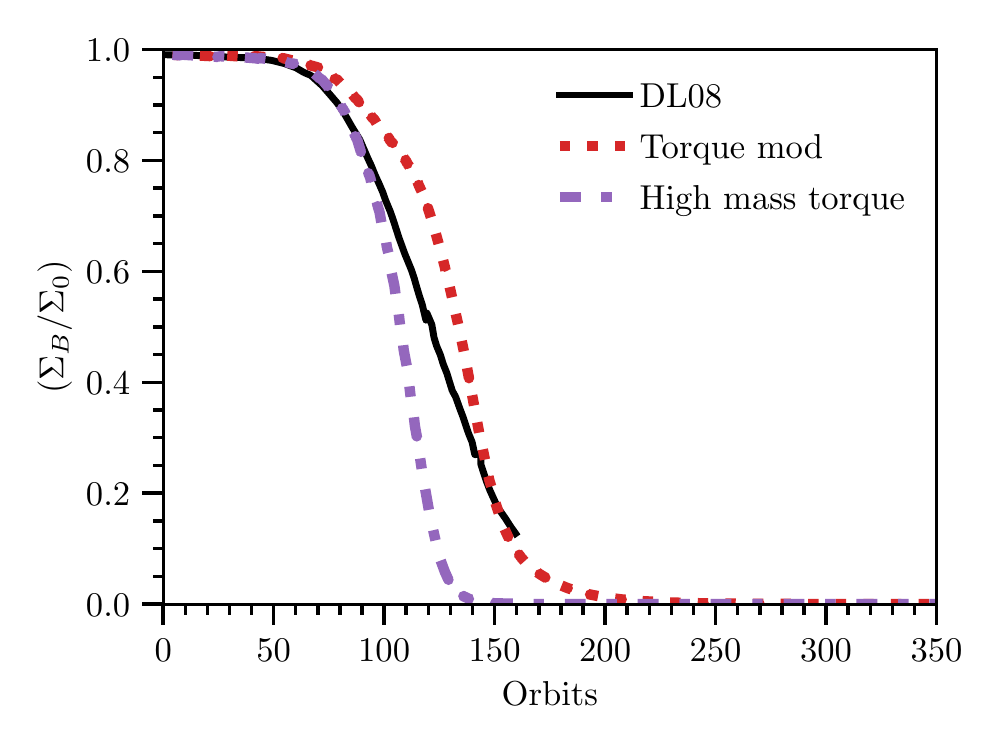}
  \end{subfigure}
  \caption{Same as  Fig.~\ref{fig_base}, but with  five times higher initial surface density (\SI{500}{g.cm.^{-2}} at \SI{5.2}{au}).}
  \label{fig_uhsg}
\end{figure*}

In this section we present the results when the disc's initial surface density is increased by a factor of five relative to the baseline case ($\SI{500}{g.cm^{-2}}$ at $\SI{5.2}{au}$). The results are presented in Fig.~\ref{fig_uhsg}.

For this simulation \citetalias{2008ApJ...685..560D} do not provide the mass evolution of the planet as a function of time, so it is missing in Fig.~\ref{fig_uhsg}. They indicate it reaches $\si{1}{\mj}$ after $\num{130}$ orbits, which is shown with the black cross in the top right panel of the figure. We find that the mass evolution agrees well with this value and with the analytic fitting formula (top left panel). The mass reached after \num{130} orbits is $2 \%$ (high mass torque) and $8 \%$ (torque mod) lower compared to \citetalias{2008ApJ...685..560D}. In the migration track there is some disagreement. The torque mod model shows almost no departure from type~I migration during the first $\num{125}$ orbits. It only does so after reaching approximately \num{0.65} times the initial separation (not seen in the figure).  The high mass torque model agrees qualitatively with the hydrodynamic simulation, though slowing down slightly faster. The behaviour of the surface density around the planet (bottom right panel) is very similar to the first two cases we discussed.

\subsection{Lower temperature}
\label{sec_resltmp}

\begin{figure*}
  \begin{subfigure}[pt]{0.49\textwidth}
  \includegraphics[width=\linewidth]{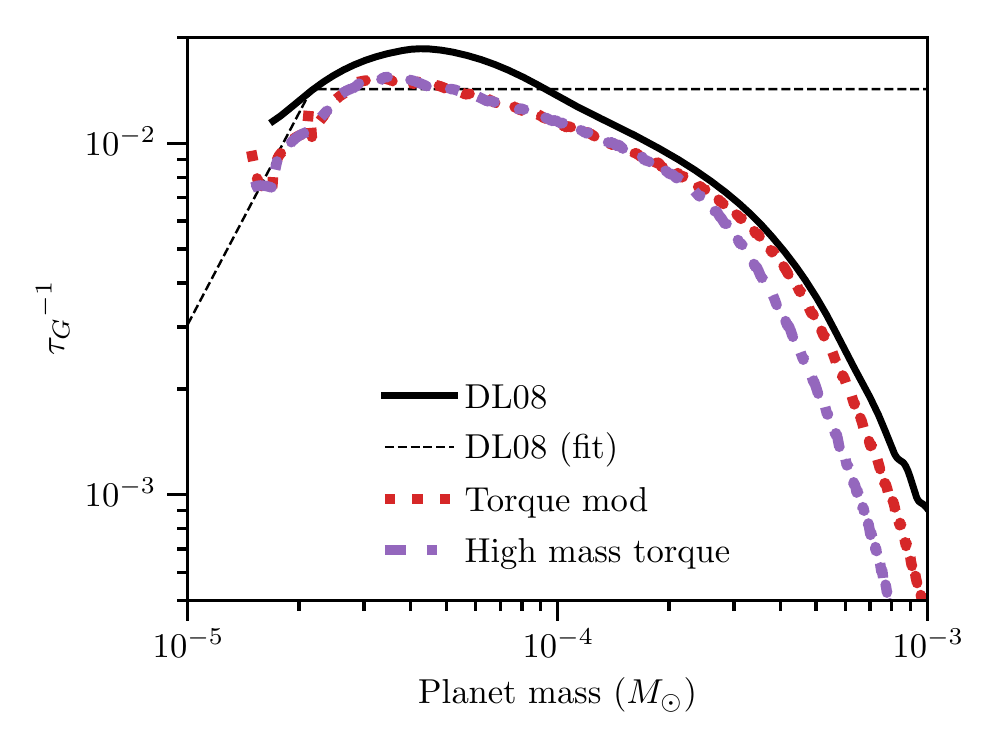}
  \end{subfigure}
  \begin{subfigure}[pt]{0.49\textwidth}
  \includegraphics[width=\linewidth]{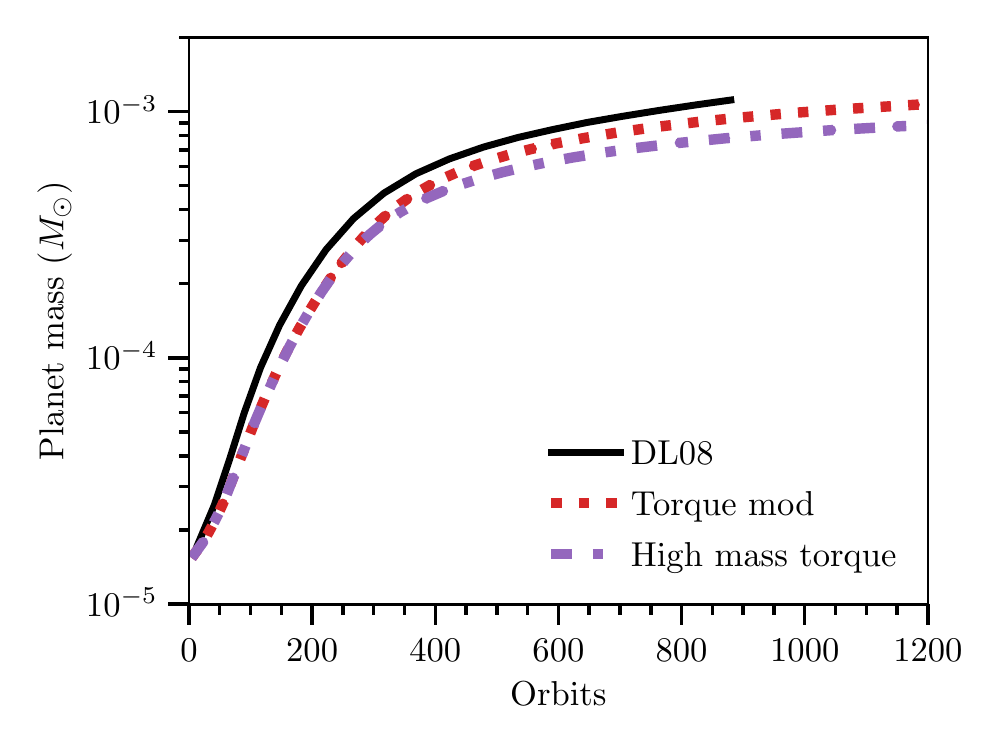}
  \end{subfigure}
  \begin{subfigure}[pt]{0.49\textwidth}
  \includegraphics[width=\linewidth]{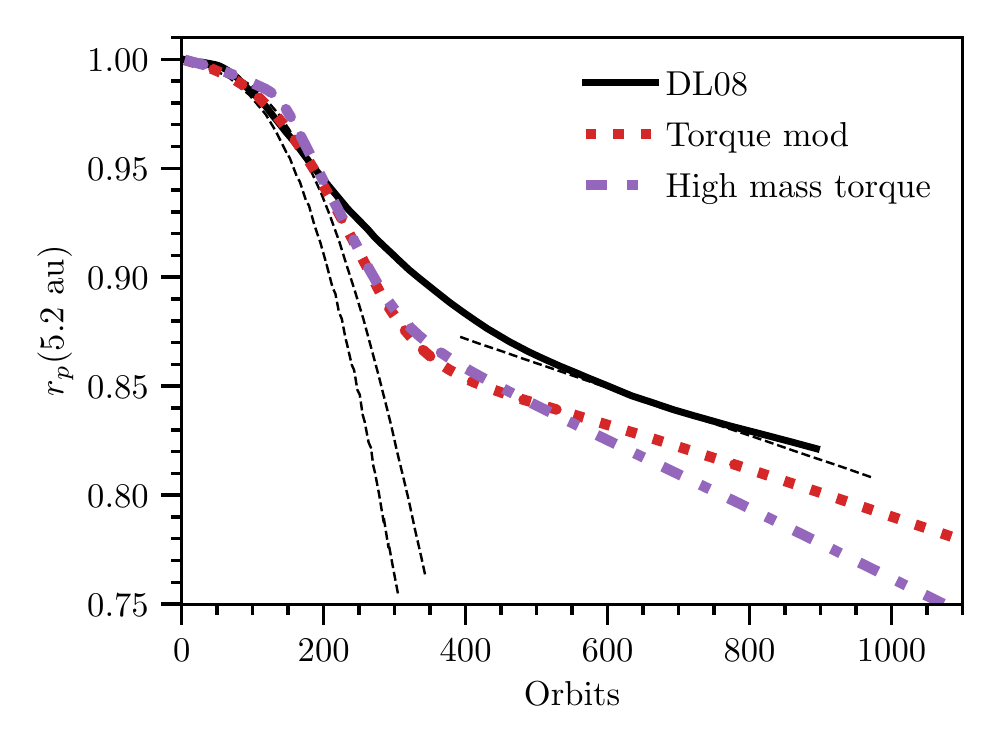}
  \end{subfigure}
  \begin{subfigure}[pt]{0.49\textwidth}
  \includegraphics[width=\linewidth]{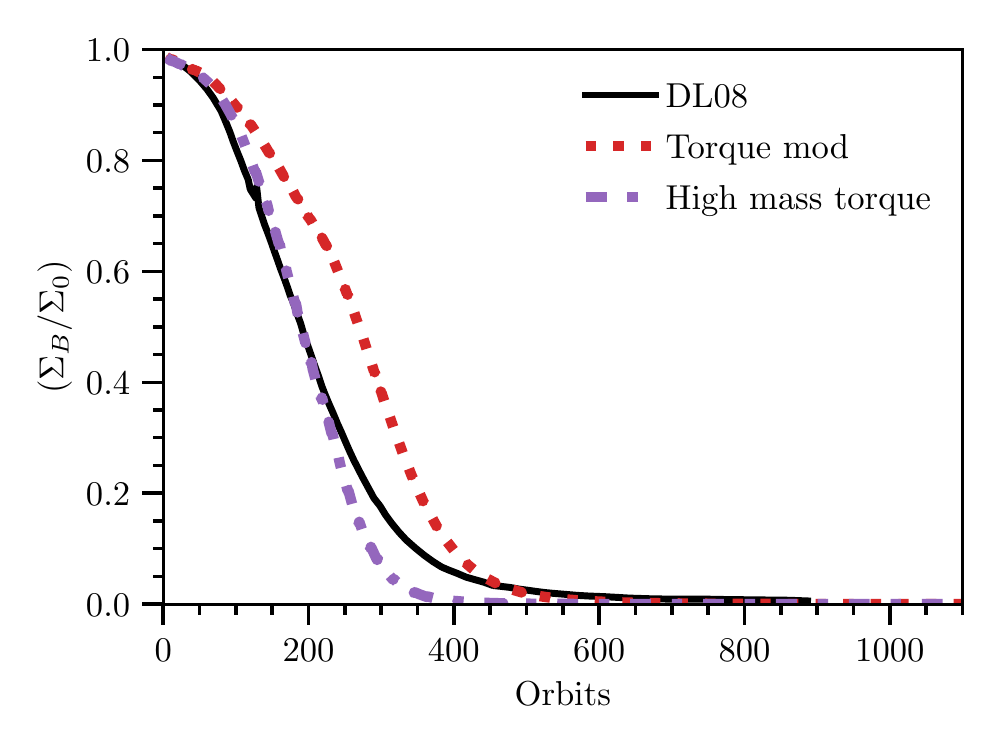}
  \end{subfigure}
  \caption{Same as  Fig.~\ref{fig_base}, but for  reduced temperature ($H/r = 0.04$).}
\label{fig_ltmp}
\end{figure*}

In this section we look at a variant calculation with a lower disc temperature ($H/r = 0.04$, this corresponds to an initial temperature of $\approx \SI{76}{K}$ at \SI{5.2}{au}). The results are presented in Fig.~\ref{fig_ltmp}.

The mass growth is faster here in comparison to the baseline case, due to the higher Bondi accretion rate, but  it slows down quickly as the Hill regime is reached. Consequently, inward migration is also faster until a gap starts to open after $\sim \num{300}$ orbits. The agreement of our model with the calculation from \citetalias{2008ApJ...685..560D} is good for both of our torque models shown. The planet migrates in marginally further, but then transitions to type~II migration at a very similar rate (bottom left panel). The evolution of $\Sigma_B$ is again slightly different between the torque mod and high mass torque simulations. The latter matches more closely the result from the hydrodynamic calculation (bottom right panel). The mass evolution is also similar (top panels), though the  mass reached at the end of the hydrodynamical simulation after about \num{850} orbits is $15 \%$ (high mass torque) and $20 \%$ (torque mod) lower compared to \citetalias{2008ApJ...685..560D}.

\subsection{Higher viscosity}
\label{sec_reshvsc}

\begin{figure*}
  \begin{subfigure}[pt]{0.49\textwidth}
  \includegraphics[width=\linewidth]{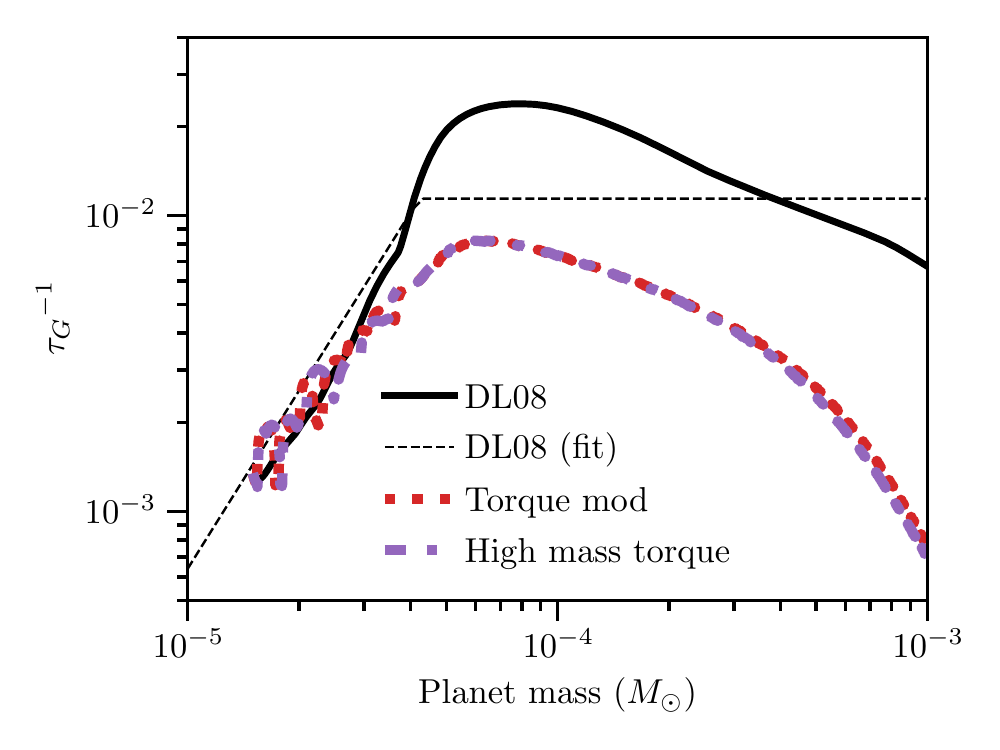}
  \end{subfigure}
  \begin{subfigure}[pt]{0.49\textwidth}
  \includegraphics[width=\linewidth]{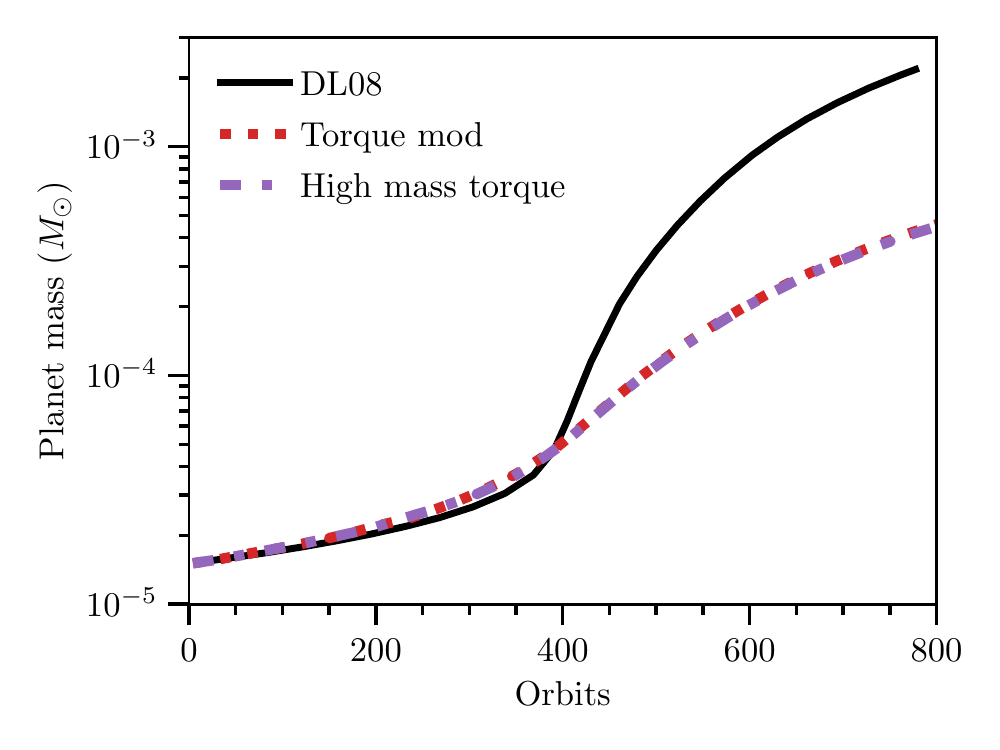}
  \end{subfigure}
  \begin{subfigure}[pt]{0.49\textwidth}
  \includegraphics[width=\linewidth]{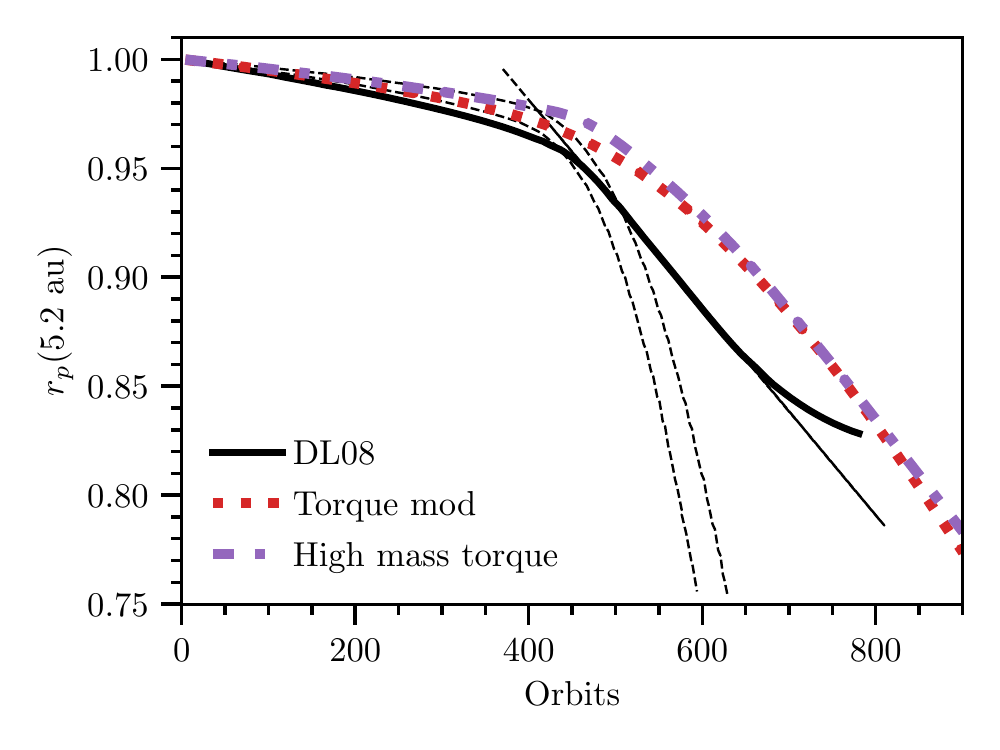}
  \end{subfigure}
  \begin{subfigure}[pt]{0.49\textwidth}
  \includegraphics[width=\linewidth]{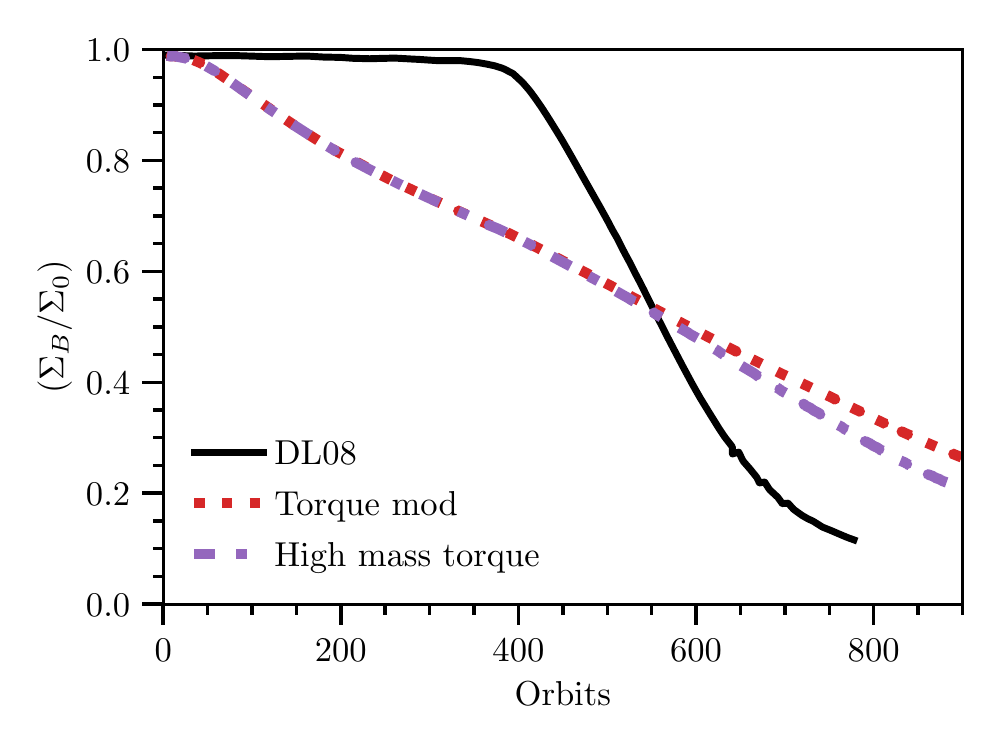}
  \end{subfigure}
  \caption{Same as  Fig.~\ref{fig_base}, but for  ten times higher viscosity ($\SI{e16}{cm^2.s^{-1}}$).}
\label{fig_hvsc}
\end{figure*}

In the simulation presented in this section, the disc viscosity is higher by an order of magnitude ($\SI{e16}{cm^2.s^{-1}}$, corresponding to $\alpha = 0.04$ at \SI{5.2}{au}) compared to the other cases. A precise agreement with the hydrodynamic simulations from \citetalias{2008ApJ...685..560D} or their analytic fit is not expected here, as we  discuss below. The result are shown in Fig.~\ref{fig_hvsc}.

The top left panel of Fig.~\ref{fig_hvsc} shows that our Hill accretion rate does not reach the value predicted by the analytic fit. Conversely, the hydrodynamic simulation reaches a value higher by more than a factor two compared with the fit. This leads to a planetary mass that is higher by a  factor of five in comparison to our model (top right panel). 

The migration tracks are not very different (bottom left panel), though in \citetalias{2008ApJ...685..560D} the planet becomes slower than expected from type~II theory after $\approx \num{500}$ orbits, likely due to the fast increase in mass. In our simulation the migration continues according to the type~II slope.

The difference in mass accretion is also seen very clearly in the evolution of $\Sigma_B$ (bottom right panel). The disc remains almost unperturbed for $\num{400}$ orbits for \citetalias{2008ApJ...685..560D}, while it starts declining steadily much earlier in our model.

These different results are caused by the global disc evolution due to the high viscosity in combination with differences in boundary conditions. The viscous timescale ($r^2/\nu$) is only $\approx \SI{19}{kyr}$, or less than 1600 orbits at the planet's initial location. The consequence is a rapid decline in the disc mass in our simulation, unlike in the other cases we consider. This has a substantial influence on the planetary evolution: mass growth is reduced due to the flow of disc material across the inner and outer boundary. This explains the difference to the result from \citetalias{2008ApJ...685..560D}: they only model a part of the disc in radial direction. Their boundary conditions at the outer edge of the grid does not allow outflow (their Section~2.1.3).\footnote{They investigated the influence of the boundary conditions and conclude that the effect is small; however, they conducted these tests using the nominal case, not the high-viscosity case.}
In this case our results are likely to be more realistic than those presented in  \citetalias{2008ApJ...685..560D} because  we model the entire disc and our boundary conditions allow for outflow of disc material. In Sect.~\ref{sec_reslase} we study a case of a similar viscosity and show that our model performs reasonably well.
Our results for the high-viscosity case are very similar for both of the torque models discussed.

\subsection{Impulse approximation}
\label{subs_ia}

We also investigated how the classical impulse approximation \citep{1979MNRAS.188..191L,1979MNRAS.186..799L,1986ApJ...309..846L} and an improved version \citep{2002MNRAS.334..248A} compare to the simulations from \citetalias{2008ApJ...685..560D} in the context of our model. The impulse approximation explicitly neglects co-rotation torques from material inside the horseshoe region. Therefore, it should only be applied when a gap is already present. However, since it is widely used in the literature, it is still interesting to study how it behaves in our test cases.
For conciseness, these results are discussed in Appendix~\ref{app_ia}.

It is found that both prescriptions give worse agreement for accretion, migration, and gap depth than  torque mod and high mass torque. The modern approach with torque densities derived from hydrodynamic simulations should thus clearly be preferred.

\subsection{Type~I migration}

Recently, a new interpretation of type~II migration has been proposed. \citet{2018ApJ...861..140K} argue that it is actually the same as type~I migration, but using the surface density in the gap. We cannot directly investigate to what degree this approach would allow  type~II migration in our model since \citet{2018ApJ...861..140K} do not consider accretion. Here we studied a similar approach instead. For this test we turned off the tidal interaction between disc and planet. The migration of the planet is then calculated as a type~I torque based on the surface density at the planet location. This analysis can also be found in Appendix~\ref{app_ia}. In summary, we find that this approach gives good agreement with the results from \citetalias{2008ApJ...685..560D} early in the planet's evolution and when gap formation starts. Migration is slowed down significantly through the reduction of $\Sigma$ by mass accretion alone. However, migration typically does not slow down enough because no deep gap forms due to the lack of a type~II torque acting on the disc. Therefore, this approach does not seem adequate to model the long-term evolution of planets in the framework of our model. Nevertheless our investigation shows the importance of the accretion process on the planet's migration and gives surprisingly good agreement with \citetalias{2008ApJ...685..560D}  for accretion and for migration. We note that this approach does not conserve angular momentum.

\subsection{Large separation}
\label{sec_reslase}

So far our analysis has concentrated on initially low-mass planets that were released at $\SI{5.2}{au}$. \citetalias{2019MNRAS.486.4398F} perform a code comparison project where they compare the performance of a number of hydrodynamical codes. Their focus is on massive planets in the outer disc, such as would be expected from disc fragmentation (see \citealt{2016ARA&A..54..271K} for a review). However, they  only study discs that are not self-gravitating (Toomre-Q-parameter $> 2$, \citealt{1964ApJ...139.1217T}). It is found that the seven hydrodynamic codes they compared agree qualitatively on the outcome, but differ quantitatively. The authors also include a comparison with existing population synthesis studies and find that the agreement among them is worse than among the hydrodynamic codes.

\begin{figure*}
  \begin{subfigure}[pt]{0.49\textwidth}
  \includegraphics[width=\linewidth]{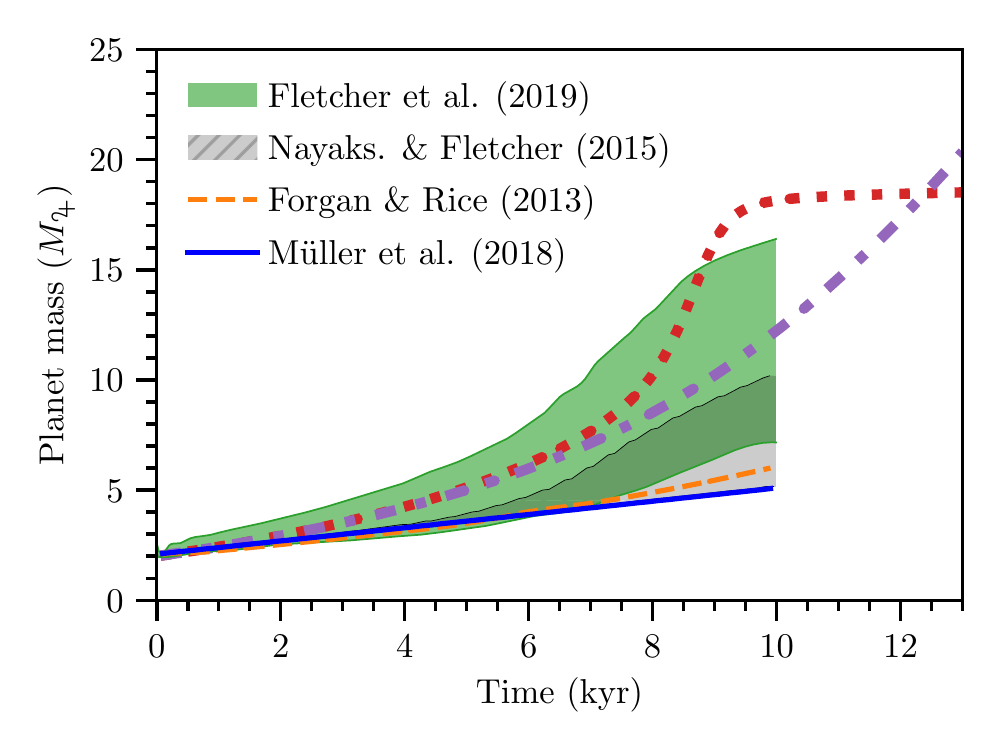}
  \end{subfigure}
  \begin{subfigure}[pt]{0.49\textwidth}
  \includegraphics[width=\linewidth]{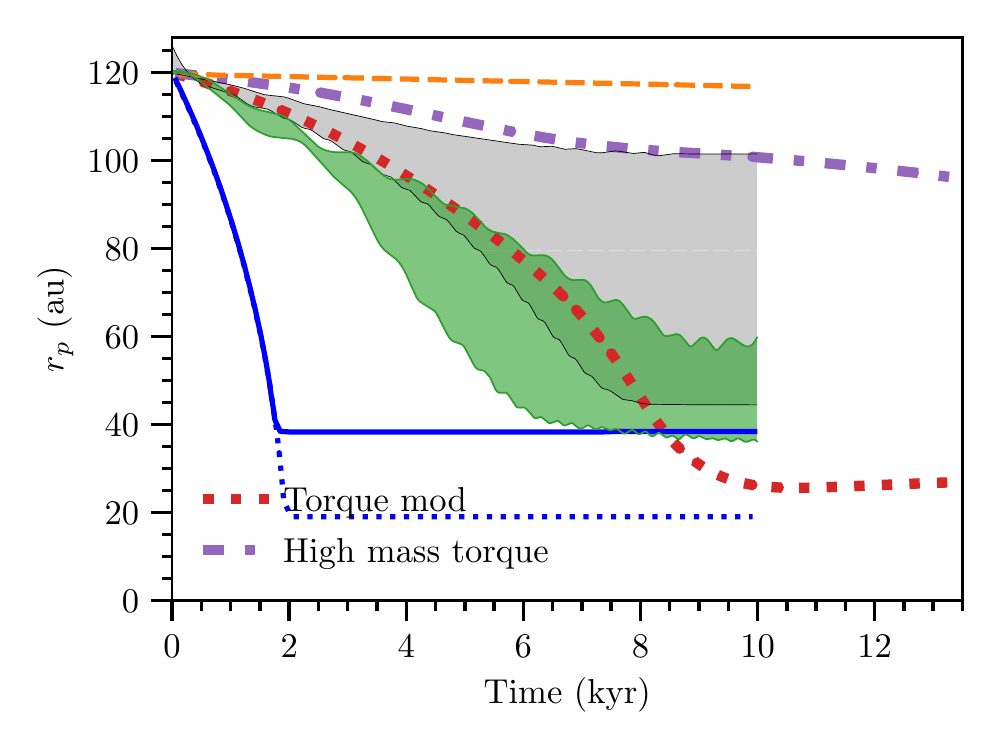}
  \end{subfigure}
  \caption{Comparison of our model to results from the code comparison project in \citetalias{2019MNRAS.486.4398F}. Evolution of an initially \SI{2}{\mj} planet inserted at $\SI{120}{au}$. Left: Mass vs time; right: Separation vs time. The green shaded region represents the region of parameter space covered by six different hydrodynamic codes. The results from our calculation are shown as thick dotted and dash-dotted lines (legend in the right panel). Results from different disc instability population synthesis codes are also shown (legend in the left panel). The grey shaded--hatched region corresponds to the region of parameter space covered by \citet{2015MNRAS.452.1654N}. The orange dashed line shows the result from \citet{2013MNRAS.432.3168F}. The result from \citet{2018ApJ...854..112M} is shown in blue. The dotted line (covered by the solid line in the left panel) represents a different gap opening criterion (see details in Sect.~\ref{sec_reslase}). }
  \label{fig_FNS}
\end{figure*}

In order to assess whether our method is also usable for large initial separations, we repeat one of their two test cases that include accretion. A planet of $\SI{2}{\mj}$ is inserted in the disc at $\SI{120}{au}$ and left to accrete and migrate. The results are shown  in Fig.~\ref{fig_FNS}.
\citetalias{2019MNRAS.486.4398F} also perform comparison runs with three different disc instability population synthesis codes: \citet{2013MNRAS.432.3168F}, \citet{2018ApJ...854..112M}, and \citet{2015MNRAS.452.1654N}. These results are also shown in the figure (digitised from their Fig.~9).

We do not expect a precise agreement with the simulations from \citetalias{2019MNRAS.486.4398F}.
The goal of this study is to compare the performance of different hydrodynamic codes. To this end, \citetalias{2019MNRAS.486.4398F}  choose highly idealised initial conditions. In particular, high-mass planets are inserted into an unperturbed disc. In a more realistic case, and if the planet had formed through core accretion, the disc would be significantly perturbed, and possibly already have a gap  in the presence of such a massive planet. Furthermore, accretion is modelled with a sink particle prescription.

Nevertheless, we assessed whether we can find a qualitative agreement with their work. For this comparison, we used the same initial surface density and temperature profiles as in  \citetalias{2019MNRAS.486.4398F}  (digitised from their Figure~1). For the torque mod torque density we used different $p_i$ parameters for Eq.~\ref{eq_f}, appropriate for the initial slopes  in \citetalias{2019MNRAS.486.4398F}: $(\beta,\zeta) = (1,0.5)$. The values are given in the rightmost column of Table~\ref{table_p}.

The results are shown in Fig.~\ref{fig_FNS}. The green shaded region represents the range of numerical results found in \citetalias{2019MNRAS.486.4398F}. As is immediately evident from the figure, our high mass torque prescription does not work well in this case. Migration proceeds much slower, despite the reduced or inverted torque density (right panel). Mass growth does not stop due to the high mass reservoir available at large separations (left panel).
The torque mod prescription fares much better. The mass evolution in this case is similar to that found by \citepalias{2019MNRAS.486.4398F}, though accretion stops later and more abruptly. Consequently the final mass reached in our simulation is approximately $15 \%$ higher than the highest value reached in \citetalias{2019MNRAS.486.4398F} after $\SI{10}{kyr}$ (data from \citetalias{2019MNRAS.486.4398F} is only available until $\SI{10}{kyr}$).
Migration proceeds in line with the slowest code (GIZMO-MFM) applied in \citetalias{2019MNRAS.486.4398F} during the first $\sim \SI{6}{kyr}$. It slows down later, leading to a somewhat smaller separation after $\SI{10}{kyr}$. This result is likely influenced by the different temperatures at small separations, as well as the higher planet mass in our simulation (see Sect.~\ref{sec:param}).

The difference in the migration tracks for our two torque models in this large separation case is noteworthy. In the other comparisons we performed the migration tracks were very similar. 
The reason for the discrepancy is that the surface density is unperturbed at the beginning of the simulation. In this scenario, the torque mod gives higher migration rates than the high mass torque since the amplitude of the latter is reduced (see Sect.~\ref{sect:Model}). When starting with a low initial planet mass, the surface density near the planet is gradually reduced, and by the time the high mass torque is reduced in amplitude significantly, the corresponding region of the disc is already partly depleted and contributes little to the overall torque.
We therefore expect the torque densities in the simulations of \citetalias{2019MNRAS.486.4398F} to differ from our high mass torque. This cannot be confirmed since the torque densities are not provided in \citetalias{2019MNRAS.486.4398F}.
Clearly, the high mass torque should not be used with massive planets in unperturbed discs. We note, however, that it may still be applicable in fragmenting discs if the fragment mass is removed from the disc (which is not done in \citetalias{2019MNRAS.486.4398F}.  In this case the surface density would be reduced automatically in the relevant region.

The population synthesis models also shown in Fig.~\ref{fig_FNS} exhibit a much larger spread in the final semi-major axis reached than what is seen in the hydrodynamic simulations. In the model of \citet{2013MNRAS.432.3168F} the planet opens a gap immediately, thus staying on a wide orbit. Conversely, the model from \citet{2018ApJ...854..112M} predicts very fast migration initially, and a gap is opened rather abruptly, when the planet reaches the inner disc. In order for a gap to open, the authors demand that the time to cross the gap is \num{100} times (solid line) or \num{1000} times (dotted line in Fig.~\ref{fig_FNS})  longer than the time it takes to open the gap. We note that \citet{2018ApJ...854..112M} use the migration timescale from \citet{2011MNRAS.416.1971B}, which is valid in self-gravitating discs. The initially much faster migration rates are therefore expected.
The result from the population synthesis code of \citet{2015MNRAS.452.1654N} is shown as the grey hatched region in Fig.~\ref{fig_FNS}. According to Sect.~4 of \citetalias{2019MNRAS.486.4398F}, the authors assumed that the viscosity $\alpha$-parameter is a random variable taking values between \num{0.005} and \num{0.05}. \citetalias{2019MNRAS.486.4398F} show the two extreme values in their Fig.~9. Our Fig.~\ref{fig_FNS} shows the entire range. The lower end of the grey shaded region in the right panel corresponds to $\alpha = 0.05$. Since the viscosity is close to this value in \citetalias{2019MNRAS.486.4398F}, we expect the model from \citet{2015MNRAS.452.1654N} to agree quite well with the hydrodynamic calculations when using the same viscosity. 

The mass accretion rate  also differs substantially between the codes. It is typically much lower in the population synthesis codes compared to the hydrodynamic simulations. The exception is the model from \citet{2015MNRAS.452.1654N}, when a high value for $\alpha$ is used (corresponding to the upper end of the hatched region in the left panel of Figure~\ref{fig_FNS}). This result is in line with the results of  \citetalias{2019MNRAS.486.4398F}. We note that accretion was neglected in \citet{2015MNRAS.452.1654N} (and  in \citealt{2013MNRAS.432.3168F}), and is  only added for the purpose of comparison in \citetalias{2019MNRAS.486.4398F}.

\section{Comparison with published prescriptions}\label{Sect:CompLit}
\begin{figure}
  \includegraphics[width=\linewidth]{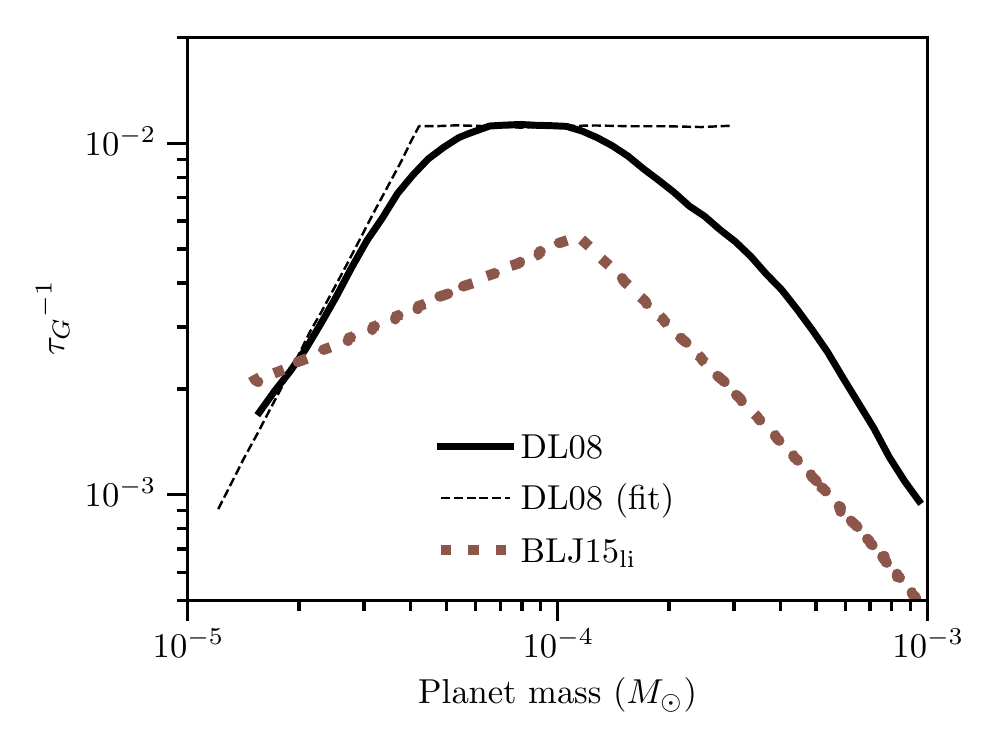}
  \caption{Inverse growth timescale, as in the top left panel of Fig.~\ref{fig_base}. The brown dotted line shows the result from the locally isothermal version of the model in \citetalias{2015A&A...582A.112B}.}
    \label{fig_rev_macc}
\end{figure}

\begin{figure*}
  \begin{subfigure}[pt]{0.49\textwidth}
  \includegraphics[width=\linewidth]{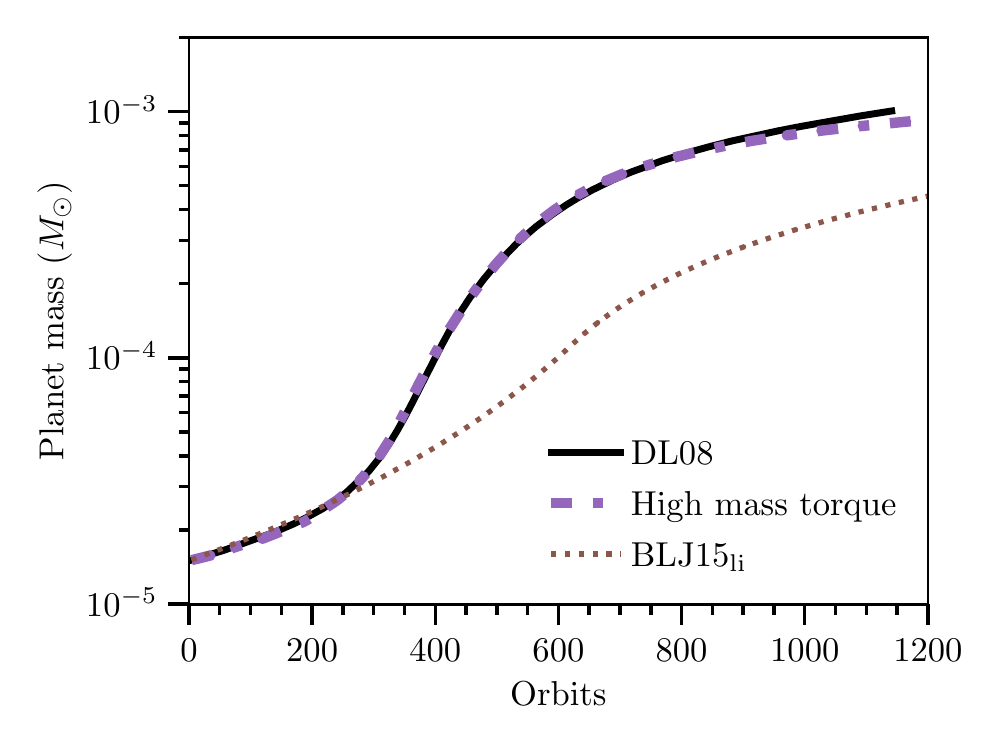}
  \end{subfigure}
  \begin{subfigure}[pt]{0.49\textwidth}
  \includegraphics[width=\linewidth]{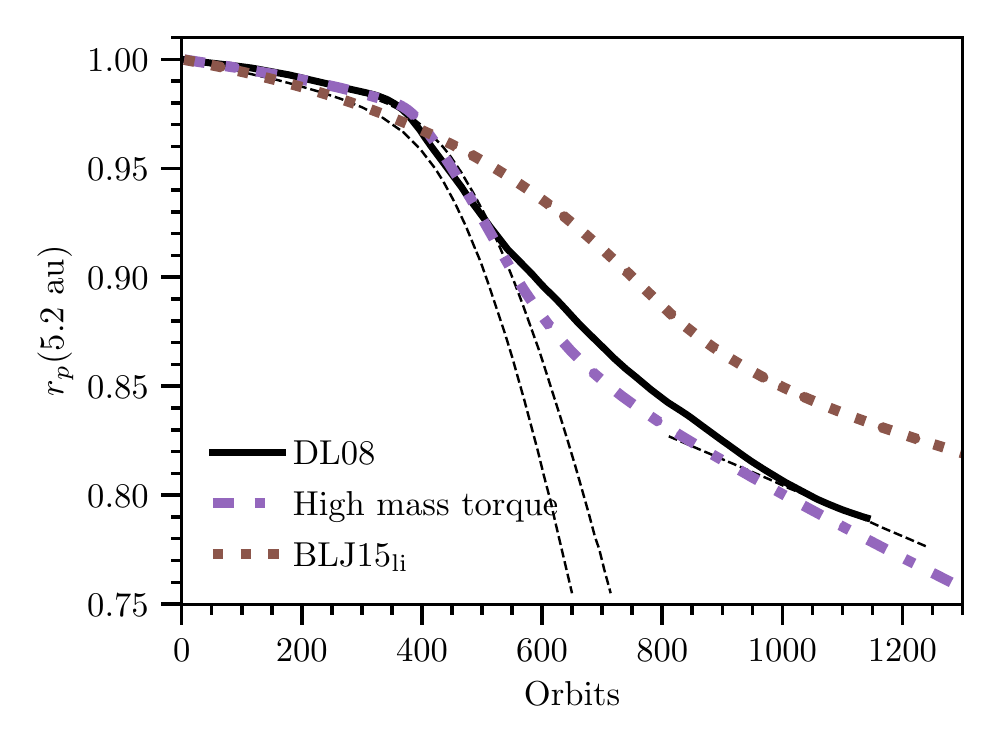}
  \end{subfigure}
  \caption{Same as Fig.~\ref{fig_base}, but for a comparison with \citetalias{2015A&A...582A.112B}$_\mathrm{li}$. The brown dotted line represents the results when using their prescription for migration and accretion (see  Appendix~\ref{app_resbase}). The results from our high mass torque model are shown for reference.}
\label{fig_rev_base}
\end{figure*}

In Sections \ref{sec:param} and \ref{sect:Results} we describe our model and identify its advantages over existing approaches. We also demonstrate how different iterations of our prescription compare to 3D hydrodynamic simulations.
In this section we compare our accretion model combined with the high mass torque to other existing prescriptions. We note that our model was specifically developed  to be used in situations where existing prescriptions are inapplicable (Sect.~\ref{sect:Introduction}). Nevertheless, it is interesting to compare different approaches in a regime where various published prescriptions are valid.
For this comparison we adopted the model presented in \citet{2015A&A...582A.112B} (BLJ15). The authors study accretion and migration of protoplanets in evolving discs in the pebble-based core accretion scenario. Since the accretion model presented here considers the disc-limited regime of gas accretion, we neglected the effects related to solid accretion in the comparison. For rapid gas accretion \citetalias{2015A&A...582A.112B} use the approach from \citet{machidakokubo2010}. Migration is described by the torque formula from \citet{Paardekooper2011b}, which includes non-isothermal and saturation effects. The discs in \citetalias{2008ApJ...685..560D} are locally isothermal. Therefore, in order to have a reasonable comparison, we applied the formula for the locally isothermal limit of  \citet{Paardekooper2010}. This means we are applying a modified version of the model in \citetalias{2015A&A...582A.112B}, denoted $\mathrm{BLJ15_{li}}$ for locally isothermal. In order to account for the planetary growth, including the formation of a gap and the transition into the type~II regime of migration, \citetalias{2015A&A...582A.112B} apply a reduction on the type~I timescale based on the gap's depth \citep{2007MNRAS.377.1324C}, a reduction of the type~II timescale \citep{Baruteau2014} and an additional smoothing between the two regimes. We use the same formulae and run the comparison for all the cases we studied in Sects.~\ref{sec_resbase} to \ref{sec_reshvsc}. The details can be found in Appendix~\ref{app_rev}. 
We find that the model used by \citetalias{2015A&A...582A.112B} gives lower gas accretion rates because the underlying accretion model of \citet{machidakokubo2010} already predicts lower accretion. In Fig.~\ref{fig_rev_macc} we show the inverse growth timescale for the $\mathrm{BLJ15_{li}}$ model applied to the baseline case (Sect.~\ref{sec_resbase}). The predicted accretion rates are lower by up to a factor of two. At the end of the simulations, the resulting planetary masses are therefore lower than those from \citetalias{2008ApJ...685..560D}.
This is seen in the left panel of Fig.~\ref{fig_rev_base}. It shows the time evolution of the planet's mass in the baseline case discussed in Sect.~\ref{sec_resbase}.
We find that gas accretion is significantly lower  in the Bondi and the  Hill regimes. The planet's mass reaches $\approx \SI{4e-4}{\msun}$, $\num{60} \%$ less compared to \citetalias{2008ApJ...685..560D} after $\num{1100}$ orbits.
The lower planetary masses also lead to slower migration in the type~I regime. As a result the planet remains further away from the star (right panel). Migration proceeds somewhat slower than seen in \citetalias{2008ApJ...685..560D}, although the slope of the  semi-major axis is similar (right panel) after migration slows down due to gap formation after around \num{800} orbits (i.e. fully in the type~II regime). The main reason for the slower migration is the lower planetary mass at earlier times (the type~I torque is proportional to $M_p$). We note that in the description of \citetalias{2015A&A...582A.112B}, the surface density is not perturbed by the planet. For the comparison, we also mimicked this behaviour in our model: the accreted gas is not removed from the disc (no conservation of mass), and the disc does not feel the tidal interaction with the planet. Therefore, the planet continues accreting even if a gap is formed, and may eventually reach a higher mass than  we see in the simulations, where accretion is modelled self-consistently.

\begin{figure*}
  \begin{subfigure}[pt]{0.49\textwidth}
  \includegraphics[width=\linewidth]{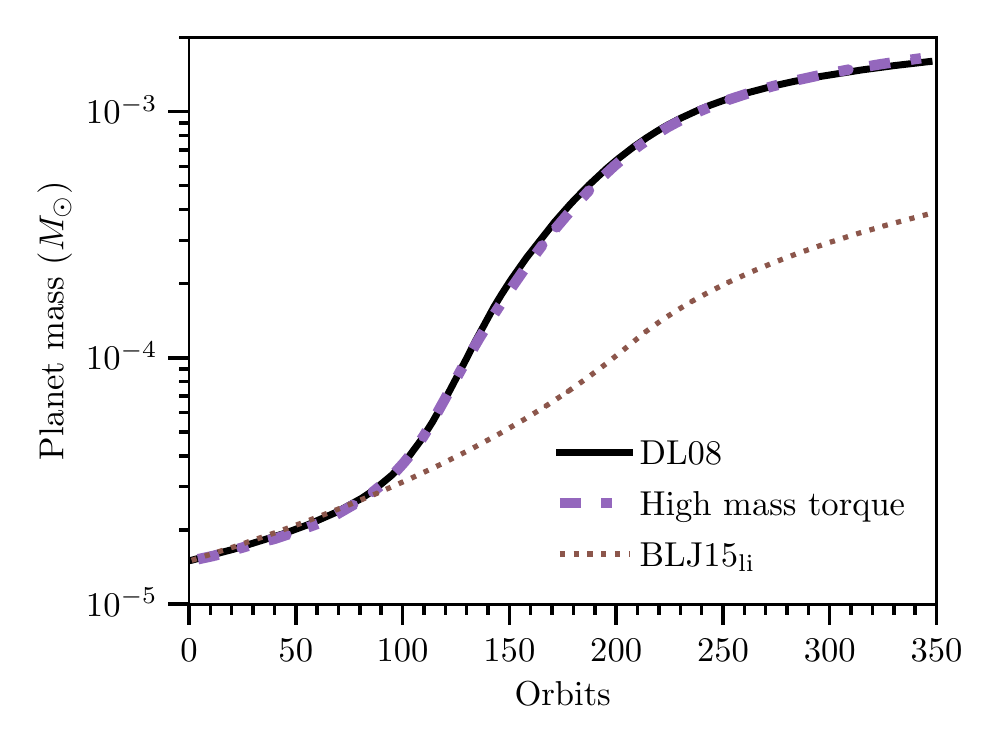}
  \end{subfigure}
  \begin{subfigure}[pt]{0.49\textwidth}
  \includegraphics[width=\linewidth]{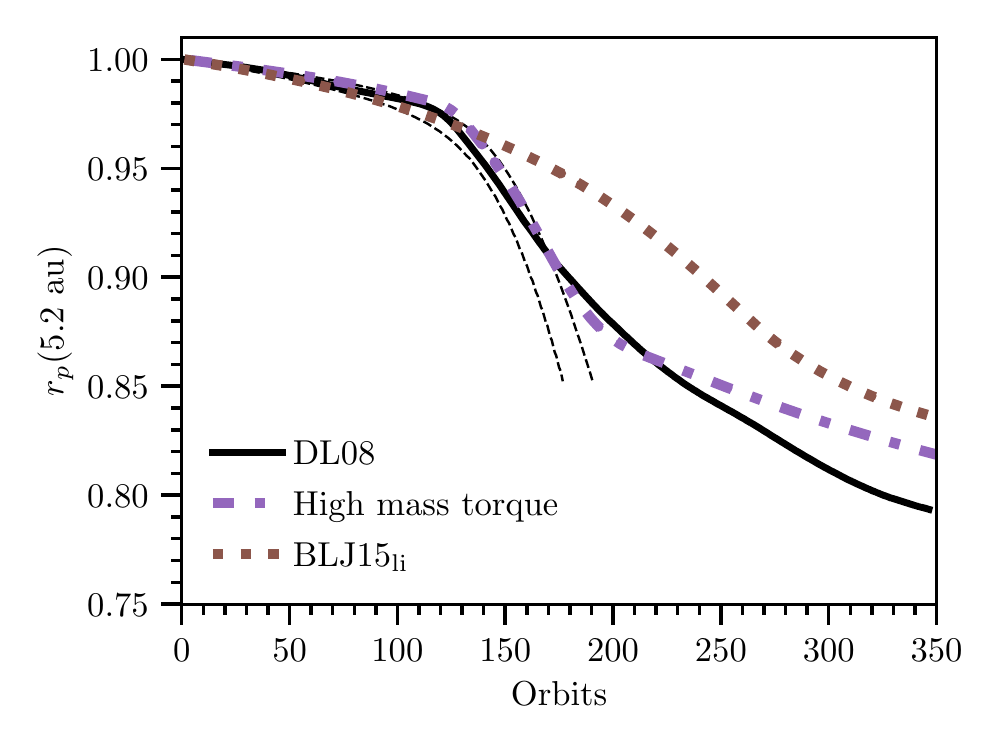}
  \end{subfigure}
  \caption{Same as Fig.~\ref{fig_rev_base}, but for  increased surface density (\SI{300}{g.cm^{-2}} at \SI{5.2}{au}).}
\label{fig_rev_hsig}
\end{figure*}

Figure~\ref{fig_rev_hsig} shows the planet's mass and semi-major axis as a function of time for the case with an increased initial surface density. Again a slower accretion leads to less migration compared to \citetalias{2015A&A...582A.112B}$_\mathrm{li}$, as seen in the left and right panels. Migration slows down due to gap formation at around \num{270} orbits.
The comparison with the other two cases from Sect.~\ref{sect:Results} can be found in Appendix~\ref{app_rev}.
Given the differences seen among hydrodynamic simulations of gas accretion themselves (see e.g.  Sect.~5 in \citealt{machidakokubo2010}), the agreement between our model and that used in \citetalias{2015A&A...582A.112B} is still reasonable.

\section{Discussion}\label{sect:Discussion}

Our comparisons with \citetalias{2008ApJ...685..560D} show reasonable agreement in four out of five cases. The discrepancy in the high-viscosity case, where mass accretion is much lower in our model, is not surprising. The corresponding hydrodynamic simulation is dominated by global effects. Our result seems reasonable in this case as well.
The agreement seen in the comparison with \citetalias{2008ApJ...685..560D} when using the high mass torque prescription indicates that applying torque densities obtained from hydrodynamic simulations in 1D models is a  promising approach for modelling migration and accretion simultaneously, including gap formation.
This is attractive since angular momentum is conserved in this approach.
Using the same torque densities for the high-mass wide-separation planets studied in \citetalias{2019MNRAS.486.4398F} did not give a good agreement. The high mass torque should therefore not be used with initially massive planets in unperturbed discs. However, the torque mod prescription gives reasonable agreement even in this case.

A limitation of our model is that it was studied only in discs that are not self-gravitating. This is also true for all the hydrodynamic simulations we compared our results to. If discs are self-gravitating, additional effects will have to be taken into account. Migration may be much faster \citep{2011MNRAS.416.1971B}, and accretion may be significantly reduced \citep{2013ApJ...767...63S}. A way to model migration in a self-gravitating disc and still conserve angular momentum is suggested in Sect.~3 of \citet{2015MNRAS.454...64N}. It could be applied in our model as well.

We also did not investigate the long-term evolution of our systems, mainly because no data is available for comparison from \citetalias{2008ApJ...685..560D} and \citetalias{2019MNRAS.486.4398F}. Generally we expect a good agreement also on longer timescales both for accretion and migration. At the end of the simulations, accretion has almost ceased because the mass reservoir is depleted, and the slopes of our migration tracks agree well with those in \citetalias{2008ApJ...685..560D}. Also, once a deep gap has opened, the exact shape of the torque density near the planet no longer has a strong influence on migration. There is one exception: once a deep gap has opened, mass flow through the gap is essentially stopped in 1D models. This is not seen in hydrodynamic models. On the contrary, significant mass transport is still observed even across deep gaps \citep{2015A&A...574A..52D,2017A&A...598A..80D,2011ApJ...729...47Z}. A possible way to account for this is described in \citet{lubowdangelo2006b,2016ApJ...828...33D}.

Our simulations also show that the surface density in the gap drops to extremely low values ($\ll \SI{e-10}{g.cm^{-2}}$) at some point. This effect is not seen in hydrodynamic simulations. For example in figures \ref{fig_base}, \ref{fig_hsig}, \ref{fig_app_base}, and \ref{fig_app_hsig}, $\Sigma_B / \Sigma_0$ remains at a small (positive) value in \citetalias{2008ApJ...685..560D}. This does not influence our results in terms of planet mass and semi-major axis, but this topic should be investigated further in the future.

Clearly, more work is needed to study the applicability of torque densities from complex simulations in 1D models. Future studies should focus on comparisons in a wide parameter space, realistic thermodynamics, and ideally also include the self-gravitating regime. Our work is a only first step in this direction.

\section{Summary, conclusions, and outlook}\label{sect:Conclusions}

We developed a 1D prescription for migration and accretion of planets in discs and calibrated it with results from hydrodynamic simulations. Its performance was assessed with different initial surface densities, temperatures, viscosities and planet locations, comparing to the results of two different hydrodynamic simulations. The accretion model is based on the Bondi--Hill accretion given in \citetalias{2008ApJ...685..560D}, refined and adapted to a 1D vertically integrated disc model. This model for accretion seems fairly robust, given the good agreement in our comparisons.

Migration is modelled by means of torque densities. We used the torque density formula from \citet{2010ApJ...724..730D} and implemented a modification inspired by \citet{2017MNRAS.469.3813H}. In a second approach we corrected for the influence of the mass growth of the planet by directly applying torque densities for high-mass planets from \citet{2010ApJ...724..730D}. Our results from the two approaches are in good agreement with the hydrodynamic simulations of \citetalias{2008ApJ...685..560D}. Our first modification also works well when applied to a $\SI{2}{\mj}$ planet inserted at a large separation.
We conclude the following:
\begin{itemize}
    \item Our proposed model for gas accretion can be applied in 1D models for a range of different parameters.
    \item Our torque mod migration model gives reasonable predictions both for initially low-mass planets at small separation and initially massive planets on wide orbits. However, it does not conserve angular momentum, and the high mass torque model should thus be preferred.
    \item The high mass torque model we propose provides a description of migration that works well in the range of parameters we studied. Together with our accretion model, it describes accurately the mass and semi-major axis evolution of initially small planets that undergo runaway accretion and open a gap. It is self-consistent by conserving mass and angular momentum. We therefore recommend this combination for application in 1D models of planet formation.
\end{itemize}

Our model for gas accretion works  reasonably well given the differences of  at most $15 \%$ using the high mass torque model and $30 \%$ when applying the torque mod  compared to \citetalias{2008ApJ...685..560D}. The result of the torque mod model gives $\sim 10-20 \%$ higher masses than the highest result in \citetalias{2019MNRAS.486.4398F} for an initially massive planet on a wide orbit. Nevertheless, it is important to study this topic further. Gas accretion is a challenging topic to study, also in hydrodynamic simulations. It may be influenced by the formation of circumplanetary discs that require high-resolution simulations to be described accurately (e.g. \citealt{2012ApJ...746..110Z,2017MNRAS.464.3158S,2020A&A...644A..41O}). More such simulations, covering a wide parameter space, are necessary to improve our understanding of accretion. This is particularly important if the discs are self-gravitating. In this regime, accretion is thought to be inhibited significantly due to the truncation of the circumplanetary disc \citep{2013ApJ...767...63S}.

Migration is also a challenge for hydrodynamic models. A proper treatment of thermodynamics is key to predicting accurate migration rates (see e.g. \citealt{2020MNRAS.496.1598R}). Even so,  only simulations applying simplified thermodynamics can run for long enough for a large number of initial conditions. Our results indicate that applying torque densities from hydrodynamic models in 1D codes is feasible, though more work is needed to test this in a wide parameter space.
It is still unclear to what degree torque densities from hydrodynamic simulations that consider more elaborate thermodynamics can be applied in 1D models. 
In the future, torque densities should be measured in radiation-hydrodynamic simulations for different planet masses, separations, and viscosities. Their applicability in 1D models should be reassessed.

Future observations may give essential clues about the migration process. For instance, kinematical detections of exoplanets may help us understand the gap formation process further, since they can probe the gas dynamics in protoplanetary discs with high precision \citep{2019Natur.574..378T}.

The prescription for migration based on the high mass torque and our  accretion scheme can be easily implemented in 1D  planet formation models. Details on the implementation are presented in Appendix~\ref{app_prac}.

Our study demonstrated the complexity of the topic and it is clear that further work is needed. 
In the future, we plan to implement our prescriptions in population synthesis studies of planet formation. Our aim is to add a treatment of clump evolution to our population synthesis of discs \citep{2021A&A...645A..43S}. The prescriptions provided in this paper will be an important part of this. The model for migration and accretion derived here should be useful for 1D models in general. Being able to describe accretion, migration, and gap formation self-consistently, relying only on local disc parameters, can improve predictions from planet population synthesis models and our understating of planet formation.

\begin{acknowledgements}
We thank the
anonymous referee for valuable comments. We also thank Gennaro D'Angelo and Clément Baruteau for the insightful discussions; and
Bertram Bitsch, Aurélien Crida, Tom Hands and Lucio Mayer for helpful comments.
This work has been carried out within the framework of the National Centre of Competence in Research PlanetS supported by the Swiss National Science Foundation. The authors acknowledge the financial support of the SNSF. O.S. and C.M. acknowledge the support from the Swiss National Science Foundation under grant 200021\_204847 ``PlanetsInTime''.
RH acknowledges support from SNSF grant \texttt{\detokenize{200020_188460}}.
\end{acknowledgements}

\section*{ORCID iDs}
Oliver\,Schib\,\orcidicon{}\\ \url{https://orcid.org/0000-0001-6693-7910}\\
Christoph\,Mordasini\,\orcidicon{}\\ \url{https://orcid.org/0000-0002-1013-2811}\\
Ravit\,Helled\,\orcidicon{}\\ \url{https://orcid.org/0000-0001-5555-2652}\\

\bibliographystyle{aa}
\bibliography{library.bib}

\begin{thebibliography}{71}
\expandafter\ifx\csname natexlab\endcsname\relax\def\natexlab#1{#1}\fi

\bibitem[{Alibert {et~al.}(2005)Alibert, Mordasini, Benz, \&
  Winisdoerffer}]{Alibert2005}
Alibert, Y., Mordasini, C., Benz, W., \& Winisdoerffer, C. 2005, Astronomy and
  Astrophysics, 434, 343

\bibitem[{Anderson {et~al.}(2018)Anderson, Bouchy, Brown, {Collier Cameron},
  Delrez, Gillon, {Gonz{\'{a}}lez Hern{\'{a}}ndez}, Hellier, Jehin, Lendl,
  Maxted, Neveu-VanMalle, Nielsen, Pepe, Perger, Pollacco, Queloz, Rey,
  S{\'{e}}gransan, Smalley, Toledo-Padr{\'{o}}n, Triaud, Turner, Udry, \&
  West}]{2018arXiv181209264A}
Anderson, D., Bouchy, F., Brown, D., {et~al.} 2018, arXiv e-prints,
  arXiv:1812.09264

\bibitem[{Armitage {et~al.}(2002)Armitage, Livio, Lubow, \&
  Pringle}]{2002MNRAS.334..248A}
Armitage, P.~J., Livio, M., Lubow, S., \& Pringle, J. 2002, Monthly Notices of
  the RAS, 334, 248

\bibitem[{Ayliffe \& Bate(2009)}]{ayliffebate2009}
Ayliffe, B.~A. \& Bate, M.~R. 2009, Monthly Notices of the RAS, 393, 49

\bibitem[{Baruteau {et~al.}(2016)Baruteau, Bai, Mordasini, \&
  Molli{\`{e}}re}]{2016SSRv..205...77B}
Baruteau, C., Bai, X., Mordasini, C., \& Molli{\`{e}}re, P. 2016, Space Science
  Reviews, 205, 77

\bibitem[{Baruteau {et~al.}(2014)Baruteau, Crida, Paardekooper, Masset, Guilet,
  Bitsch, Nelson, Kley, \& Papaloizou}]{Baruteau2014}
Baruteau, C., Crida, A., Paardekooper, S.-J., {et~al.} 2014, Protostars and
  Planets VI, 3

\bibitem[{Baruteau \& Masset(2013)}]{2013LNP...861..201B}
Baruteau, C. \& Masset, F. 2013, {Recent Developments in Planet Migration
  Theory}, ed. J.~Souchay, S.~Mathis, \& T.~Tokieda, Vol. 861, 201

\bibitem[{Baruteau {et~al.}(2011)Baruteau, Meru, \&
  Paardekooper}]{2011MNRAS.416.1971B}
Baruteau, C., Meru, F., \& Paardekooper, S.-J. 2011, Monthly Notices of the
  RAS, 416, 1971

\bibitem[{Birnstiel {et~al.}(2010)Birnstiel, Dullemond, \&
  Brauer}]{2010A&A...513A..79B}
Birnstiel, T., Dullemond, C., \& Brauer, F. 2010, Astronomy and Astrophysics,
  513, A79

\bibitem[{Bitsch {et~al.}(2015)Bitsch, Lambrechts, \&
  Johansen}]{2015A&A...582A.112B}
Bitsch, B., Lambrechts, M., \& Johansen, A. 2015, Astronomy and Astrophysics,
  582, A112

\bibitem[{Crida \& Morbidelli(2007)}]{2007MNRAS.377.1324C}
Crida, A. \& Morbidelli, A. 2007, Monthly Notices of the RAS, 377, 1324

\bibitem[{Crida {et~al.}(2006)Crida, Morbidelli, \&
  Masset}]{2006Icar..181..587C}
Crida, A., Morbidelli, A., \& Masset, F. 2006, Icarus, 181, 587

\bibitem[{D'Angelo \& Bodenheimer(2016)}]{2016ApJ...828...33D}
D'Angelo, G. \& Bodenheimer, P. 2016, Astrophysical Journal, 828, 33

\bibitem[{D'Angelo \& Lubow(2008)}]{2008ApJ...685..560D}
D'Angelo, G. \& Lubow, S.~H. 2008, Astrophysical Journal, 685, 560

\bibitem[{D'Angelo \& Lubow(2010)}]{2010ApJ...724..730D}
D'Angelo, G. \& Lubow, S.~H. 2010, Astrophysical Journal, 724, 730

\bibitem[{Dawson \& Johnson(2018)}]{2018ARA&A..56..175D}
Dawson, R.~I. \& Johnson, J.~A. 2018, Annual Review of Astron and Astrophys,
  56, 175

\bibitem[{Dittkrist {et~al.}(2014)Dittkrist, Mordasini, Klahr, Alibert, \&
  Henning}]{2014A&A...567A.121D}
Dittkrist, K.~M., Mordasini, C., Klahr, H., Alibert, Y., \& Henning, T. 2014,
  Astronomy and Astrophysics, 567, A121

\bibitem[{D{\"{u}}rmann \& Kley(2015)}]{2015A&A...574A..52D}
D{\"{u}}rmann, C. \& Kley, W. 2015, Astronomy and Astrophysics, 574, A52

\bibitem[{D{\"{u}}rmann \& Kley(2017)}]{2017A&A...598A..80D}
D{\"{u}}rmann, C. \& Kley, W. 2017, Astronomy and Astrophysics, 598, A80

\bibitem[{Emsenhuber {et~al.}(2021)Emsenhuber, Mordasini, Burn, Alibert, Benz,
  \& Asphaug}]{2021A&A...656A..69E}
Emsenhuber, A., Mordasini, C., Burn, R., {et~al.} 2021, Astronomy and
  Astrophysics, 656, A69

\bibitem[{Fletcher {et~al.}(2019)Fletcher, Nayakshin, Stamatellos, Dehnen,
  Meru, Mayer, Deng, \& Rice}]{2019MNRAS.486.4398F}
Fletcher, M., Nayakshin, S., Stamatellos, D., {et~al.} 2019, Monthly Notices of
  the Royal Astronomical Society, 486, 4398

\bibitem[{Forgan \& Rice(2013)}]{2013MNRAS.432.3168F}
Forgan, D. \& Rice, K. 2013, Monthly Notices of the RAS, 432, 3168

\bibitem[{Fung {et~al.}(2014)Fung, Shi, \& Chiang}]{2014ApJ...782...88F}
Fung, J., Shi, J.-M., \& Chiang, E. 2014, Astrophysical Journal, 782, 88

\bibitem[{Goldreich \& Ward(1973)}]{goldreichward1973}
Goldreich, P. \& Ward, W.~R. 1973, Astrophysical Journal, 183, 1051

\bibitem[{Hallam \& Paardekooper(2017)}]{2017MNRAS.469.3813H}
Hallam, P. \& Paardekooper, S.~J. 2017, Monthly Notices of the RAS, 469, 3813

\bibitem[{Hueso \& Guillot(2005)}]{2005A&A...442..703H}
Hueso, R. \& Guillot, T. 2005, Astronomy and Astrophysics, 442, 703

\bibitem[{Ida {et~al.}(2020)Ida, Muto, Matsumura, \&
  Brasser}]{2020MNRAS.494.5666I}
Ida, S., Muto, T., Matsumura, S., \& Brasser, R. 2020, Monthly Notices of the
  RAS, 494, 5666

\bibitem[{Ikoma {et~al.}(2001)Ikoma, Emori, \& Nakazawa}]{2001ApJ...553..999I}
Ikoma, M., Emori, H., \& Nakazawa, K. 2001, Astrophysical Journal, 553, 999

\bibitem[{Kanagawa {et~al.}(2017)Kanagawa, Tanaka, Muto, \&
  Tanigawa}]{2017PASJ...69...97K}
Kanagawa, K.~D., Tanaka, H., Muto, T., \& Tanigawa, T. 2017, Publications of
  the ASJ, 69, 97

\bibitem[{Kanagawa {et~al.}(2018)Kanagawa, Tanaka, \&
  Szuszkiewicz}]{2018ApJ...861..140K}
Kanagawa, K.~D., Tanaka, H., \& Szuszkiewicz, E. 2018, Astrophysical Journal,
  861, 140

\bibitem[{Kley {et~al.}(2009)Kley, Bitsch, \& Klahr}]{kleybitsch2009}
Kley, W., Bitsch, B., \& Klahr, H. 2009, Astronomy and Astrophysics, 506, 971

\bibitem[{Kley \& Crida(2008)}]{2008A&A...487L...9K}
Kley, W. \& Crida, A. 2008, Astronomy and Astrophysics, 487, L9

\bibitem[{Kley \& Nelson(2012)}]{2012ARA&A..50..211K}
Kley, W. \& Nelson, R. 2012, Annual Review of Astron and Astrophys, 50, 211

\bibitem[{Kratter \& Lodato(2016)}]{2016ARA&A..54..271K}
Kratter, K. \& Lodato, G. 2016, Annual Review of Astronomy and Astrophysics,
  54, 271

\bibitem[{Lin \& Papaloizou(1979{\natexlab{a}})}]{1979MNRAS.188..191L}
Lin, D. \& Papaloizou, J. 1979{\natexlab{a}}, Monthly Notices of the RAS, 188,
  191

\bibitem[{Lin \& Papaloizou(1979{\natexlab{b}})}]{1979MNRAS.186..799L}
Lin, D. \& Papaloizou, J. 1979{\natexlab{b}}, Monthly Notices of the RAS, 186,
  799

\bibitem[{Lin \& Papaloizou(1986)}]{1986ApJ...309..846L}
Lin, D. \& Papaloizou, J. 1986, Astrophysical Journal, 309, 846

\bibitem[{Lubow \& D'Angelo(2006)}]{lubowdangelo2006b}
Lubow, S.~H. \& D'Angelo, G. 2006, Astrophysical Journal, 641, 526

\bibitem[{Lubow \& Ida(2010)}]{lubowida2010}
Lubow, S.~H. \& Ida, S. 2010, arXiv, astro-ph.E

\bibitem[{Lubow {et~al.}(1999)Lubow, Seibert, \& Artymowicz}]{lubowseibert1999}
Lubow, S.~H., Seibert, M., \& Artymowicz, P. 1999, Astrophysical Journal, 526,
  1001

\bibitem[{L{\"{u}}st(1952)}]{1952ZNatA...7...87L}
L{\"{u}}st, R. 1952, Zeitschrift Naturforschung Teil A, 7, 87

\bibitem[{Lynden-Bell \& Pringle(1974)}]{1974MNRAS.168..603L}
Lynden-Bell, D. \& Pringle, J. 1974, Monthly Notices of the RAS, 168, 603

\bibitem[{Machida {et~al.}(2010)Machida, Kokubo, Inutsuka, \&
  Matsumoto}]{machidakokubo2010}
Machida, M.~N., Kokubo, E., Inutsuka, S.-I., \& Matsumoto, T. 2010, arXiv,
  astro-ph.S

\bibitem[{Malik {et~al.}(2015)Malik, Meru, Mayer, \&
  Meyer}]{2015ApJ...802...56M}
Malik, M., Meru, F., Mayer, L., \& Meyer, M. 2015, Astrophysical Journal, 802,
  56

\bibitem[{Masset \& Papaloizou(2003)}]{2003ApJ...588..494M}
Masset, F. \& Papaloizou, J. 2003, Astrophysical Journal, 588, 494

\bibitem[{Masset \& Casoli(2010)}]{massetcasoli2010}
Masset, F.~S. \& Casoli, J. 2010, Astrophysical Journal, 723, 1393

\bibitem[{Mayor \& Queloz(1995)}]{1995Natur.378..355M}
Mayor, M. \& Queloz, D. 1995, Nature, 378, 355

\bibitem[{Mordasini {et~al.}(2012)Mordasini, Alibert, Klahr, \&
  Henning}]{mordasinialibert2012b}
Mordasini, C., Alibert, Y., Klahr, H., \& Henning, T. 2012, Astronomy and
  Astrophysics, 547, 111

\bibitem[{Mordasini {et~al.}(2014)Mordasini, Klahr, Alibert, Miller, \&
  Henning}]{2014A&A...566A.141M}
Mordasini, C., Klahr, H., Alibert, Y., Miller, N., \& Henning, T. 2014,
  Astronomy and Astrophysics, 566, A141

\bibitem[{Movshovitz {et~al.}(2010)Movshovitz, Bodenheimer, Podolak, \&
  Lissauer}]{movshovitzbodenheimer2010a}
Movshovitz, N., Bodenheimer, P.~H., Podolak, M., \& Lissauer, J.~L. 2010,
  arXiv, astro-ph.E

\bibitem[{M{\"{u}}ller {et~al.}(2018)M{\"{u}}ller, Helled, \&
  Mayer}]{2018ApJ...854..112M}
M{\"{u}}ller, S., Helled, R., \& Mayer, L. 2018, Astrophysical Journal, 854,
  112

\bibitem[{Nakamoto \& Nakagawa(1994)}]{1994ApJ...421..640N}
Nakamoto, T. \& Nakagawa, Y. 1994, Astrophysical Journal, 421, 640

\bibitem[{Nayakshin(2015)}]{2015MNRAS.454...64N}
Nayakshin, S. 2015, Monthly Notices of the RAS, 454, 64

\bibitem[{Nayakshin \& Fletcher(2015)}]{2015MNRAS.452.1654N}
Nayakshin, S. \& Fletcher, M. 2015, Monthly Notices of the Royal Astronomical
  Society, 452, 1654

\bibitem[{Oliva \& Kuiper(2020)}]{2020A&A...644A..41O}
Oliva, G.~A. \& Kuiper, R. 2020, Astronomy and Astrophysics, 644, A41

\bibitem[{Paardekooper {et~al.}(2010)Paardekooper, Baruteau, Crida, \&
  Kley}]{Paardekooper2010}
Paardekooper, S.~J., Baruteau, C., Crida, A., \& Kley, W. 2010, Monthly Notices
  of the Royal Astronomical Society, 401, 1950

\bibitem[{Paardekooper {et~al.}(2011)Paardekooper, Baruteau, \&
  Kley}]{Paardekooper2011b}
Paardekooper, S.~J., Baruteau, C., \& Kley, W. 2011, Monthly Notices of the
  Royal Astronomical Society, 410, 293

\bibitem[{Paardekooper \& Mellema(2006)}]{2006A&A...459L..17P}
Paardekooper, S.~J. \& Mellema, G. 2006, Astronomy and Astrophysics, 459, L17

\bibitem[{Papaloizou \& Lin(1984)}]{1984ApJ...285..818P}
Papaloizou, J. \& Lin, D. 1984, Astrophysical Journal, 285, 818

\bibitem[{Pollack {et~al.}(1996)Pollack, Hubickyj, Bodenheimer, Lissauer,
  Podolak, \& Greenzweig}]{Pollack1996}
Pollack, J.~B., Hubickyj, O., Bodenheimer, P., {et~al.} 1996, Icarus, 124, 62

\bibitem[{Rowther \& Meru(2020)}]{2020MNRAS.496.1598R}
Rowther, S. \& Meru, F. 2020, Monthly Notices of the RAS, 496, 1598

\bibitem[{Schib {et~al.}(2021)Schib, Mordasini, Wenger, Marleau, \&
  Helled}]{2021A&A...645A..43S}
Schib, O., Mordasini, C., Wenger, N., Marleau, G.~D., \& Helled, R. 2021,
  Astronomy and Astrophysics, 645, A43

\bibitem[{Shabram \& Boley(2013)}]{2013ApJ...767...63S}
Shabram, M. \& Boley, A.~C. 2013, Astrophysical Journal, 767, 63

\bibitem[{Shakura \& Sunyaev(1973)}]{1973A&A....24..337S}
Shakura, N. \& Sunyaev, R. 1973, Astronomy and Astrophysics, 500, 33

\bibitem[{Szul{\'{a}}gyi {et~al.}(2017)Szul{\'{a}}gyi, Mayer, \&
  Quinn}]{2017MNRAS.464.3158S}
Szul{\'{a}}gyi, J., Mayer, L., \& Quinn, T. 2017, Monthly Notices of the RAS,
  464, 3158

\bibitem[{Tanaka {et~al.}(2002)Tanaka, Takeuchi, \& Ward}]{2002ApJ...565.1257T}
Tanaka, H., Takeuchi, T., \& Ward, W.~R. 2002, The Astrophysical Journal, 565,
  1257

\bibitem[{Teague {et~al.}(2019)Teague, Bae, \& Bergin}]{2019Natur.574..378T}
Teague, R., Bae, J., \& Bergin, E.~A. 2019, Nature, 574, 378

\bibitem[{Toomre(1964)}]{1964ApJ...139.1217T}
Toomre, A. 1964, Astrophysical Journal, 139, 1217

\bibitem[{Ward(1997)}]{1997Icar..126..261W}
Ward, W.~R. 1997, Icarus, 126, 261

\bibitem[{Zhu {et~al.}(2012)Zhu, Hartmann, Nelson, \&
  Gammie}]{2012ApJ...746..110Z}
Zhu, Z., Hartmann, L., Nelson, R.~P., \& Gammie, C.~F. 2012, Astrophysical
  Journal, 746, 110

\bibitem[{Zhu {et~al.}(2011)Zhu, Nelson, Hartmann, Espaillat, \&
  Calvet}]{2011ApJ...729...47Z}
Zhu, Z., Nelson, R.~P., Hartmann, L., Espaillat, C., \& Calvet, N. 2011,
  Astrophysical Journal, 729, 47

\end{thebibliography}

\newpage

\begin{appendix}

\section{Evolution of the surface density}
\label{app_sd}
Here we present the evolution of the surface density for the cases not shown in Fig.~\ref{fig_surfdens}.
The top left panel shows the case of a higher initial surface density (Sect.~\ref{sec_reshsig}, $\SI{300}{g.cm^{-2}}$ at $\SI{5.2}{au}$). The gap opening proceeds qualitatively similarly to the baseline case, but much faster, as is expected from the faster increase of the planet's mass.
In the top right panel the very high surface density (Sect.~\ref{sec_reshsig}, initially $\SI{500}{g.cm^{-2}}$ at $\SI{5.2}{au}$) is shown. We note that the scaling of the y-axis is different from the other panels. Gap opening is even faster here due to the rapid  increase in the planet mass.
The bottom left panel shows the low-temperature case (Sect~\ref{sec_resltmp}, $H/r = 0.04$). The evolution of the surface density is qualitatively very similar to the baseline case, though much faster due to the stronger torque at lower temperature.
In the bottom right panel we finally show the high-viscosity case (Sect~\ref{sec_reshvsc}, $\SI{e16}{cm^2.s^{-1}}$). Here the evolution of the disc is important, as discussed. The background surface density drops by more than a factor of two during the first \num{500} orbits. Due to the high viscosity, it takes a very long time  for a gap to open. We note that the time interval shown here is much longer than in the simulation of \citetalias{2008ApJ...685..560D}.
\begin{figure*}
  \begin{subfigure}[pt]{0.49\textwidth}
  \includegraphics[width=\linewidth]{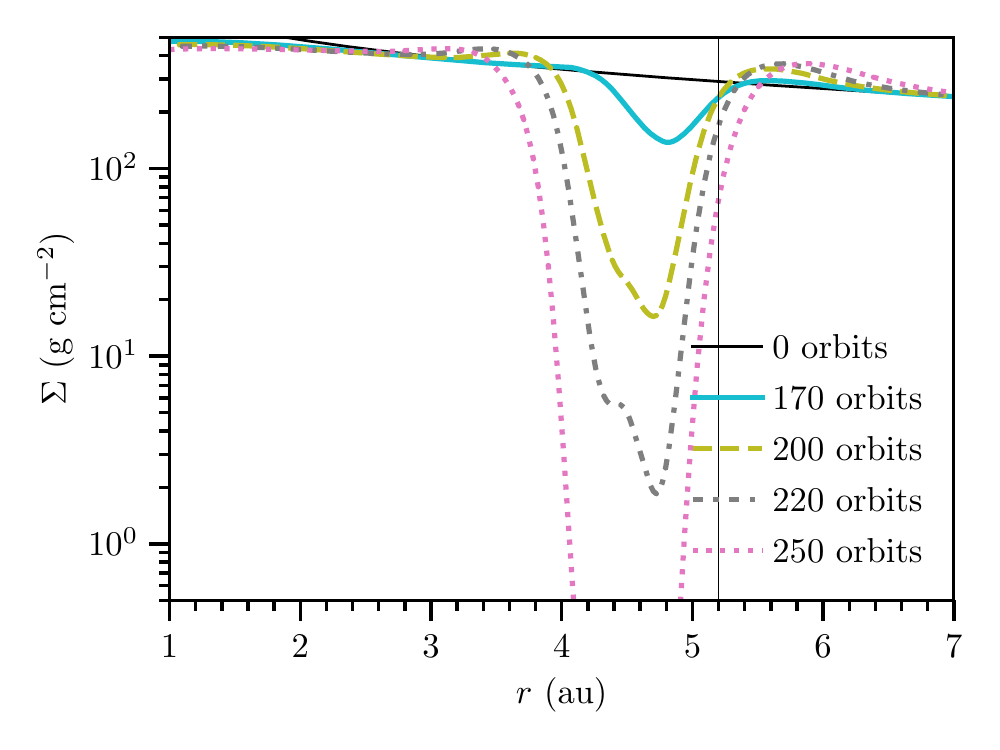}
  \end{subfigure}
  \begin{subfigure}[pt]{0.49\textwidth}
  \includegraphics[width=\linewidth]{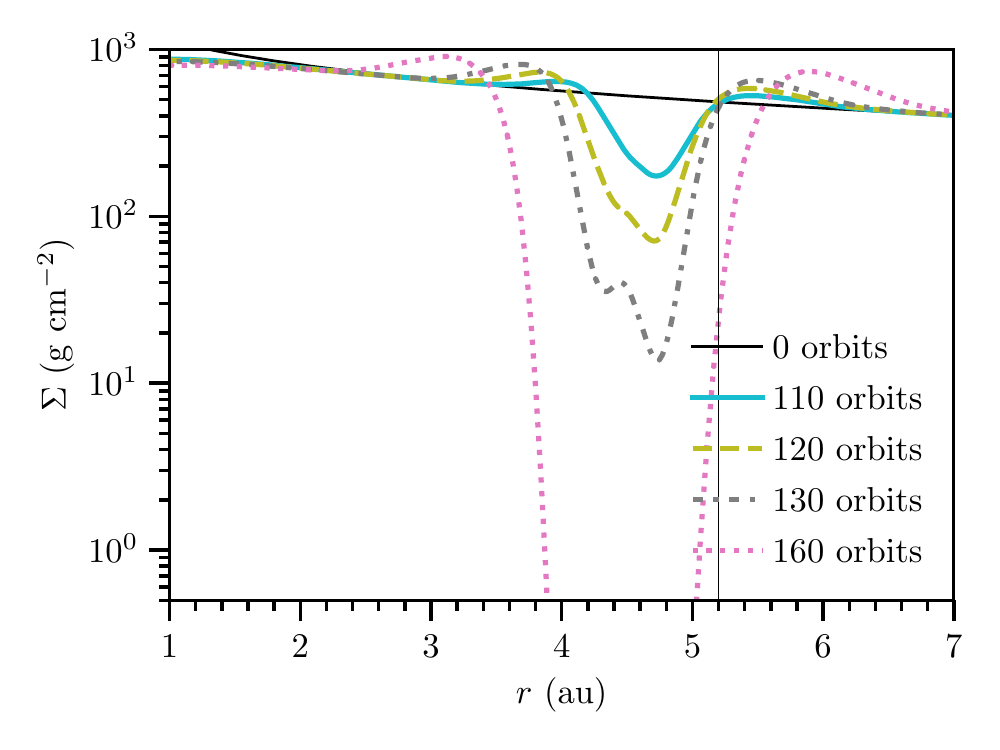}
  \end{subfigure}
  \begin{subfigure}[pt]{0.49\textwidth}
  \includegraphics[width=\linewidth]{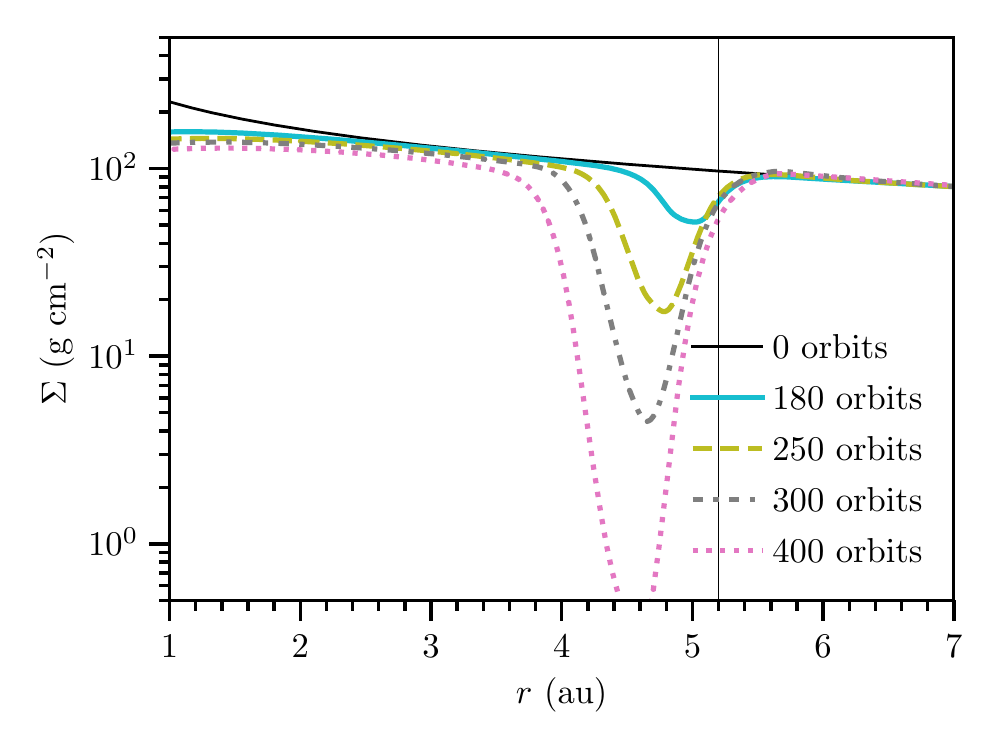}
  \end{subfigure}
  \begin{subfigure}[pt]{0.49\textwidth}
  \includegraphics[width=\linewidth]{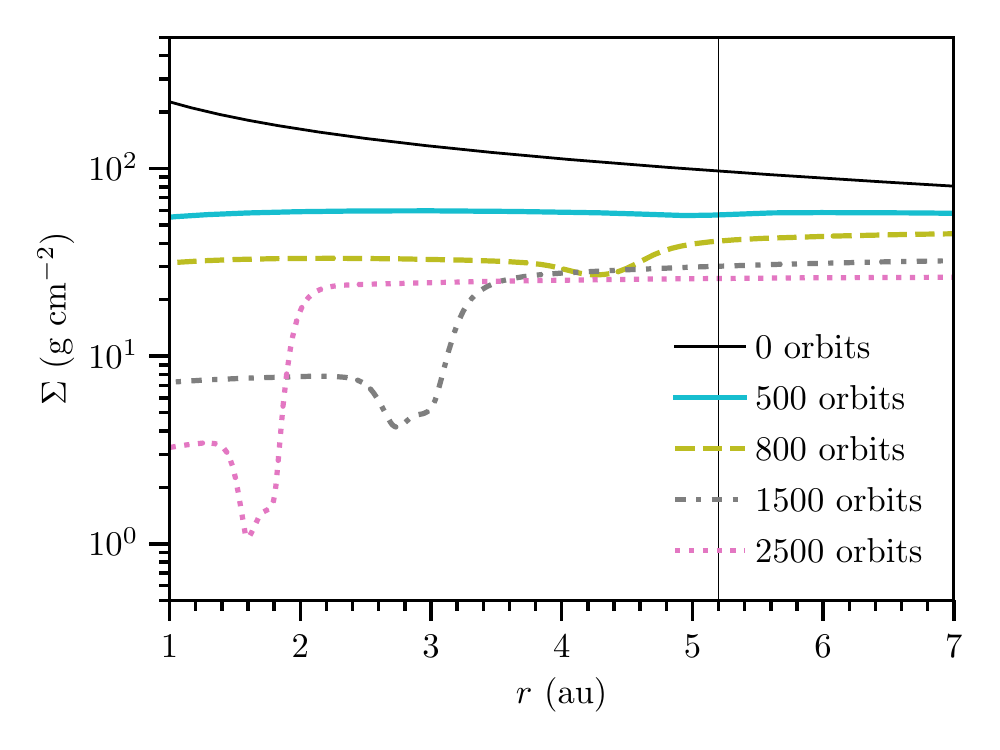}
  \end{subfigure}
  \caption{Evolution of the surface density for the cases not shown in Fig.~\ref{fig_surfdens}. Top left: Higher surface density (initially $\SI{300}{g.cm^{-2}}$ at $\SI{5.2}{au}$). Top right: Very high surface density (initially $\SI{500}{g.cm^{-2}}$ at $\SI{5.2}{au}$). Bottom left: Lower temperature ($H/r = 0.04$). Bottom right: Higher viscosity ($\SI{e16}{cm^2.s^{-1}}$).}
  \label{fig_sigall}
\end{figure*}

\FloatBarrier

\section{Impulse approximation and type~I migration}
\label{app_ia}

In this section we investigate how some other migration prescriptions used in the literature compare to our test cases. One of them is the classic impulse approximation \citep{1979MNRAS.188..191L,1979MNRAS.186..799L,1986ApJ...309..846L}
\begin{equation}
\label{eq_ia}
    \frac{d \mathcal{T}}{dm} = \sign(r_p-r) \frac{f}{2} \Omega(r_p)^2 r_p^2 \left(\frac{M_p}{\mstar}\right)^2 \left(\frac{r}{\left|\Delta_p \right|}\right)^4,
\end{equation}
where the term $\left| \Delta_p \right| = \max(H,\left| r - r_p \right|)$ excludes material closer than one scale height from the planet. The factor $f$ is an order of unity constant taken to be \num{0.23} \citep{1984ApJ...285..818P}. The function $\sign(r_p-r)$ is replaced by $(r_p-r)/H$ inside of one scale height to prevent a discontinuity at $r = r_p$ \citep{1986ApJ...309..846L}. We  call Eq.~\ref{eq_ia} the  \citetalias{1986ApJ...309..846L} formula in the following.

Another torque density commonly used is that given in \citet{2002MNRAS.334..248A}:

\begin{equation}
    \frac{d \mathcal{T}}{dm} =
    \begin{cases}-\frac{1}{2} \Omega(r_p)^2 r_p^2 \left(\frac{M_p}{\mstar}\right)^2 \left(\frac{r}{\left|\Delta_p \right|}\right)^4, & r < r_p\\
    \frac{1}{2} \Omega(r_p)^2 r_p^2 \left(\frac{M_p}{\mstar}\right)^2 \left(\frac{r_p}{\left|\Delta_p \right|}\right)^4, & r > r_p.
    \end{cases}
\end{equation}
It is similar to Eq.~\ref{eq_ia}, but it uses a modification to give a symmetric treatment for the disc inside and outside the planet's orbit. We  call it the  \citetalias{2002MNRAS.334..248A} formula from now on. The \citetalias{1986ApJ...309..846L} formula and the \citetalias{2002MNRAS.334..248A} formula are shown in Fig.~\ref{fig_torque}.

Recently, it has been proposed that type~II migration is nothing other than type~I migration that uses the reduced surface density in the gap \citep{2018ApJ...861..140K}. The authors did not include gas accretion by the planet. We study a modification of this scenario, which we call `type~I', by applying a standard type~I migration timescale $\tau_I$ \citep{2002ApJ...565.1257T}
\begin{equation}
    \tau_I = (2.7 + 1.1~\beta)^{-1} \frac{\mstar}{M_p} \frac{\mstar}{\Sigma_p r_p^2} \left( \frac{c_\mathrm{s}}{r_p \Omega_p}\right) \Omega_p^{-1}
\end{equation}
without applying a torque to the disc.
This violates the conservation of angular momentum, but it allows us  to study an interesting effect. The surface density near the planet is still reduced due to accretion. The migration of the planet is slowed down exclusively through this reduction.

We do not discuss the case with increased viscosity again here. As noted in Sect.~\ref{sec_reshvsc}, this case is dominated by global effects.
As for the earlier cases, the parameter $C_\mathrm{H}$ is chosen in such a way that the inverse growth timescale agrees with the analytic formula from \citetalias{2008ApJ...685..560D}. In the case of the type~I torque we also revert  to using a feeding zone radius of $1 R_\mathrm{H}$ to prevent too high accretion rates. The numerical values chosen for these parameters are given in Table~\ref{table_config}.

\subsection{Baseline case}
\label{app_resbase}

\begin{figure*}
  \begin{subfigure}[pt]{0.49\textwidth}
  \includegraphics[width=\linewidth]{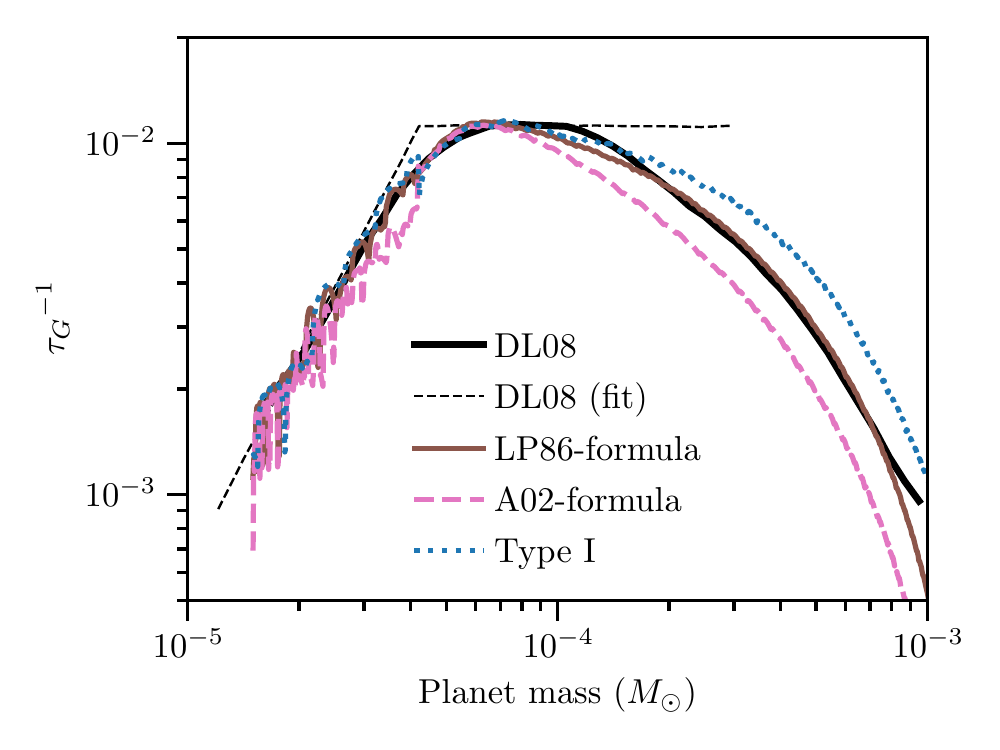}
  \end{subfigure}
  \begin{subfigure}[pt]{0.49\textwidth}
  \includegraphics[width=\linewidth]{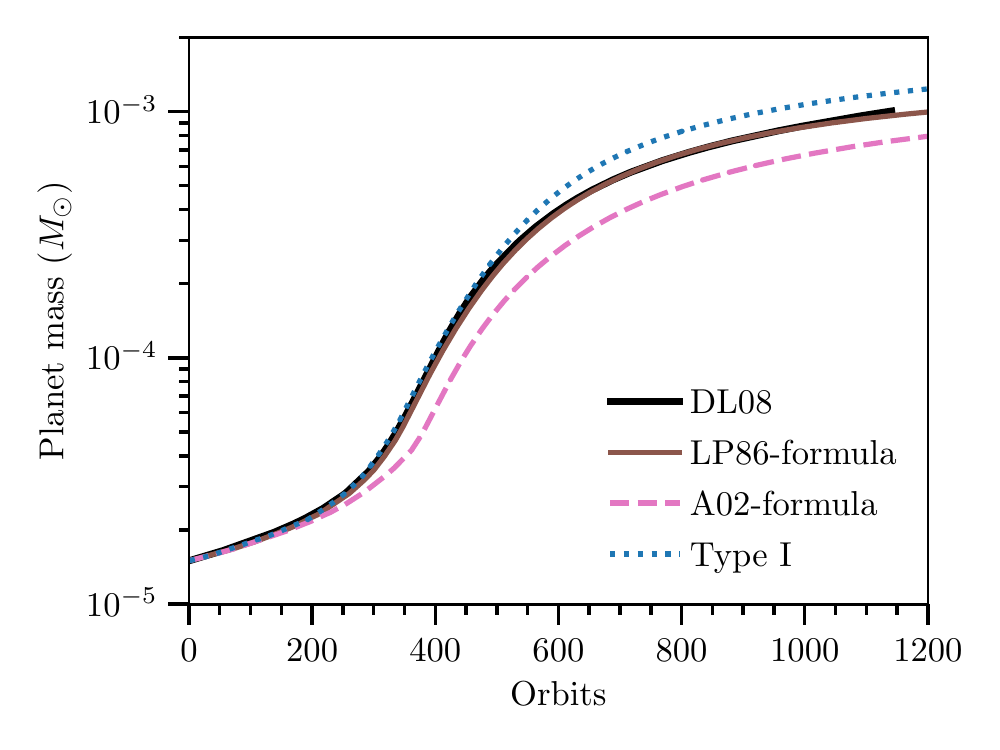}
  \end{subfigure}
  \begin{subfigure}[pt]{0.49\textwidth}
  \includegraphics[width=\linewidth]{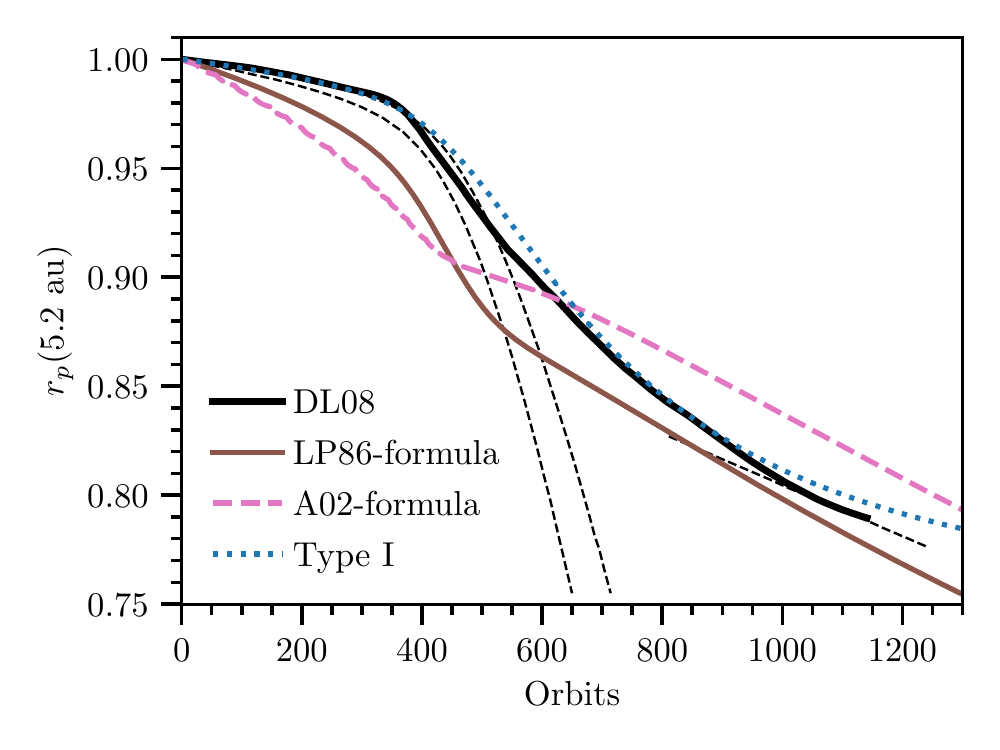}
  \end{subfigure}
  \begin{subfigure}[pt]{0.49\textwidth}
  \includegraphics[width=\linewidth]{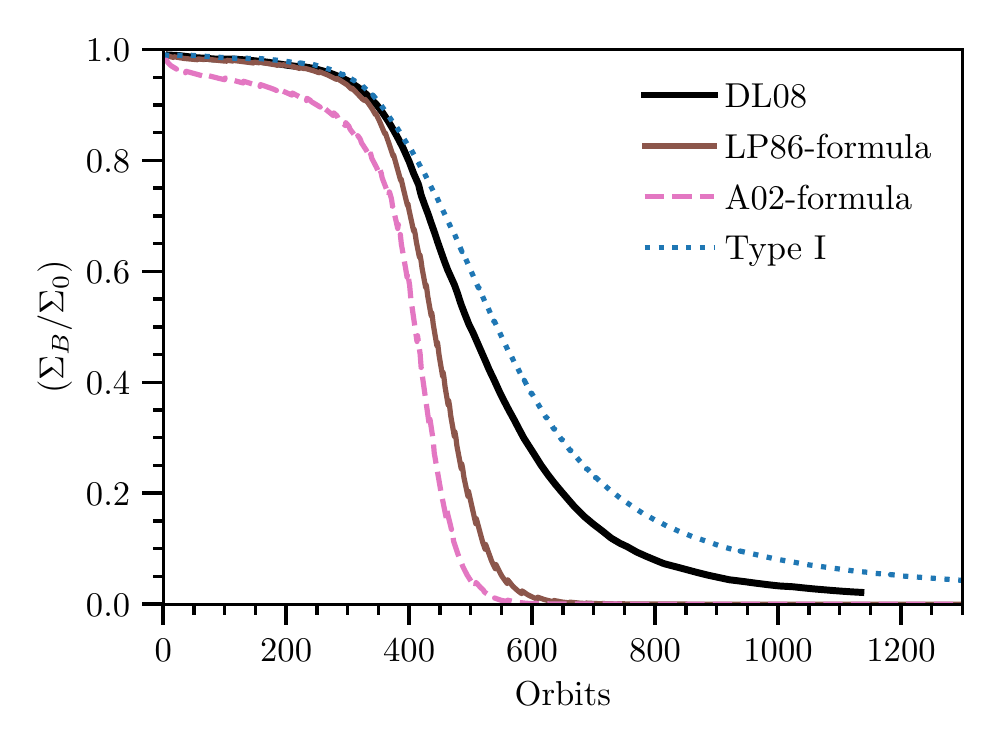}
  \end{subfigure}
  \caption{Same as Fig.~\ref{fig_base}, but for alternative torque models. The solid brown line represents the results from the \citetalias{1986ApJ...309..846L} formula. The dashed pink line shows the results based on the \citetalias{2002MNRAS.334..248A} formula. The blue dotted line gives the results when applying only a type~I torque using the surface density at the planet location.}
  \label{fig_app_base}
\end{figure*}

Figure \ref{fig_app_base} shows the results for the three alternative migration treatments in the baseline case (Sect.~\ref{sec_resbase}). With the \citetalias{1986ApJ...309..846L} formula, the mass accretion (top left panel) is very similar to the result from \citetalias{2008ApJ...685..560D} and our earlier results. When we apply the \citetalias{2002MNRAS.334..248A} formula, the accretion rate drops earlier despite the slightly increased $C_\mathrm{H}$ (Table \ref{table_config}), leading to a lower final mass (top right panel). In the type~I prescription the accretion rate drops later, which leads to a somewhat higher final mass. In this case we used a smaller $R_f$ and a higher $C_\mathrm{H}$ in order to match the Hill accretion rate given in \citetalias{2008ApJ...685..560D}. Clearly, in the absence of a tidal torque, less material is removed from the planet location. This is also seen in the bottom right panel: the surface density near the planet is depleted more slowly than seen in \citetalias{2008ApJ...685..560D}. On the contrary, the depletion is much too fast when applying the \citetalias{1986ApJ...309..846L} formula , and even faster with the \citetalias{2002MNRAS.334..248A} formula.
All three prescriptions lead to a mass accretion that is in reasonable agreement (within $25 \%$) with that found by \citetalias{2008ApJ...685..560D}. However, only the `type~I' method closely follows their migration track during the first $\sim\num{500}$ orbits. The approximate agreement in the other two at later times may be a coincidence.

\subsection{Higher surface density}
\label{app_reshsig}

\begin{figure*}
  \begin{subfigure}[pt]{0.49\textwidth}
  \includegraphics[width=\linewidth]{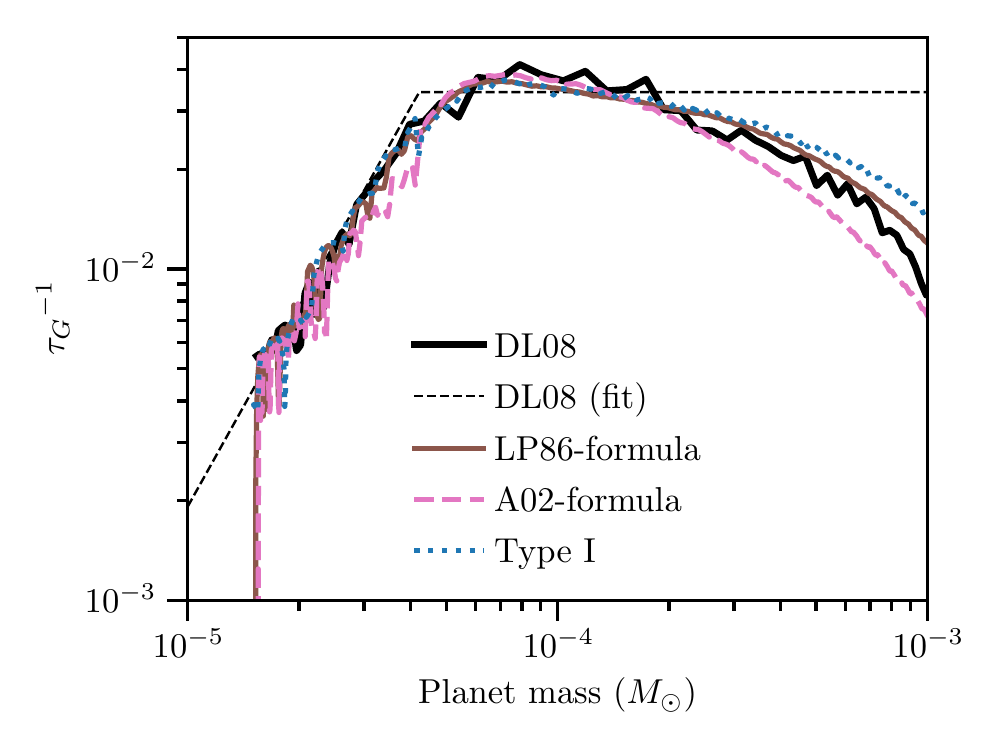}
  \end{subfigure}
  \begin{subfigure}[pt]{0.49\textwidth}
  \includegraphics[width=\linewidth]{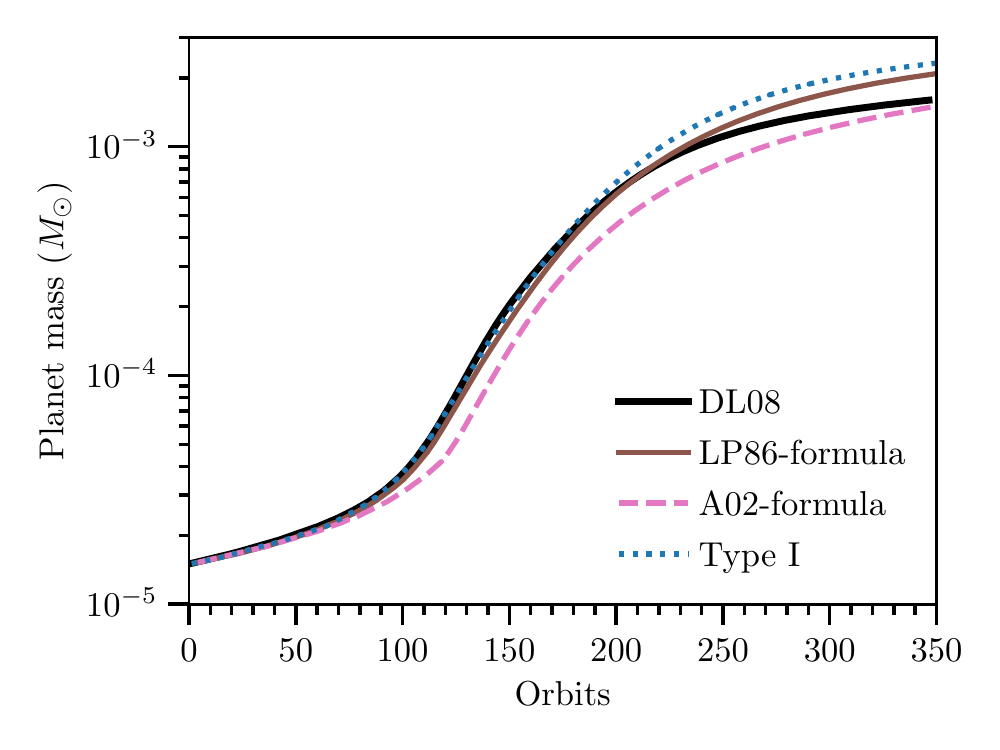}
  \end{subfigure}
  \begin{subfigure}[pt]{0.49\textwidth}
  \includegraphics[width=\linewidth]{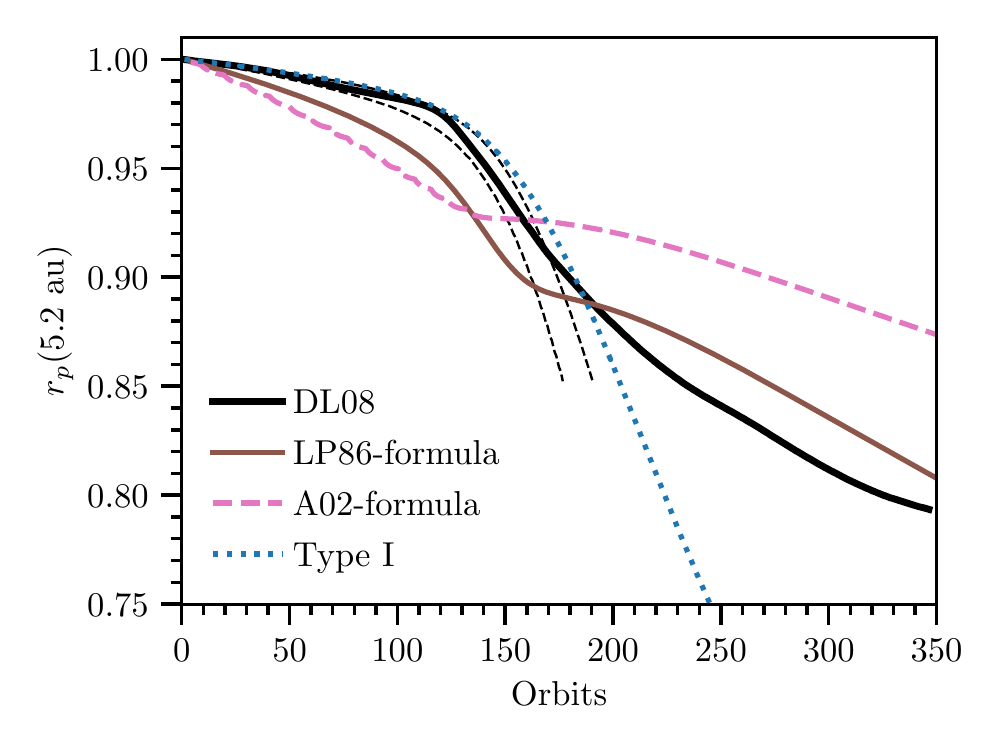}
  \end{subfigure}
  \begin{subfigure}[pt]{0.49\textwidth}
  \includegraphics[width=\linewidth]{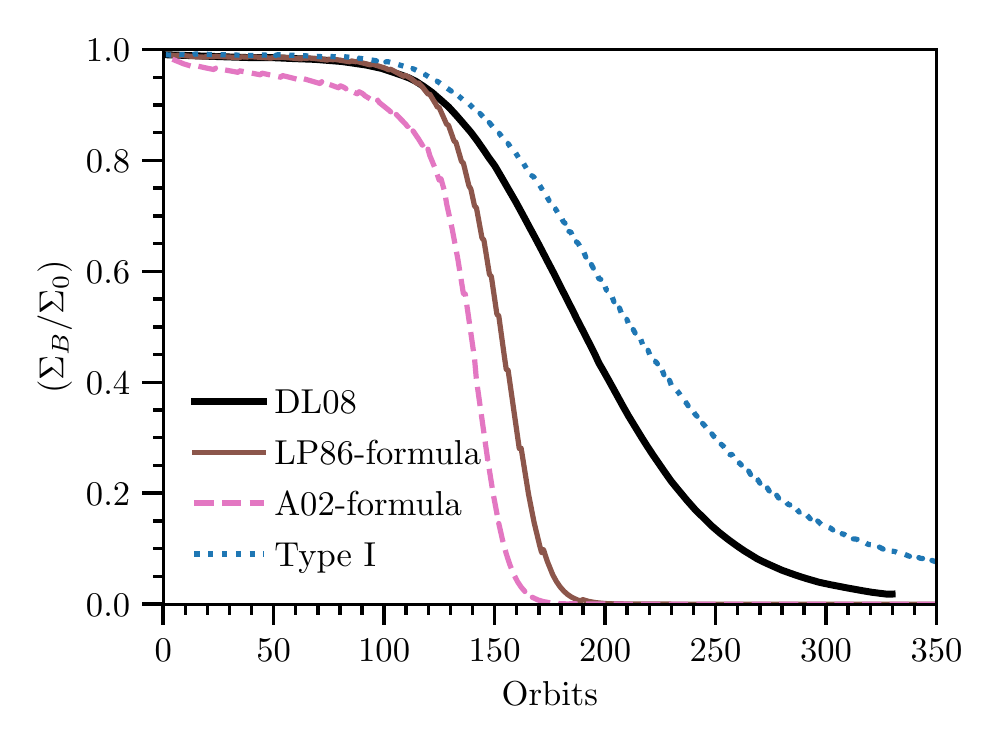}
  \end{subfigure}
  \caption{Same as Fig.~\ref{fig_app_base}, but for  increased surface density (\SI{300}{g.cm^{-2}} at \SI{5.2}{au}).}
  \label{fig_app_hsig}
\end{figure*}

Here we discuss the \citetalias{1986ApJ...309..846L} formula, the \citetalias{2002MNRAS.334..248A} formula, and our `type I' approach in the case with an increased initial surface density. The results are given in Fig.~\ref{fig_app_hsig}.
The mass accretion (top panels) proceeds qualitatively similarly to the baseline case, with the \citetalias{2002MNRAS.334..248A} formula giving somewhat too little and  `type~I' giving too high accretion. This time, the \citetalias{1986ApJ...309..846L} formula also gives higher accretion at late times than found in \citetalias{2008ApJ...685..560D}.
Migration (bottom left panel) is surprisingly similar to \citetalias{2008ApJ...685..560D} with the \citetalias{1986ApJ...309..846L} formula, though migration is too slow and gap formation too abrupt. The \citetalias{2002MNRAS.334..248A} formula shows a migration track that slows down too early and too abruptly. In this case, the `type~I' prescription shows almost no slowdown during the first $\sim 250$ orbits. A significant slowdown only occurs after $\sim 400$ orbits when the planet reaches less than \num{.6} times its initial semi-major axis (not seen).
The depletion of the surface density near the planet (bottom right panel) proceeds in an analogous fashion to the baseline case (\ref{app_resbase}).

\subsection{Very high surface density}
\label{app_resuhsg}

\begin{figure*}
  \begin{subfigure}[pt]{0.49\textwidth}
  \includegraphics[width=\linewidth]{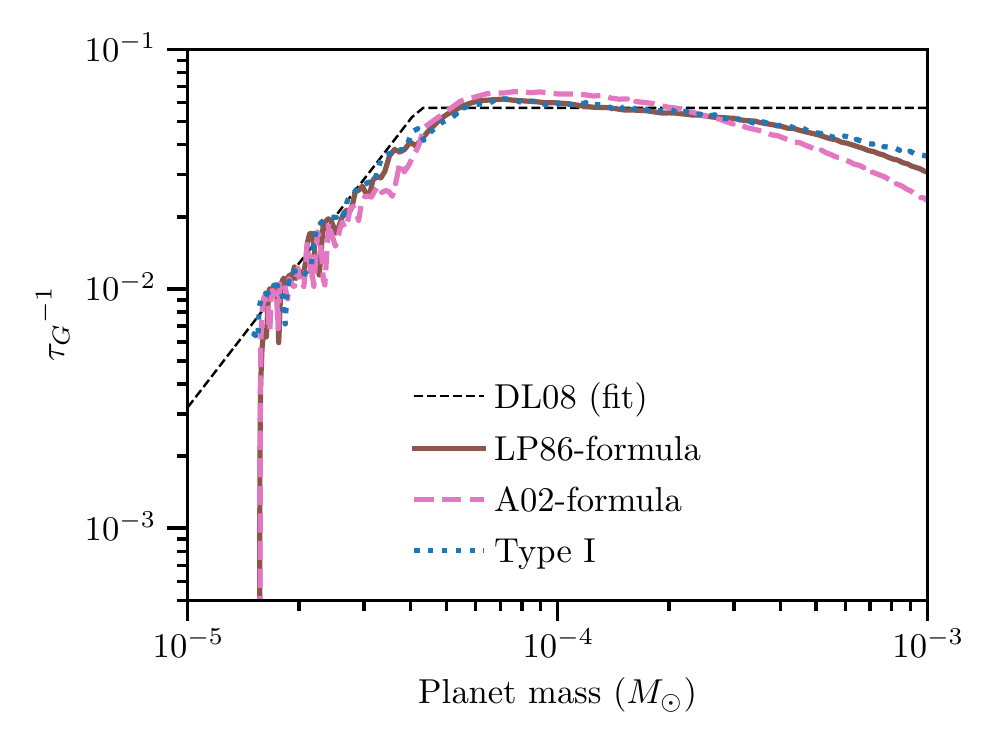}
  \end{subfigure}
  \begin{subfigure}[pt]{0.49\textwidth}
  \includegraphics[width=\linewidth]{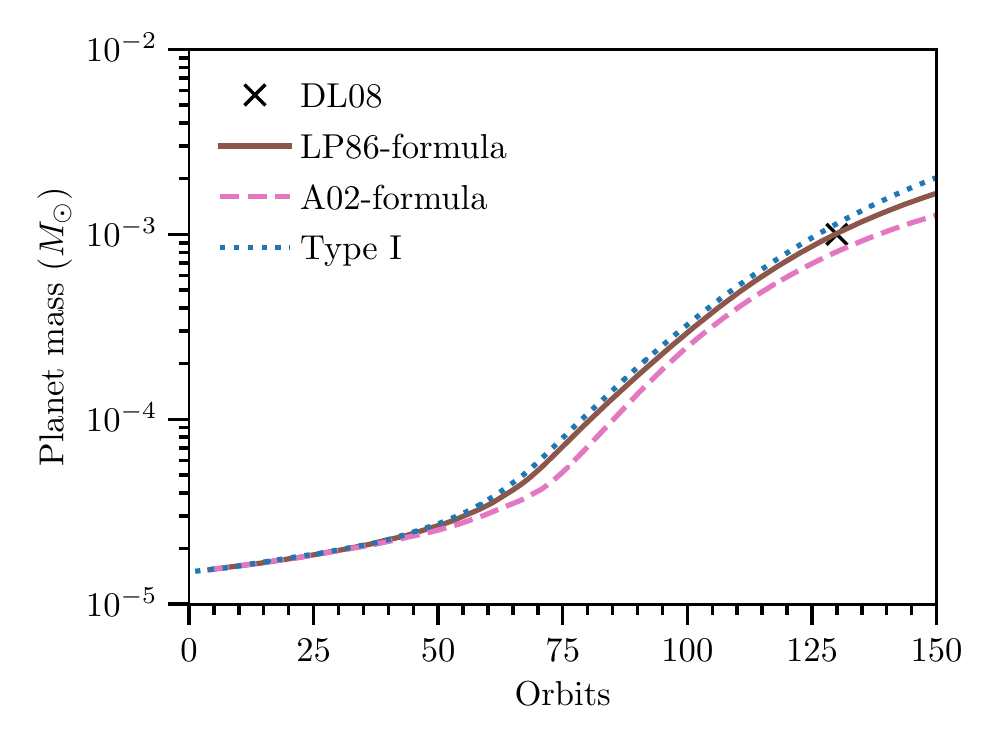}
  \end{subfigure}
  \begin{subfigure}[pt]{0.49\textwidth}
  \includegraphics[width=\linewidth]{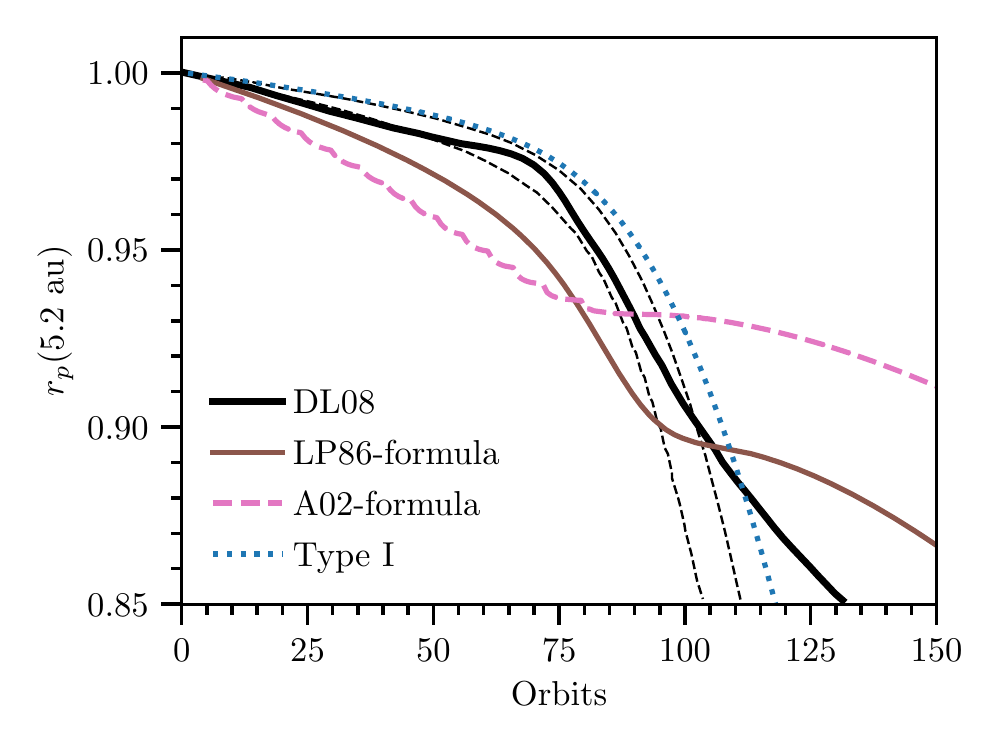}
  \end{subfigure}
  \begin{subfigure}[pt]{0.49\textwidth}
  \includegraphics[width=\linewidth]{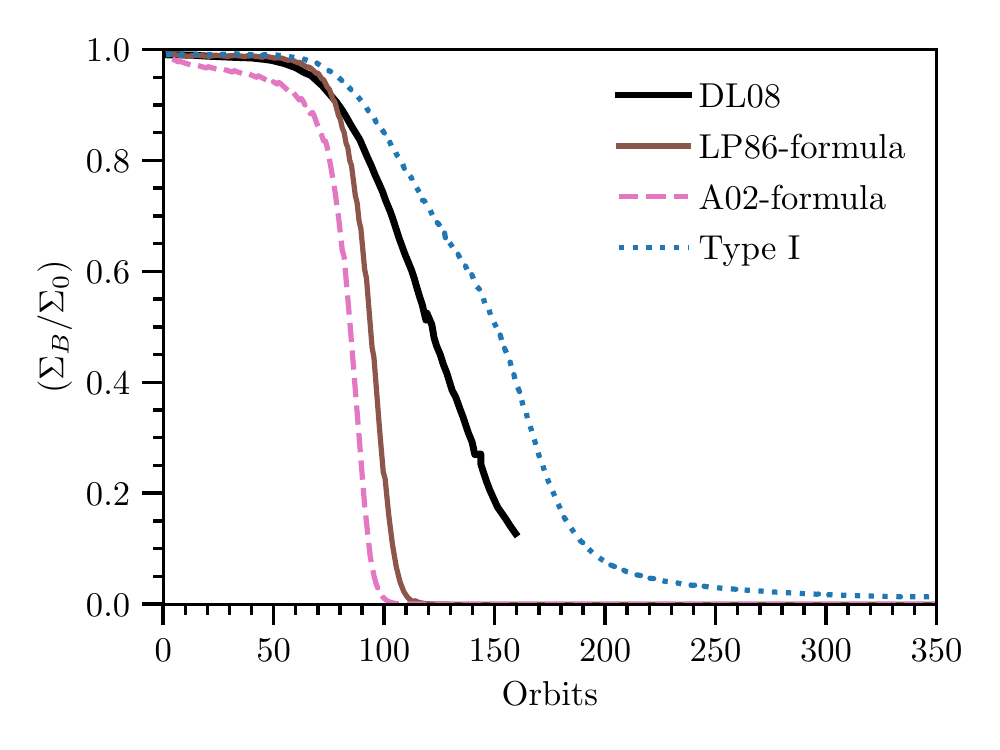}
  \end{subfigure}
  \caption{Same as Fig.~\ref{fig_app_base}, but for  initial surface density of $\SI{500}{g.cm.^{-2}}$ at $\SI{5.2}{au}$.}
  \label{fig_app_uhsg}
\end{figure*}

In this section we discuss the alternative prescriptions in the case of an initial surface density of \SI{500}{g.cm^{-1}}. The results can be found in Fig.~\ref{fig_app_uhsg}. Mass accretion and depletion of the surface density near the planet once more proceed qualitatively similarly as above (top and bottom right panels).
In this case none of the prescriptions give a migration track that agrees well with \citetalias{2008ApJ...685..560D}. They either open a gap too far out or too far in, as seen in the bottom left panel.

\subsection{Lower temperature}
\label{app_resltmp}

\begin{figure*}
  \begin{subfigure}[pt]{0.49\textwidth}
  \includegraphics[width=\linewidth]{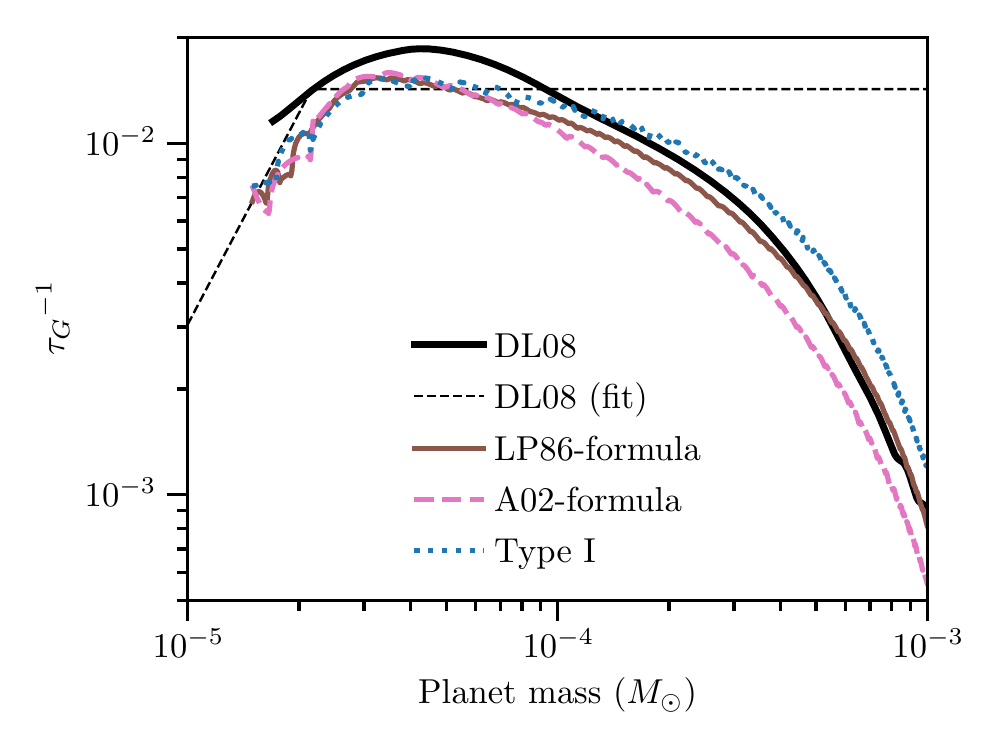}
  \end{subfigure}
  \begin{subfigure}[pt]{0.49\textwidth}
  \includegraphics[width=\linewidth]{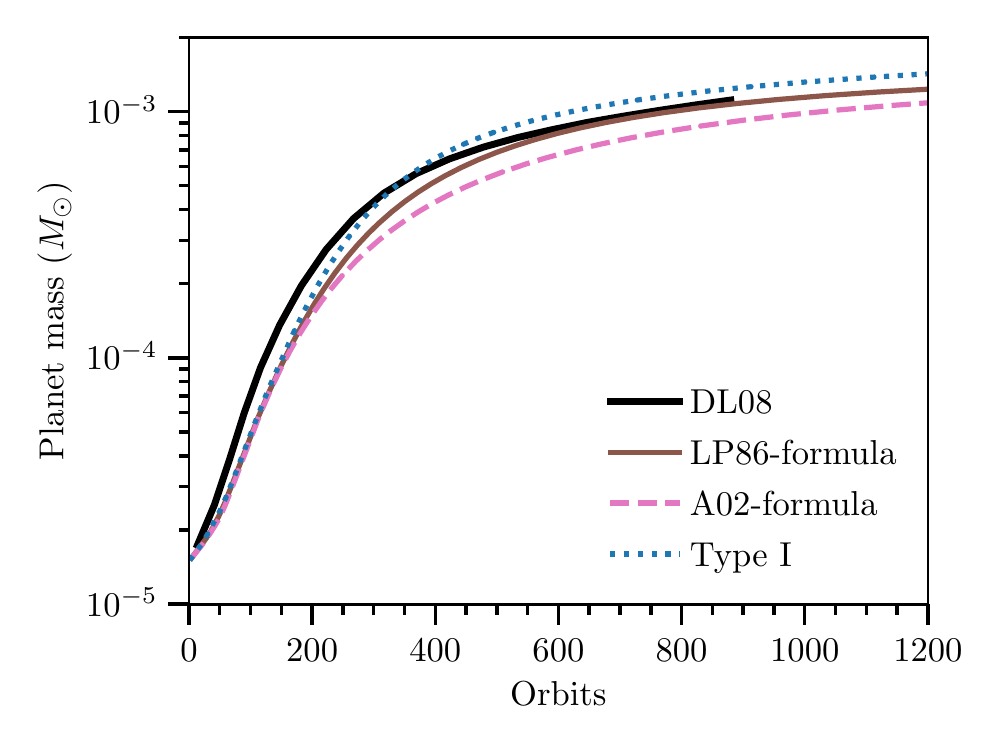}
  \end{subfigure}
  \begin{subfigure}[pt]{0.49\textwidth}
  \includegraphics[width=\linewidth]{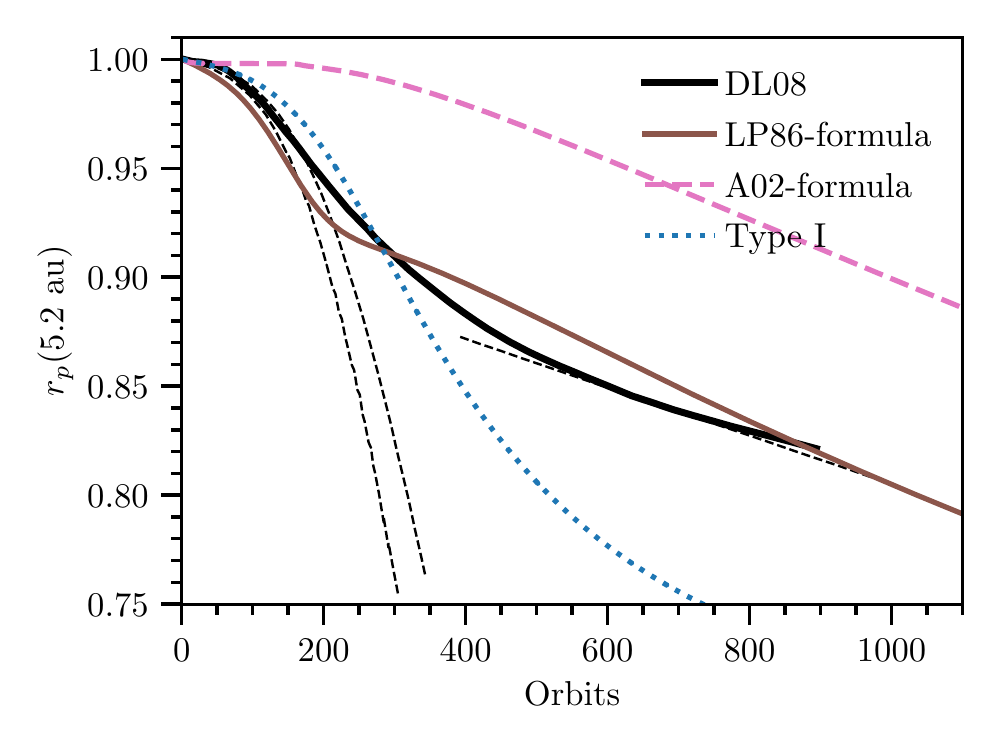}
  \end{subfigure}
  \begin{subfigure}[pt]{0.49\textwidth}
  \includegraphics[width=\linewidth]{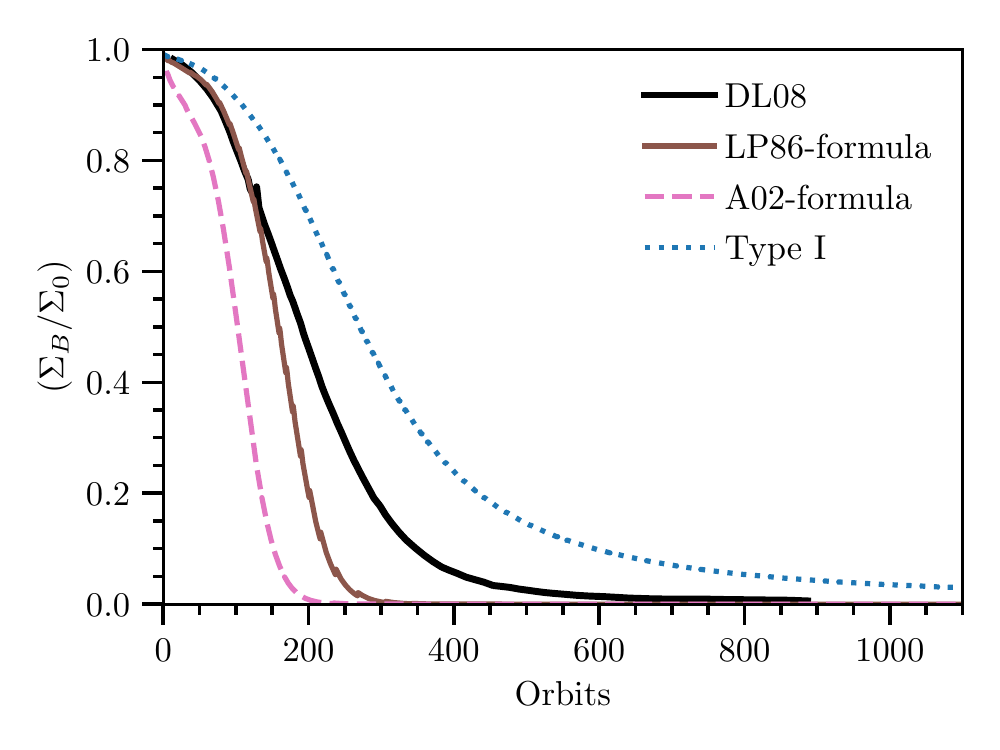}
  \end{subfigure}
  \caption{Same as Fig.~\ref{fig_app_base}, but for  reduced temperature ($H/r = 0.04$).}
\label{fig_app_ltmp}
\end{figure*}

Here, we look at how the alternative torque models compare in the case of reduced temperature. The results can be found in Fig.~\ref{fig_app_ltmp}.
Mass accretion proceeds again similarly (top panels). The difference between the torque models in this case are moderate.
The surface density near the planet (bottom right panel) is affected differently again by the different torque densities, as described above.
Due to the lower temperature, gap opening is much faster. This is most pronounced when using the \citetalias{2002MNRAS.334..248A} formula. In this case a gap starts to open immediately, leaving the planet too far out (bottom left panel). When using the `type~I' prescription the planet migrates in line with \citetalias{2008ApJ...685..560D}, but slows down less, hence migrating much further in. The \citetalias{1986ApJ...309..846L} formula gives a reasonable agreement with \citetalias{2008ApJ...685..560D} for the migration track.

In summary, we find that none of the alternative prescriptions we studied in Appendix~\ref{app_ia} gives an  agreement to the results from \citetalias{2008ApJ...685..560D} similar to our torque mod and high mass torque prescriptions discussed in the main text.  The \citetalias{1986ApJ...309..846L} formula and the \citetalias{2002MNRAS.334..248A} formula both open gaps too early in at least some of the cases, and deplete the surface density around the planet too much. The `type~I' recipe, on the other hand, gives good agreement for the semi-major axis at early times, but migration slows down much too late in some cases. This is related to the evolution of the surface density near the planet, which is depleted at a rate slower than seen in \citetalias{2008ApJ...685..560D}. It is worth noting that this prescription still causes a reduction in the surface density near the planet by more than an order of magnitude through accretion alone, and consequently an eventual slowdown of migration.

\FloatBarrier

\section{Sensitivity of migration on surface density and temperature slopes}
\label{app_slope}
\begin{figure}
  \includegraphics[width=\linewidth]{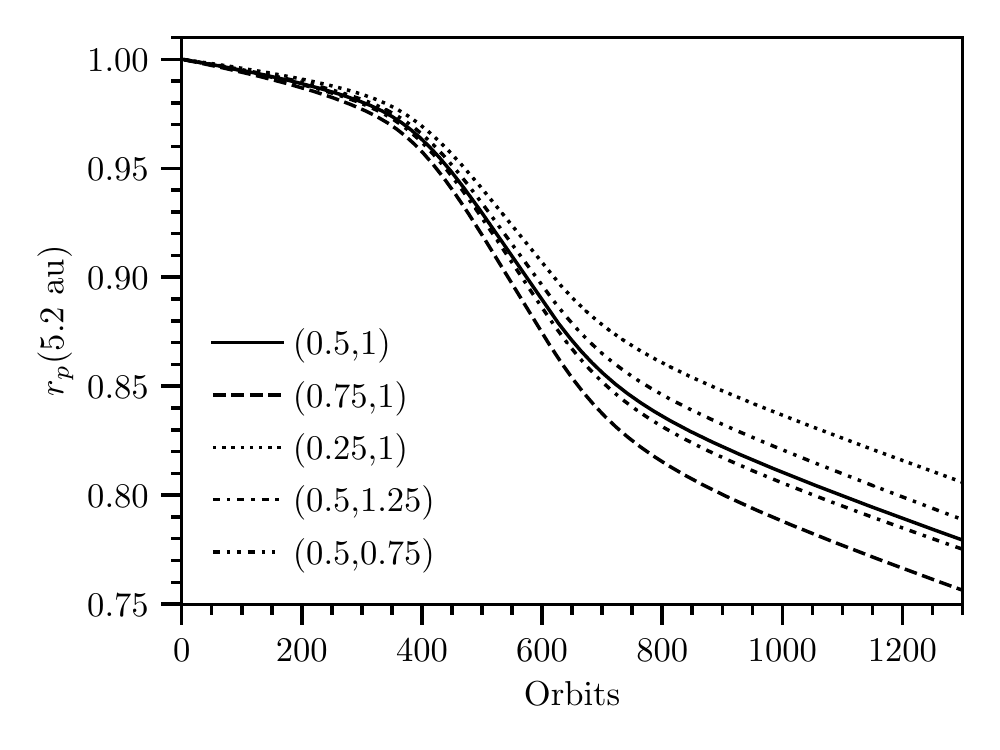}
  \caption{Evolution of the semi-major axis for different values of the slopes of surface density and temperature (see legend and the text in Appendix~\ref{app_slope}).}
    \label{fig_app_slopetest}
\end{figure}

In Section~\ref{subs_mig} we argued that using torque densities based on the work of \citetalias{2010ApJ...724..730D} is a reasonable approach even if the slopes of surface density and temperature, $\beta$ and $\zeta$, vary slightly. Here we present a sensitivity calculation to support this statement. For this test we used the torque mod, since it can be applied to an arbitrary torque density covered by the grid from \citet{2010ApJ...724..730D}. Figure~\ref{fig_app_slopetest} shows the time evolution of the planet's semi-major axis when different functions $\mathcal{F} \left(x, \beta, \zeta\right)$ are used. Either $\beta$ or $\zeta$ is varied by \num{0.25} in both directions.
The deviation from the nominal run with parameters $(0.5,1)$ is very small when the temperature slope $\zeta$ is varied. The change in semi-major axis (\SI{5.2}{au} minus $r_p$ after \num{1300} orbits) is at most $4 \%$ different. The difference is larger, but still moderate when changing $\beta$ (at most $12 \%$).
The $p_i$ parameters used for the calculations shown in Fig.~\ref{fig_app_slopetest} are listed in Table~\ref{table_p_app}
\begin{table}
\small
\addtolength{\tabcolsep}{-3pt}
\centering
\begin{tabular}{cllll}
\hline\hline
\begin{tabular}[c]{@{}c@{}}($\beta$,$\zeta$)\end{tabular} &
\begin{tabular}[c]{@{}c@{}}(0.75,1)\end{tabular} &
\begin{tabular}[c]{@{}c@{}}(0.25,1)\end{tabular} &
\begin{tabular}[c]{@{}c@{}}(0.5,1.25)\end{tabular} &
\begin{tabular}[c]{@{}c@{}}(0.5,0.75)\end{tabular} \\
\hline
$p_1$ & 0.02907625 & 0.0307695 & 0.028806875 & 0.0303808875 \\
$p_2$ & 1.1285525 & 1.083705 & 1.10836125 & 1.09367 \\
$p_3$ & 0.92678075 & 0.953953 & 0.95779025 & 0.939862 \\
$p_4$ & 0.0432845 & 0.041157 & 0.0429604625 & 0.0413219 \\
$p_5$ & 0.884929 & 0.916865 & 0.900614875 & 0.901030375 \\
$p_6$ & 1.0423025 & 1.040685 & 1.025077375 & 1.06267375 \\
$p_7$ & 0.016941975 & 0.15997565 & 0.181304225 & 0.055344575  \\
$p_8$ & 4.52963 & 4.591765 & 4.800585 & 4.29918125\\
\hline
\end{tabular}
\caption{Values used for the $p_i$ parameters in Eq.~\ref{eq_f} for Appendix~\ref{app_slope} (see legend and Sect.~\ref{subs_mig}).}
\label{table_p_app}
\end{table}
The $p_i$ for these combinations of $\beta$ and $\zeta$ are not directly available from \citetalias{2010ApJ...724..730D}. Therefore, we interpolated the values for (0.75,1), (0.25,1), (0.5,1.25), and 0.5,0.75) by using the values for [(0.5,1), (1.5,1)], [(0.5,1), (0,1)], [(0.5,1), (0,2), (1,2)], and [(0,0), (1,0), (0.5,1)], respectively, given in their Table~1.

\FloatBarrier

\section{Accretion with autogravitation}
\label{app_ag}

Here we derive the expression for $\dot{M}_\mathrm{gas}$ (see Sect.~\ref{subs_acc}) when the disc's autogravitation is taken into account. We follow the approach presented in \citet{2005A&A...442..703H}. In this case equations \ref{eq_vert} and \ref{eq_H} need to be replaced by
\begin{equation}
\rho(r,z)=\rho_0(r) \exp \bigg(-\bigg(\frac{\left|z\right|}{H_0}+\bigg(\frac{z}{H_1}\bigg)^2\bigg)\bigg),
\label{eq_vert_ag}
\end{equation}
and
\begin{equation}
H=\frac{H_1}{\sqrt{2}}\exp\bigg(\frac{H_1^2}{4H_0^2}\bigg)\bigg(1-\erf\bigg(\frac{H_1}{2H_0}\bigg)\bigg),
\end{equation}
where
\begin{equation}
H_0=\frac{c_\mathrm{s}^2}{4\pi G\Sigma},
\end{equation}
\begin{equation}
H_1=\frac{\sqrt{2}c_\mathrm{s}}{\sqrt{\frac{\mathrm{G} \mstar}{r^3}}}.
\end{equation}
Plugging Eq.~\ref{eq_vert_ag} into Eq.~\ref{eq_macc} then yields
\begin{equation}\label{eq_mdot_ag}
\begin{aligned}
    \dot{M}_\mathrm{gas} & = \sqrt{\pi} C_\mathrm{B,H} \int_{r_p-R_f}^{r_p+R_f}\!\rho_0(r) H_1 \exp{\left(\frac{H_1^2}{4 H_0^2}\right)} \times\\
    &\left(\erf{\left[\frac{\sqrt{R_f^2-(r-r_p)^2}}{H_1} + \frac{H_1}{2 H_0}\right]}-\erf{\left[\frac{H_1}{2 H_0}\right]}\right)\,v_\mathrm{rel}\,\mathrm{d}r.
\end{aligned}
\end{equation}
The expression for the angular frequency also changes when autogravitation is considered. For a more detailed description, see Section ~2.2 in \citet{2021A&A...645A..43S}.
 
 \FloatBarrier

\section{Comparison}
\label{app_rev}

In this section we present the results of our comparison with the migration--accretion prescription presented in \citetalias{2015A&A...582A.112B}. The authors use the torque formula from \citet{Paardekooper2011b} for type~I migration. This formula was derived for non-isothermal type~I migration. Since the simulations in \citetalias{2008ApJ...685..560D}, on which our work is based, are locally isothermal, we used the torque formula for the locally isothermal isothermal limit given in \citet{Paardekooper2010} instead as it is more appropriate for our comparison:
\begin{equation}
    \tau_\mathrm{lociso} = (1.7 + 2~\beta + 1.8 \zeta)^{-1} \frac{\mstar}{M_p} \frac{\mstar}{\Sigma_p r_p^2} \left( \frac{c_\mathrm{s}}{r_p \Omega_p}\right) \Omega_p^{-1}.
\end{equation}
In the type~II regime, migration is assumed to proceed on a viscous timescale $\tau_\mathrm{visc} = r_p^2/\nu $ with a slowdown  for massive planets \citep{Alibert2005,Baruteau2014}. This leads to a type~II timescale of 
\begin{equation}
    \tau_\mathrm{II} = \tau_\mathrm{visc} \times \max \left( 1, \frac{M_p}{4 \pi \Sigma r_p^2} \right),
\end{equation}
where the (unperturbed) surface density is evaluated at the planet's location. In order to account for the slowdown in migration related to the formation of a gap, the migration rate is multiplied by the function $f(\mathcal{P})$:
\begin{equation}
    f(\mathcal{P}) =
    \begin{cases}
        \frac{\mathcal{P}-0.541}{4} & \mathrm{if}~\mathcal{P} < 2.4646 \\
        1-\exp \left( - \frac{\mathcal{P}^{3/4}}{3} \right) & \mathrm{otherwise}.
    \end{cases}
\end{equation}
Here the function $\mathcal{P}$ describes the gap opening  and is defined as \citep{2006Icar..181..587C}
\begin{equation}
    \mathcal{P} = \frac{3}{4} \frac{H}{R_\mathrm{H}} + \frac{50}{q \mathcal{R}},
\end{equation}
where $q = M_p/\mstar$ and $\mathcal{R} = r_p^2 \Omega_p / \nu$ the Reynolds number.
A gap is assumed to be fully open (surface density reduced by a factor of ten) when $\mathcal{P} \approx \num{1}$. In this case the type~I migration timescale calculated as described here may still differ from $\tau_\mathrm{II}$. As a result,  \citetalias{2015A&A...582A.112B} interpolate between the type~I and type~II timescales to ensure a smooth transition. They use a linear interpolation in $\mathcal{P}$  starting at $\mathcal{P} = 2.4646$ (fully type~I) going to $\mathcal{P} = 1$ (fully type II, private communication with B.~Bitsch).
Rapid gas accretion is modelled using the  fitting formula presented in \citet{machidakokubo2010} that gives the accretion rate for two regimes,
\begin{equation}\label{eq_gBl}
    \dot{M}_\mathrm{gas,low} = 0.83 \Omega_p \Sigma H^2 \left( \frac{R_\mathrm{H}}{H} \right)^{9/2}
\end{equation}
and
\begin{equation}\label{eq_gBh}
    \dot{M}_\mathrm{gas,high} = 0.14 \Omega_p \Sigma H^2
\end{equation}
 for low-mass and high-mass planets, respectively. The lower of the two is used as the  effective accretion rate.
In Equations~\ref{eq_gBl} and \ref{eq_gBh}, $\Sigma$ and $H$ are evaluated at the planet's location. No gas is removed from the disc, so $\Sigma$ is not perturbed.
We implemented this model into our code and applied it to the systems described in Sect.~\ref{sec_resbase} to \ref{sec_reshvsc}.

\begin{figure*}
  \begin{subfigure}[pt]{0.49\textwidth}
  \includegraphics[width=\linewidth]{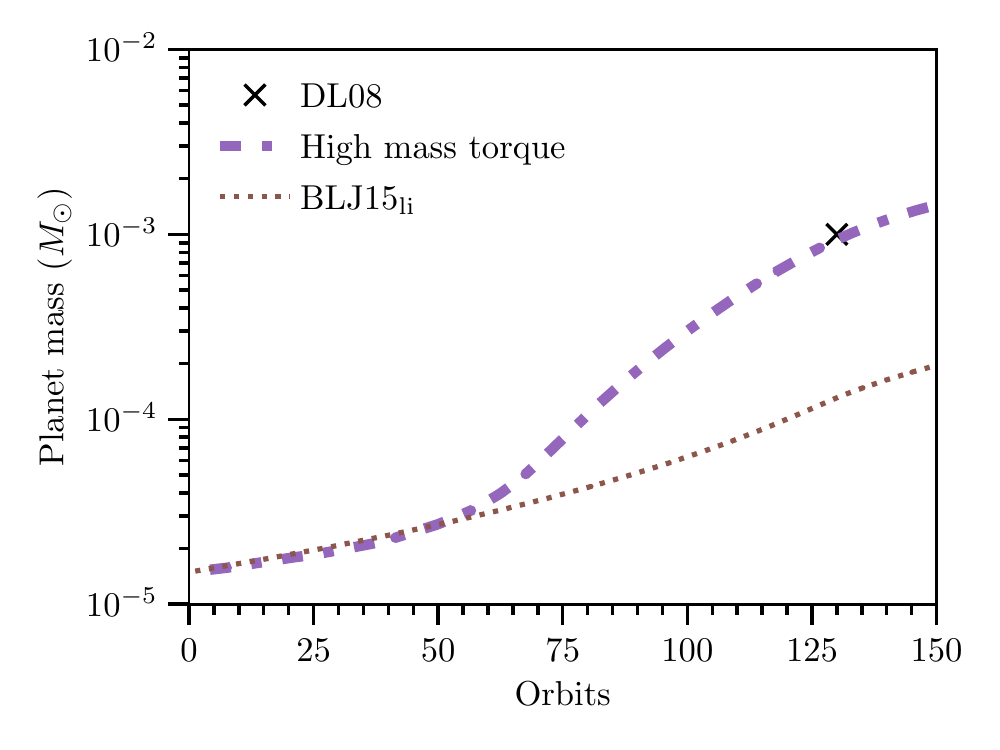}
  \end{subfigure}
  \begin{subfigure}[pt]{0.49\textwidth}
  \includegraphics[width=\linewidth]{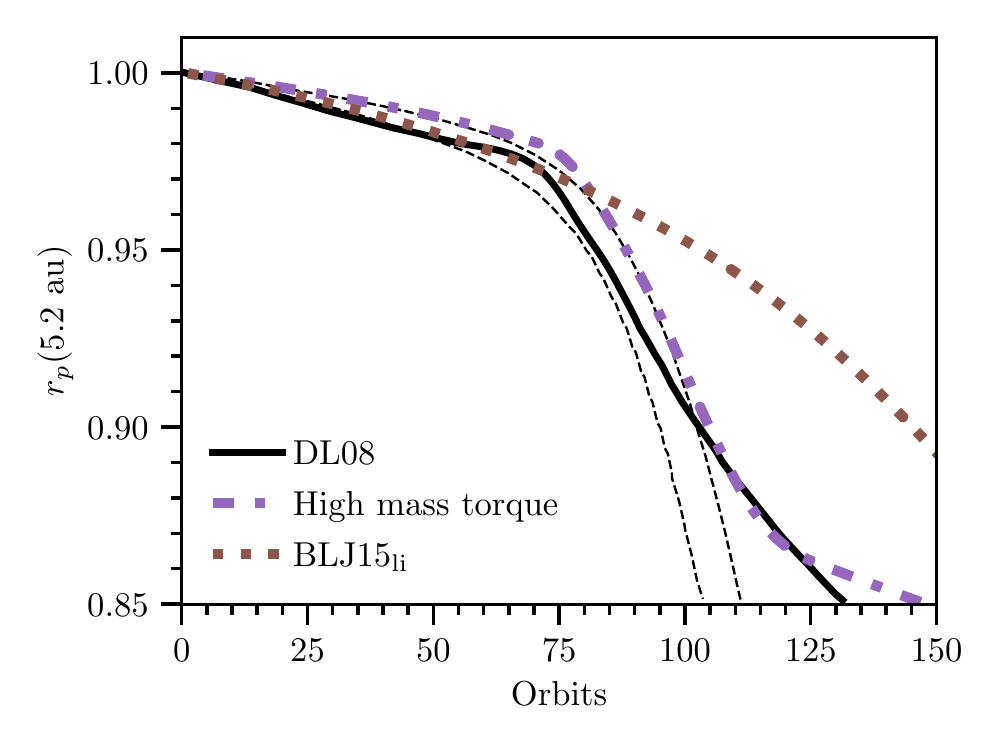}
  \end{subfigure}
  \caption{Same as Fig.~\ref{fig_rev_base} , but with a five times higher initial surface density (\SI{500}{g.cm.^{-2}} at \SI{5.2}{au}).}
\label{fig_rev_uhsg}
\end{figure*}

The results for the comparison with the baseline case and with the case with higher surface density are given in Sect.~\ref{Sect:CompLit}. Here we discuss two additional cases.
Figure.~\ref{fig_rev_uhsg} shows the planetary mass and semi-major axis for the case of the fivefold increase in  initial surface density. Again, gas accretion is much lower than seen in the hydrodynamic simulation. Consequently migration is also slower. In this case we see no slowdown due to gap formation during the first $\num{150}$ orbits.

\begin{figure*}
  \begin{subfigure}[pt]{0.49\textwidth}
  \includegraphics[width=\linewidth]{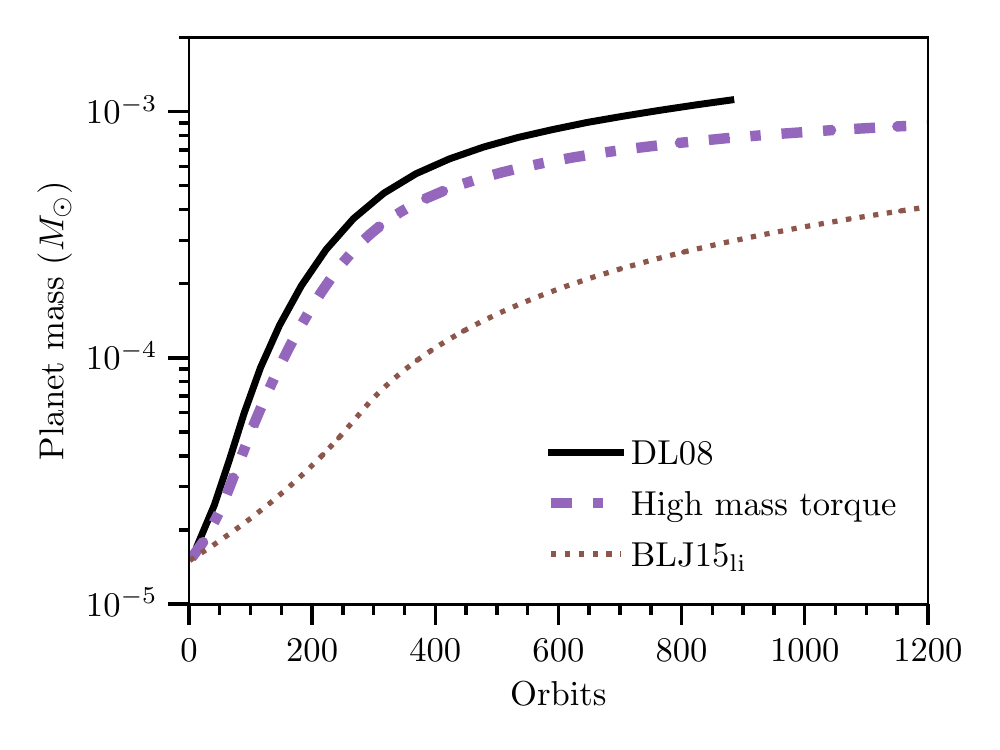}
  \end{subfigure}
  \begin{subfigure}[pt]{0.49\textwidth}
  \includegraphics[width=\linewidth]{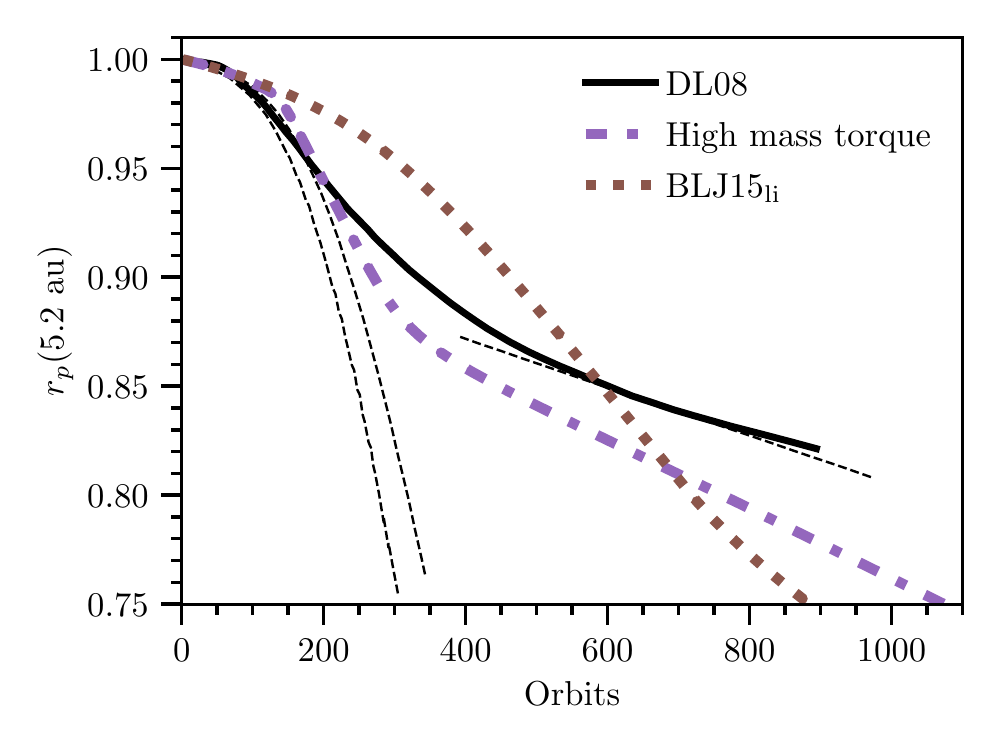}
  \end{subfigure}
  \caption{Same as Fig.~\ref{fig_rev_base}, but for  reduced temperature ($H/r = 0.04$).}
\label{fig_rev_ltmp}
\end{figure*}

Finally, figure~\ref{fig_rev_ltmp} shows the comparison with the low-temperature case, where  gas accretion is also much lower. The behaviour of migration is somewhat different from the other cases. While it is also slower early due to the lower planetary mass, migration does not slow down significantly due to the (partial) opening of the gap. Therefore, the planet actually migrates further in during the first \num{900} orbits than it does in \citetalias{2008ApJ...685..560D}.
In summary, we find that our Bondi--Hill accretion model combined with the high mass torque gives a better agreement with the simulations of \citetalias{2008ApJ...685..560D}. This is not surprising since we calibrated our accretion model with one of their simulations.
We also applied the accretion rates from Eq.~\ref{eq_gBl} and \ref{eq_gBh} to the test case from the code comparison project (Sect.~\ref{sec_reslase}). In this case it also leads to low accretion rates compared to all other models. The accretion rate is more than a factor of two lower than the lowest-mass case shown in the left panel of Fig.~\ref{fig_FNS}. We note that \citet{machidakokubo2010} did not consider planets more massive than \SI{1}{\mj}, and therefore their fitting formula may not be applicable here.
Nevertheless it is unclear whether the accretion rates given by Eqs.~\ref{eq_gBl} and \ref{eq_gBh} or the values we use in our model are more realistic, as the models involved are very different (locally isothermal in \citetalias{2008ApJ...685..560D}, beta-cooling in \citetalias{2019MNRAS.486.4398F}, and  radiation-hydrodynamic in \citet{machidakokubo2010}).

\FloatBarrier

\section{Practical use}\label{app_prac}

\begin{figure}
  \includegraphics[width=\linewidth]{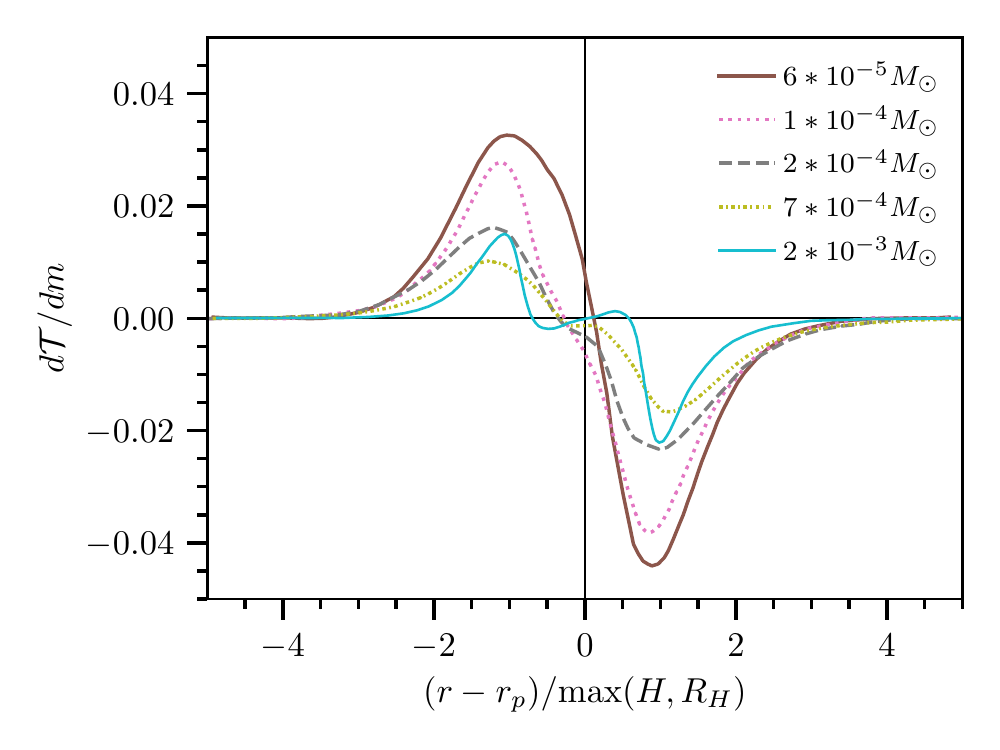}
  \caption{Torque densities in units of $\Omega^2 r_p^2 (M_p/M_*)^2 (r_p/H)^4$ for different planet masses taken from \citet{2010ApJ...724..730D}. The inversion developing as the mass increases is clearly  seen  near the origin. The high mass torque is obtained by an interpolation (see text Appendix~\ref{app_prac}).}
    \label{fig_app_torque}
\end{figure}

In this section we describe how our prescription can be implemented in 1D models. We recommend using the high mass torque in combination with the Bondi--Hill accretion scheme, as given in Table~\ref{table_config}: with a feeding zone radius of $\SI{3}{R_\mathrm{H}}$, $C_\mathrm{B} = 10.0$, and $C_\mathrm{H}$ = 0.19.
The torque density for the high mass torque is obtained by an interpolation of the digitised torque densities given in \citet{2010ApJ...724..730D}.
Some examples are shown in Fig.~\ref{fig_app_torque} and the  data can be downloaded from the CDS.
The interpolation at the lowest masses is done by using $\mathcal{F}(x,0.5,1)$ as zero mass. No extrapolation is perfomed above the highest mass.
The high mass torque is reliable as long as the gradients of the surface density and temperature do not deviate from $(\beta,\zeta) = (0.5,1)$ by more than a few 10\%.
The main limitation is related to the migration of low-mass planets (tens of $\SI{}{\mearth}$) in non-isothermal discs. The condition for non-isothermality is given by \citep{2014A&A...567A.121D} $\tau_\mathrm{cool} \gg \tau_\mathrm{U-turn}$, where
\begin{equation}
    \tau_\mathrm{cool} = \frac{l_\mathrm{cool} \rho C_\mathrm{V}}{8 \sigma T^3} \left( 8 \rho \kappa l_\mathrm{cool} + \frac{1}{\rho \kappa l_\mathrm{cool}} \right),
\end{equation}
\begin{equation}
    \tau_\mathrm{U-turn} = \frac{64 x_\mathrm{s} h^2}{9 q r_p \Omega_p}.
\end{equation}
In the above equations
\begin{equation}
    x_\mathrm{s} = 1.16 r_p \sqrt{\frac{q}{h \sqrt{\gamma}}}
\end{equation}
is the width of the co-orbital region, $q=M_p/\mstar$, $h=H/r_p$, $l_\mathrm{cool} = \min (H,x_s)$ is a cooling length, $\rho = \Sigma/(\sqrt{2 \pi} H)$ is the midplane density, $\gamma$ is the adiabatic index, and $C_\mathrm{V}$ is the heat capacity at constant volume.
In this regime non-isothermal effects may slow down or even invert migration, an effect that is not included in our prescription.
As long as no torque densities in this regime are available, we recommend applying the torque calculated from the torque formula in \citet{Paardekooper2011b} to the planet (while still applying the high mass torque density to the disc) as long as the departure from isothermality is strong. This violates conservation of angular momentum, but the effect at low masses is moderate.

In order to prevent a discontinuity in the accretion rate when transitioning from the Bondi regime to the Hill regime, we apply a smoothing in $M_p$. We set the effective accretion rate $\dot{M}_{\mathrm{eff},i}$ in each grid cell to
\begin{equation}\label{eq_smooth}
    \dot{M}_{\mathrm{eff},i} = \dot{M}_i \times
    \begin{cases}
        1 - 0.25 \left( 0.5 \tanh{\left\{ \frac{\left( M_p - M_t \right)}{M_t/3}\right\}} + 0.5 \right)
        & M_p < M_t \\
        0.25 + 0.75 \left( 0.5 \tanh{\left\{ \frac{\left( M_p - M_t \right)}{M_t/3}\right\}} + 0.5 \right)
         & M_p \ge M_t,
    \end{cases}
\end{equation}
where $\dot{M_i}$ is the accretion rate in the $i$-th grid cell obtained from Eq.~\ref{eq_mdot} and $M_t$ is given in Eq.~\ref{eq_mt}.
We note that the function given in Eq.~\ref{eq_smooth} is not continuous at $M_p = M_t$. This is necessary to obtain a continuous global inverse growth timescale.

In this paper we did not investigate the regime where the disc is self-gravitating. Applying the torque densities we describe in this work for self-gravitating discs would require  further modifications. This is also true for accretion, as we discuss in Sect.~\ref{sect:Discussion}.
\end{appendix}

\end{document}